\documentclass[a4paper, reqno]{amsart}
\usepackage{graphicx}
\usepackage[normalem]{ulem}
\usepackage{color,xcolor,graphicx}
\usepackage{amsmath,amsfonts,amsthm,amssymb}
\usepackage{latexsym,mathtools,csquotes}
\usepackage{amstext}
\usepackage{enumerate}
\usepackage{cancel}
\newcommand\can{\mathrm{can}}
\newcommand\gcan{\mathrm{gcan}}

\newcommand\ir{\mathrm{ir}}

\newcommand{\kb}{{\mathbf k}}

\newcommand\cH{\mathcal{H}}

\renewcommand{\bar}{\overline}
\newcommand{\Bog}{{\operatorname{bg}}}

\renewcommand{\i}{{\operatorname{i}}}
\newcommand{\e}{\operatorname{e}}

\newcommand{\hc}{{\operatorname{h.c.}}}

\newcommand{\ext}{{\operatorname{ext}}}

\newcommand{\s}{{\operatorname{s}}}
\newcommand{\eff}{{\operatorname{eff}}}
\renewcommand{\d}{{\operatorname{d}}}

\newcommand{\Ran}{{\operatorname{Ran}}}

\renewcommand{\Re}{{\operatorname{Re}}}

\newcommand{\zz}{{\mathbb{Z}}}

\newcommand{\cc}{{\mathbb{C}}}
\newcommand{\rr}{{\mathbb{R}}}
\newcommand{\x}{\mathbf{x}}

\newcommand{\y}{\mathbf{y}}
\newcommand{\0}{{\bf 0}}
\newcommand{\kk}{{\bf k}}

\newcommand{\q}{{\bf q}}
\newcommand{\rrr}{{\bf r}}
\newcommand{\pp}{{\bf p}}
\newcommand{\qq}{{\bf q}}
\newcommand{\p}{\mathbf{p}}
\renewcommand{\y}{\mathbf{y}}
\renewcommand{\l}{\mathbf{l}}

\newcommand{\shf}{{\operatorname{shift}}}

\newcommand{\mai}{{\operatorname{main}}}

\def\bbbone{{\mathchoice {\rm 1\mskip-4mu l} {\rm 1\mskip-4mu l}
		{\rm 1\mskip-4.5mu l} {\rm 1\mskip-5mu l}}}
\def\one{\bbbone}

\theoremstyle{plain}
\newtheorem{thm}{Theorem}[section]
\newtheorem{prop}[thm]{Proposition}
\newtheorem{lemma}[thm]{Lemma}
\newtheorem{cor}[thm]{Corollary}

\theoremstyle{definition}

\newtheorem{defn}[thm]{Definition}

\newtheorem{assump}[thm]{Assumption}

\newtheorem{remark}[thm]{Remark}

\usepackage{pgfplots}
\usepackage{tikz}
\usepackage{setspace}
\usepackage{caption}
\usepackage{subcaption}
\usetikzlibrary{matrix}
\pgfplotsset{width=11cm,compat=1.9}
\usepgfplotslibrary{external}
\usetikzlibrary{hobby}
\usetikzlibrary{mindmap}
\usetikzlibrary{shadows.blur}

\pgfplotsset{compat=newest}

\usetikzlibrary{patterns}
\usetikzlibrary{arrows.meta,positioning}

\newcommand\beq{\begin{equation}}
	\newcommand\eeq{\end{equation}}
\renewcommand\k{\mathbf{k}}
\renewcommand\p{\mathbf{p}}
\renewcommand\q{\mathbf{q}}
\newcommand{\ben}{\begin{arabicenumerate}}  
	\newcommand{\een}{\end{arabicenumerate}}
\newcounter{smallarabics}
\newenvironment{arabicenumerate}
{\begin{list}{{\normalfont\textrm{(\arabic{smallarabics})}}}
		{\usecounter{smallarabics}\setlength{\itemindent}{0cm}
			\setlength{\leftmargin}{5ex}\setlength{\labelwidth}{4ex}
			\setlength{\topsep}{0.75\parsep}\setlength{\partopsep}{0ex}
			\setlength{\itemsep}{0ex}}}
	{\end{list}}

\numberwithin{equation}{section}
\numberwithin{figure}{section}

\title[Zero]{Phonon Dispersion Relation in Weakly Coupled Bose Gas Beyond Bogoliubov Approximation}

\author{J. Derezi\'{n}ski}
\author{L. Pettinari}
\address[J. Derezi\'{n}ski]
{Dept. of Math. Methods in Phys., University of Warsaw\\ 
	Pasteura 5, 02-093 Warszawa, Poland} 
\email{Jan.Derezinski@fuw.edu.pl}
\address[L. Pettinari]{Dipartimento di Matematica, Universit\`a di Trento and INFN-TIFPA and INdAM, Via Sommarive 14, I-38123 Povo, Italy}
\email{lorenzo.pettinari@unitn.it}

\date{\today}

\begin{document}
	
	\begin{abstract}
We study the quasiparticle dispersion relation for a homogeneous Bose gas
in a large cubic box with periodic boundary conditions at zero
temperature. Inspired by the Bogoliubov method, we rearrange the usual
Bose gas Hamiltonian in a form
adapted to perturbation theory in the weak-coupling regime, both in
the canonical and grand-canonical setting. We compute 
the first non-trivial corrections to the dispersion relation in both
settings. We show that for small momenta  they are linear in the
absolute value of the  momentum. Thus in this approximation 
phonons have a well-defined velocity, which we compute.
	\end{abstract}

	\maketitle
	\tableofcontents
	
\section{Introduction}
Modern developments in the theory of Bose gases largely depend on the
original argument proposed by Bogoliubov in 1947 \cite{B_47}. Within
his approach, the Bose gas is approximated by a gas of non-interacting
quasiparticles, whose {\em dispersion relation} is given by the famous formula\footnote{We will omit most of the time the dependence on $\nu$ in $\omega_\Bog(\k,\nu)$ and restore it when relevant to the discussion.}
\begin{equation}\label{dispe}
  \omega_\Bog(\k) \equiv\omega_{\Bog}(\k,\nu)
 : = \sqrt{\frac{|\k|^4}{4}+ \nu\frac{\hat{v}(\k)}{\hat{v}(\0)}|\k|^2}.
\end{equation}
 Here, $\hat v(\k)$ is the Fourier transform of the interaction
  potential $v(\x)$ and $\nu$ is the parameter involved in  the so-called ``c-number substitution''.
\eqref{dispe}  predicts a linear dispersion relation for
{\em phonons},  quasiparticles with  low momentum $|\k|$. For larger momenta, for some shapes of $\hat v$,  we have a
local maximum and next a local minimum---traditionally the corresponding quasiparticles
are called {\em maxons}, resp. {\em rotons}. It is remarkable that this picture,
at least qualitatively, is validated by 
experiments with ${}^4\mathrm{He}$ \cite{GBSKDFO_2021,BS_2018} and
alkali gases \cite{VXRAK_2002}.
Neutron scattering (for larger momenta) and sound
propagation (for small momenta) in ${}^4\mathrm{He}$ allow us to measure its  quasiparticle
dispersion relation, which turns out to be a rather well-defined curve
similar to its theoretical prediction. Experiments show that this 
relation is remarkably sharp.  In ${}^4\mathrm{He}$ we even see maxons and
rotons; in alkali gases they are typically absent.

 In the strict Bogoliubov approximation the quasiparticle dispersion relation has
a precise mathematical meaning: the {\em Bogoliubov Hamiltonian} can be
understood as a system of independent harmonic oscillators with energy
described by \eqref{dispe}. The Bogoliubov Hamiltonian has a
well-defined thermodynamic limit, and in this limit
$\k\mapsto \omega_\Bog(\k)$ 
describes a singular shell in the energy-momentum spectrum.

Clearly, the dispersion relation \eqref{dispe} is only an
approximation to the true one. Mathematically, this is due 
to two factors:
\begin{enumerate}\item To obtain the Bogoliubov Hamiltonian one needs
  to discard various terms from the full Hamiltonian.
  \item It is not obvious how to fix the parameter $\nu$, which
    appears in the Bogoliubov dispersion relation.
    \end{enumerate}
In our paper  we  compute leading corrections to the
dispersion relation for weak potentials. More precisely, we
 replace $v$ with $\kappa v$, obtaining a perturbative expansion
of the energy correction with the small parameter $\sqrt{\kappa}$.
We compute the lowest order correction, which  has the order $\kappa^1$.
 We  prove that
 for low momenta this correction is asymptotically linear.
 Thus  in this approximation
phonons have a well defined velocity, which we
 compute.

The first problem that we encounter is the definition
of the dispersion relation itself. For the full system  in finite volume
it is ill defined---it
is expected to  emerge only in
thermodynamic limit. 
Unfortunately, 
we do not know how to interpret the full $N$-body
Hamiltonian in thermodynamic limit if we keep the density positive.
In the traditional approach of theoretical physics this problem is
resolved by considering the Hamiltonian in a finite box of size $L$
with periodic boundary conditions. Then one considers its {\em Green
functions} (time-ordered correlation functions of some observables) and takes their limit as
$L\to\infty$. One expects to see the quasiparticle  dispersion
relation in singularities of the  Green functions   
analytically continued in
energy-momentum. This approach was pioneered by
Beliaev \cite{B_58_1,B_58}, who computed the correction to the dispersion relation in the
dilute limit.

In our paper we do not introduce Green functions. Instead, we consider 
the lowest order perturbation theory for   one-quasiparticle states in 
fixed momentum sectors in thermodynamic limit. As explained in \cite{DP_2026}, this is 
consistent with the approach based on Green functions.

Already in the lowest nontrivial order of perturbation theory
the shift in the dispersion relation has  a small nonzero  imaginary
contribution. Thus quasiparticles are unstable, and the measured dispersion
relation should have a nonzero breadth.
The imaginary shift of the dispersion relation  was
first computed by Beliaev in \cite{B_58}, and  recently confirmed  by one of
the authors (JD), Napi\'orkowski and Ben Li \cite{DLN_2023}.

In this paper we are interested in
the real part of the shift of dispersion relation. We will consider
it in two settings: {\em canonical} and
{\em grand-canonical}. We will treat the canonical setting as the main one of
our paper, even though computations in the grand-canonical setting  seem to 
be cleaner.  However, interestingly, the final result is
simpler in the canonical setting.

In the canonical formalism, we fix the density  $n$ and we consider a
sequence of systems with $n=\lim\limits_{L\to\infty}\frac{N}{L^d}$, where
$N=N(L)$ is the number of particles and $L^d$ is the volume of the box with
periodic boundary conditions. The parameter in the initial Bogoliubov
approximation is taken to be
		\begin{align}
			\nu &= \kappa n\hat{v}(\0)=\kappa \lim_{L\to+\infty} \frac{N\hat{v}(\0)}{L^d} .
		\end{align}

                In the grand canonical formalism, the number operator $N^L$ is not a constant times the identity, and the main parameter
                is $\mu$, the {\em chemical potential}.
In the Bogoliubov Hamiltonian the role of  the parameter $\nu$ is
taken by $\mu_0$ defined as
\begin{equation}
		\mu_0 = \kappa n_{0,\Bog}\hat{v}(\0)=\kappa\left\langle\frac{N_\0}{L^d}\right\rangle_\Bog \hat{v}(\0) ,
		\end{equation}
		where $\langle\cdot\rangle_\Bog$ is the mean value in
                the  Bogoliubov vacuum and $N_\0$ is the zero-mode
                number operator.  In order to
                compare the two approaches we use the identification
\begin{align}			\nu &=\kappa n_\Bog\hat v(\0)=\kappa \left\langle
                                      \frac{N^L}{L^d}\right \rangle_\Bog
                                      \hat{v}(\0) .
\end{align}
The chemical potential $\mu$ can be approximately computed from $\mu_0$.
Thus we have three distinct parameters: $\nu$, $\mu$, $\mu_0$: they
have the same leading value, but differ by a term of order $\kappa$.

One of the essential ingredients in Bogoliubov's paper is the
following idea.
 When we decompose the total number of particles into
the zero-mode and excited modes
$$
N^L=N_\0+N^>,
\qquad
N^>=\sum_{\p\neq\0}a_\p^*a_\p,
$$
we expect that $N_\0$ is macroscopic, whereas $N^>$ is of  lower order.
Following the paper of Bogoliubov, this is usually implemented by using the
{\em \textit{c}-number substitution}, where the 
zero-mode operators $a_\0, a_\0^*$ are replaced by the
operator $\sqrt{N_\0}$, which is then taken to be approximately constant. This procedure is based on the idea that, if the gas is in a condensate state, the zero-mode should be macroscopically occupied, and commutators involving $a_\0,\; a_\0^*$ should be negligible in the thermodynamic limit $N,L^d\to +\infty$.
In our paper we never use a direct $c$-number substitution. Instead,
we  treat the Bogoliubov Hamiltonian as the
unperturbed part of the full Hamiltonian.
This is done in a different way in the canonical and grand-canonical settings.

In our analysis we impose appropriate regularity hypotheses on the
                interaction potential $v$ and the parameter $\nu$, see Assumptions \ref{ass:
                  primary assumption on the potential} and \ref{ass:
                  secondary hypotheses}.
        Besides, when we justify discarding some terms from the
        Hamiltonian in the canonical setting we 
        assume the positivity of the
        condensate density $n_{0,\Bog}>0$,
(which holds automatically for a small coupling constant $\kappa$) see
Assumption \ref{ass: condensate hypotheses}.

Let us sketch our computations of the correction to the dispersion
relation in
  the canonical setting.
\begin{enumerate}
\item First we embed the physical space
with exactly $N$ particles in a larger space where the zero-mode is
allowed to take negative values. This trick goes back to Arnowitt and Girardeau
\cite{AG_59}; see also Takano \cite{T_61}. By separating the zero-mode,
one identifies the $N$-particle space with a truncated Fock space of
excitations. This approach was used by Derezi\'nski and
Napi\'orkowski~\cite{DN_2014}, and in a closely related form by Lewin,
Nam, Serfaty, and Solovej~\cite{LNSS_2015}. It allows us to use the
formalism of 2nd quantization and Fock spaces.
	\item
	We decompose the finite-volume interaction into its diagonal
	and off-diagonal parts, $V_{\text{diag}}$ and $V_{\text{off}}$;
	see Sections \ref{Separating the condensate}
	and \ref{sec: off-diagonal}. The diagonal part depends on the zero
	mode only through $N_\0$, whereas the off-diagonal part contains
	$a_\0$ and $a_\0^*$ separately. We treat the latter using the
	extended-space formalism of Section~\ref{sec: extended space},
	obtaining
	\begin{align}\label{qwer}
		H^L
		=&H_\Bog^L+\sqrt{\kappa}H_3^L
		+\kappa H_2^L+\kappa\sum_{i=1}^3H_{4,i}^L
		+\kappa\sqrt{\kappa}H_{3,x}^L\\
		&+\kappa R_{4,1}^L+\kappa R_{4,2}^L.
	\end{align}
	Here $\kappa$ is introduced by rescaling
	$\hat v\to\kappa\hat v$ while keeping $\nu$ fixed.
	The operators $H_\Bog^L$, $H_2^L$, $H_3^L$, and $H_{4,i}^L$
	are independent of $\kappa$; $H_{3,x}^L$ retains a further
	$\kappa$-dependence. Apart from the remainders $R_{4,i}^L$,
	these operators act on the excitation Fock space, made over
        the space of nonzero modes.
	\item
	A {\em Bogoliubov transformation} diagonalizes the quadratic
        Bogoliubov Hamiltonian
	$H_\Bog^L$. Its ground state is $\Omega_\Bog$, and its one-quasiparticle
	states $b_\k^*\Omega_\Bog$ have excitation energy $\omega_\Bog(\k)$.
	
	We use $H_\Bog^L$ as the unperturbed Hamiltonian in
the	Rayleigh--Schr\"odinger perturbation theory
	\cite[Chap. XII]{RS_78}. At the first nontrivial order, $\kappa$,
	the dispersion correction consists of two contributions:	
	The quartic terms gives the {\em Hellmann-Feynman term}
	\begin{align}
		\delta_4^{\can,L}(\k,\nu,\hat v)
		:={}&(b_\k^*\Omega_\Bog|
		H_{4}^L b_\k^*\Omega_\Bog)
		-(\Omega_\Bog|H_{4}^L\Omega_\Bog)
		\notag\\
		={}&(\Omega_\Bog|
		[b_\k,[H_{4}^L,b_\k^*]]\Omega_\Bog).
		\label{def: HF term}
	\end{align}
	Subtracting the ground-state correction removes the common extensive
	contribution. In the thermodynamic limit, we set
	\begin{equation}
		\delta_4^\can(\k,\nu,\hat v)
		:=\lim_{L\to+\infty}
		\delta_4^{\can,L}(\k,\nu,\hat v).
	\end{equation}
        The cubic term yields the  {\em Fermi golden rule contribution}. It has a
        real and imaginary part. 
	Its imaginary part was computed in \cite{DP_2026}; here we study
	\begin{align}
		\delta_3^\can(\k,\nu,\hat v)
		:={}&\lim_{\epsilon\searrow0}\lim_{L\to\infty}\delta_3^{\can,L}(\k,\nu,\hat v,\epsilon),\\
		\delta_3^{\can,L}(\k,\nu,\hat v,\epsilon)
		:={}&-\big(H_3^L b_\k^*\Omega_\Bog\big|
		(H_\Bog^L-E_\Bog^L-\omega_\Bog(\k)-\i\epsilon)^{-1}
		H_3^L b_\k^*\Omega_\Bog\big)
		\notag\\
		&+\big(H_3^L\Omega_\Bog\big|
		(H_\Bog^L-E_\Bog^L-\i\epsilon)^{-1}
		H_3^L\Omega_\Bog\big),
		\label{def: Fermi Golden}
	\end{align}
	where $E_\Bog^L$ is the ground-state energy of $H_\Bog^L$.
	This expression also admits a commutator representation;
	see \cite[Sec. 2.10]{DP_2026} and \eqref{forri}.
         The full 
        correction is 
        the sum: 
        \beq 
        \delta^\can(\k,\nu,\hat{v}):=\delta_4^\can(\k,\nu,\hat{v})+\delta_3^\can(\k,\nu,\hat{v}).\eeq 
	
	\item
Subsection~\ref{sec: irrelevant terms}, and Section \ref{Tail estimates},
	 justify discarding the remaining terms.
	The contributions of $R_{4,i}^L$, $H_2^L$, $H_{4,1}^L$,
	and $H_{4,3}^L$ to the order-$\kappa$ excitation-energy shift
	vanish in the thermodynamic limit, while
	$\kappa\sqrt{\kappa}H_{3,x}^L$ contributes only at higher orders.

	\item
          We compute
          the  asymptotics as $\frac{|\k|}{\sqrt{\nu}}\to0$
	of $\delta_4^\can(\k,\nu,\hat v)$ 
	and $\delta_3^\can(\k,\nu,\hat v)$. Adding them up we obtain
        the final result.
      \end{enumerate}

The first main result of our paper is the computation of the phonon
velocity correction in the canonical setting:

\begin{thm}\label{thm: shift intro}
		In the canonical formalism, the dispersion relation 
                computed up to the order $\kappa$ has the following
                low momentum
                behaviour:
                \begin{align}
			&\omega_{\Bog}(\k)+ \kappa\delta^\can(\k,\nu,\hat{v}) \\=\,& |\k|c_\mathrm{can}(\nu,\hat{v})\left(1+o\left(\left(|\k|\nu^{-\frac{1}{2}}\right)^0\right)\right) \notag 
			\text{ as }\frac{|\k|}{\sqrt{\nu}}\to 0,
                                                                        \end{align} where
	\begin{align}
		&	c_\mathrm{can}(\nu,\hat v) =
		\sqrt{\nu} - \kappa \frac{\hat{v}(\0)\sqrt{\nu}}{16}\int\frac{\d \p}{(2\pi)^3}\frac{\hat v(\p)^2}{\hat{v}(\0)^2}\frac{|\p|^4}{\omega_{\Bog}(\p)^3} . \label{eq: cbel.} 
	\end{align} 
	\end{thm} 
			
	The computations in the grand-canonical setting are  a little different:
        \begin{enumerate}
          \item We work in the 2nd quantized formalism from the beginning, without fixing the particle number. The basic object is
the {\em grand-canonical Hamiltonian} $H^L-\mu N^L$, where $N^L$ is the number operator.
 As already mentioned, the basic parameter is the chemical
  potential $\mu$.
  \item
        We separate the macroscopic zero-mode occupation from its
		fluctuations using the displacement operator of the
                zero-mode $W_\alpha$:
		\[
		W_\alpha^*a_\0 W_\alpha=a_\0-\alpha,
		\qquad
		\alpha=\sqrt{\frac{\mu_0L^d}{\kappa\hat v(\0)}}.
		\]
		The corresponding Bogoliubov state $\Omega_{\Bog}(\mu_0)$ satisfies
		$(a_\0-\alpha)\Omega_{\Bog}(\mu_0)=0$, so that its condensate density is
		$n_{0,\Bog}=\frac{\mu_0}{\kappa\hat v(\0)}$.
	\item	
		At fixed chemical potential $\mu$, we choose $\mu_0$
                by approximately minimizing
		the expectation of $H^L-\mu N^L$ in $\Omega_{\Bog}(\mu_0)$.
		The stationarity condition
		\[
		\frac{\partial}{\partial\mu_0}
		\big(\Omega_{\Bog}(\mu_0)\big|
		(H^L-\mu N^L)\Omega_{\Bog}(\mu_0)\big)=0
		\]
		gives
		\beq
		\mu_0=\mu-\kappa\mu_1+o(\kappa).
		\label{mu0}\eeq
		The resulting term $-\kappa\mu_1N^>$ is essential for the cancellation
		of the infrared singularities in the perturbative correction to
		the dispersion relation.
              \item
We decompose the grand-canonical Hamiltonian, obtaining a formula
similar to \eqref{qwer}, where now the terms $R_{4,i}^L$ and
$H_{3,x}^L$ are absent. We compute the shift---the computations are
almost identical as in the canonical setting, leading to
        \begin{equation}
        \delta^\gcan(\k,\mu,\hat{v}):=\delta_4^\gcan(\k,\mu,\hat{v})+\delta_3^\gcan(\k,\mu,\hat{v}).\end{equation}

                \item
		Finally, we adjust $\mu_0$ to reproduce the prescribed density.
		At the order considered here, the density relation is
		\[
		n=n_\Bog = n_{0,\Bog}+n_\Bog^>,
		\qquad
		\nu=\kappa\hat v(\0)n
		=\mu_0+\kappa\hat v(\0)n_\Bog^>.
		\]
              Thus
we have a relationship between $\nu$, $\mu_0$ and
  $\mu$.
                After expressing both the Bogoliubov dispersion and its perturbative
		correction in terms of $\nu$, the coefficient $c_\gcan$ agrees with
		$c_\can$ through order $\kappa$.
\end{enumerate}

The second main result of our paper yields the phonon
velocity correction in the grand-canonical setting:

	\begin{thm}\label{thm: shift intro gcan}
          In the grand-canonical formalism,
the dispersion relation 
                computed up to the order $\kappa$ has the following 
                behaviour:
\begin{align}
		&	\omega_{\Bog}(\k,\mu_0)+ \kappa\delta_4^\gcan(\k,\mu,\hat{v})+\kappa\delta_3^\gcan(\k,\mu,\hat{v})\\ = \,&|\k|c_\mathrm{gcan}(\mu,\hat{v})\left(1+o\left(\left(|\k|\mu^{-\frac{1}{2}}\right)^0\right)\right) \notag 
			\text{ as }\frac{|\k|}{\sqrt{\mu}}\to 0,
                                                                              \end{align}
                                                                              where
                                                                              $\mu_0$
                                                                              is
                                                                              computed
                                                                              from
                                                                              $\mu$
                                                                              as
                                                                              in
                                                                              \eqref{mu0} and
	\begin{align}
			c_\mathrm{gcan}(\mu,\hat v)& =
		\sqrt{\mu_0} - \frac{\kappa
           \hat{v}(\0)\sqrt{\mu_0}}{16}\int\frac{\d
           \p}{(2\pi)^3}\frac{\hat
           v(\p)^2}{\hat{v}(\0)^2}\frac{|\p|^4}{\omega_{\Bog}(\p,\mu_0)^3}  \label{eq:
           cbel..}\\
                &+\frac{\kappa\hat v(\0)}{4\sqrt{\mu_0}} \int \frac{\d \p}{(2\pi)^3}
                  \frac{\big(\frac{|\p|^2}{2}+\frac{\hat v(\p)}{\hat
                  v(\0)}\mu_0-\omega_{\Bog}(\p,\mu_0)\big)}{\omega_{\Bog}(\p,\mu_0)}.
	\end{align} 
	\end{thm} 
Let us stress that 
the grand-canonical and canonical pictures agree up
        to second order perturbation theory in $\sqrt{\kappa}$
for the phonon velocity:
        \begin{equation}
        c_\can(\nu,\hat{v}) = c_\gcan(\mu,\hat{v})(1+o(\kappa)).\end{equation}
        Note that dispersion relation is a
        much finer object than extensive quantities, for which the
        usual statement of the equivalence of the two pictures is
        usually stated.

Let us remark that when perturbation computations are performed in a naive
way, then one obtains  infrared-divergent terms of the form $\frac{1}{|\k|}$ in the small-momentum limit $\frac{|\k|}{\sqrt{\nu}}\to 0$.
This was noted  e.g. in \cite{DLN_2023,DP_2026}, where the imaginary
part of the shift was computed, which does not require such a careful treatment.
This point was already discussed in some of the older physics
literature \cite{B_58_1,B_58,GN_64,HP_59}, where it was argued that
interacting particles always exhibit a depletion of the condensate
number, so that the substitution $N_\0 \simeq \one N$ is not justified
and one omits many important contributions arising from disconnected diagrams \cite{HP_59}.

In our computations  the shifts $\delta_3^\can$ and 
$\delta_4^\can$, as well as $\delta_3^\gcan$ and 
$\delta_4^\gcan$, are given by  relatively complicated expressions.
Besides, all of them are infrared singular: they behave like
$O(k^{-1})$. This infrared blowup cancels when we add them to obtain
$\delta^\can$ and $\delta^\gcan$. Besides, the formulas miraculously
simplify, especially in the canonical approach.

In our paper we use the weak coupling approximation, because it has a
rather clean mathematical meaning.
Unfortunately, the weak coupling regime is usually a poor approximation of realistic
quantum systems---e.g. their interactions  typically possess a hard
core. For alkali gases it is more appropriate to use the {\em dilute
limit}. This is one of the reasons why this limit
is more commonly used by both  the theoretical and mathematical physics
 communities. Note, however, that the dilute limit is not a
good approximation for liquid helium. In particular, it does not
lead to maxons and rotons, which can be visible in the weak coupling approximation.

The dilute limit is controlled by the density of the gas $n$ and the
\textit{scattering length} $\mathfrak{a}$, where we take
$n\mathfrak{a}^3\to 0$.
 For weak coupling, $\mathfrak{a}$ can be expanded in a
  perturbative series $\sum_{n\geq 0}\mathfrak{a}_n\kappa^{n+1}$,
  where e.g. $\mathfrak{a}_0=\frac{\hat v(\0)}{4\pi}$.
  Using this we can recast the coefficient $c_\mathrm{can}$ in a rather elegant form: \begin{cor}\label{cor: dilute limit}
		In the limit $\kappa\to 0$, $\nu = \kappa n\hat v(\0) \to 0$ the following estimate holds \begin{equation}\label{eq: dilute limit}
			c_\mathrm{can}( \nu,\hat{v}) =  \sqrt{4\pi n_{0,\Bog}\mathfrak{a}}\left(1+ \frac{7}{6\pi^2}\sqrt{n_{0,\Bog}(4\pi\mathfrak{a})^3}\right)(1+o(\kappa^0)),\text{ as }\kappa\to 0.
		\end{equation}
	\end{cor} Clearly, since $c_\can = c_\gcan(1+o(\kappa))$, an analogous corollary holds also in the grand-canonical framework.
	The main term in Equation \eqref{eq: dilute limit} is
        universal in the potential and recovers the well known
        expression first computed by Beliaev \cite{B_58}.

        There exists a large mathematical literature about the Bose gas.
A rigorous justification of Bogoliubov’s \textit{c}–number substitution
in the mean-field setting was first established by Seiringer
\cite{S_2011}; see also \cite{LNSS_2015,DN_2014,GS2013,NS_2015} for
related developments. Corresponding results in the Gross–Pitaevskii
regime were subsequently obtained in \cite{BBCS_2019,BSS2022,NT_2023},
and extensions beyond this regime appear in \cite{BCS2022}.
        Many of the results  concerning the bottom of the
                excitation spectrum beyond Bogoliubov theory have
                recently been achieved in the Gross–Pitaevskii regime
                \cite{COSS_2025}. In the mean-field regime, analogous
                analysis was carried out in \cite{BPS2021,NN_2017}.
However, almost none of the rigorous results concern quasiparticle
dispersion relations. This is not very surprising, since the concept of
quasiparticle dispersion relation is mathematically difficult.

On the other hand, quasiparticles are physically important and visible
in experiments. Therefore, in the theoretical physics literature, they
were considered very early. The series of two remarkable papers by
Beliaev \cite{B_58_1,B_58}, who computed the shift of the dispersion relation at zero
temperature in the dilute limit, dates back to 1958. Beliaev used the
Green functions approach in the canonical setting, organizing his
computations with the help of diagrams (which were akin to, but different
from, Feynman diagrams). The grand-canonical approach has been
introduced to the Bose gas by Hugenholtz and
Pines~\cite[Sec.~7]{HP_59}. Beliaev's result was rederived with
similar methods  in 
 later papers  \cite{HG_98,MS_59}.

There exist a few papers that assume from the beginning that the
interaction is given by a hard core
potential and apply a method of
 \textit{non-Hermitian Hamiltonians}
\cite{LY_58,MM_60,MS_59,T_59}. In this setting, Mohling and Sirlin
\cite{MS_59} decomposed the interaction into \textit{diagonal} and
\textit{off-diagonal} parts, separating terms involving occupation
numbers from those involving creation and annihilation operators
individually.

One can try to compute the energy-momentum spectrum of Bose
gas using variational methods and quasifree states. This was first
done by Arnowitt-Girardeau 
\cite{AG_59}; see also \cite{T_61,CDZ_2009}.
Unfortunately, this method yields an (unphysical) gap in the
spectrum; in particular, in this method it is difficult to determine the
velocity of low momentum phonons.

As we mentioned above,  quasiparticles are expected to be visible only
in thermodynamic limit. Unfortunately, this limit is difficult to
control rigorously. An attempt to study 
                the energy–momentum spectrum in simultaneous large
                volume and mean field limit was made in \cite{DN_2014}.
The imaginary part of dispersion relation in the weak coupling regime
was studied in \cite{DLN_2023}, rederiving essentially the result of
Beliaev. These results concerned zero temperature.  Positive
temperatures pose additional conceptual difficulties. The results of
\cite{DLN_2023} were extended to low positive temperatures by the
authors of this manuscript in
\cite{DP_2026} with the help of the Araki--Woods representations.

The Bose gas was also studied using methods of
	constructive quantum field theory. In
	\cite{Benfatto_1995,Benfatto_1996}, Benfatto combined a functional
	integral representation with renormalization group methods to
        study Green functions of
	a weakly interacting three-dimensional Bose gas at zero temperature;
	see also \cite{BG_1995}. At the level of renormalized perturbation
	theory, his analysis recovered the phonon behaviour of the infrared
	two-point function, with the sound velocity whose leading term agreed
	with the Bogoliubov prediction. The formulation was grand-canonical:
	the condensate density was prescribed, while the chemical potential
was adjusted through a renormalization condition related to the
	Hugenholtz--Pines identity \cite{HP_59}. In our notation, the
	identification $\mu=\mu_0=\nu$ therefore applied only at the leading
	order; the corrections to these parameters must be distinguished
	beyond the Bogoliubov approximation.

A different strategy, based on techniques from quantum field theory, was developed by Galanda and Pinamonti in \cite{GP2025,GP2026}, where equilibrium states for an interacting Bose gas are constructed via a suitable resummation of perturbative expansions.

		$C^*$-algebraic methods have been proposed for an interacting
                Bose gas by Buchholz and collaborators in
                \cite{B_2020,BY_2025}. It is conceivable that the
                results about dispersion relations that we present in
                our paper could be recast in a more elegant form using
                those ideas.

		We now describe the organization of the paper.
                Section \ref{sec: c-number} is devoted to the
                derivation of the effective Hamiltonian used to
                compute the  the  shift of the dispersion relation in the
                canonical formalism.
		 Section \ref{sec: computation of the real part 1}
                 contains the computation of this shift, which consists of
                 the Hellmann--Feynman and the Fermi golden rule
                 terms. We also  compute the  phonon velocity and
                 compare it with the result of Beliaev for the dilute
                 limit. In this section computations are done in a
                 formal way---their rigorous justification is deferred
                 to Section \ref{Estimates on perturbative
                   corrections}.
                 	Section \ref{Grand canonical derivation}
                        proposes a different derivation of the shift,
                        based on  the grand-canonical formalism.
                        Section \ref{Tail estimates} deals with the estimates of the
        unphysical "tail terms" which appear in  the
        effective Hamiltonian in the canonical setting. In Appendix \ref{app: born series} we
        recall the definition of the scattering length.


\section{Improved c-number substitution in canonical approach}
\label{sec: c-number}

	
	In this section we will discuss how to reorganize
	the Bose gas Hamiltonian in a way useful for calculations at a
	positive density. 	
This derivation  is based on the canonical formalism
		(large but fixed number of particles). There is a second possible derivation which
		uses the
		grand-canonical formalism (variable number of particles and fixed
		chemical potential).
		It is contained in Section
		\ref{Grand canonical derivation}, and it expands the derivation described e.g. in  \cite{DLN_2023}.

                As a result of our derivation, the full Hamiltonian at
                a fixed density will be the sum of several terms. The
                main term will be the Bogoliubov Hamiltonian depending
                on the parameter $\nu$. This parameter is closely related to
                the chemical potential, and proportional to the
                density of particles.  The whole construction is a
                refined version of what is usually called the
                ``$c$-number substitution''.
	
		We will first discuss the first-principles  
		$N$-body Hamiltonian,
		in Subsections \ref{The first-principles Hamiltonian in
			infinite volume} and \ref{The first-principles
                        Hamiltonian in finite volume}. Then we
                      partition it in various tems	in
		\ref{Separating the condensate}--\ref{sec: irrelevant
                  terms}. At the end we obtain what can be called an
                ``improved $c$-number substitution Hamiltonian''.

		\subsection{The first-principles Hamiltonian in infinite volume}
		\label{The first-principles Hamiltonian in infinite volume}
	As long as possible, we will keep the dimension
                  $d$ of our system arbitrary. However, we are mostly
                  interested in $d=3$, which we will assume in our
                  concrete computations.

 The following are
                our basic hypotheses on the interaction potential  $v$
                for the preliminary analysis of the Hamiltonian. These
                hypotheses are sufficient to ensure the possibility of
                introducing the usual Bogoliubov rotation and writing
                the \textit{first-principle  Hamiltonian} as the
                \textit{improved c-number substitution
                  Hamiltonian}. 
		\begin{assump}\label{ass: primary assumption on the potential} \textbf{(Preliminary hypotheses on the potential)}
			Assume there exists some $\nu_0\in ]0,+\infty[$ such that the potential satisfies the following assumptions for all the possible choices of $\nu \in ] 0, \nu_0 ] $
		\begin{subequations}\label{eq: assumptions on potential}
			\begin{align}
				&\label{eq: continuity of Fourier1} v\in L^1(\mathbb{R}^d,\d \x),\; v(\x)\in\mathbb{R};\\
				&\label{eq: cutoff1} \hat{v}\in L^2(\mathbb{R}^d,\d \k);\\
				&\label{eq: positivity of Fourier1} \hat{v}(\0)>0,\; \hat{v}(\k) > -\hat{v}(\0)\frac{|\k|^2}{4\nu};\\
				& v(\x) = v(-\x), \text{ which implies } \hat{v}(\k) = \hat{v}(-\k). \label{eq: symmetry1} 
			\end{align} 
		\end{subequations}
				\end{assump}
				\begin{remark}
				In the following, when assuming the
                                validity of Assumption \ref{ass: primary assumption on the potential} for a certain $\nu_0>0$, we will \textbf{always take} $\nu\in ]0,
\nu_0]$ without any further mention of this fact.
\end{remark}

We would like to study 	$N$ bosons interacting with potential $v$. In
an infinite volume their   
 Hamiltonian and momentum are
		\begin{align}
			H_N&:=-\frac{1}{2}\sum_{i=1}^{N}\Delta_{i}+\sum_{1\leq i<j\leq N}
			v(\x_{i}-\x_{j}) ,\label{hamiltonian1.}\\
			\mathbf{P}_N&:=-\i\sum_{i=1}^N\nabla_i.
		\end{align}
They		act on 
		the Hilbert
		space
		\beq\label{phys-sp.}
		L^{2}_{\s}\big((\rr^d)^N\big)=\otimes_\s^N L^2(
		\rr^d).\eeq

		One can  also consider an
		arbitrary number of particles, using the formalism of second
		quantization. Then the Hamiltonian, number operator and the momentum are
		\begin{align}
			H&:=\mathop{\oplus}\limits_{N=0}^\infty H_{N} \\
			&=\int \d \x\; a_\x^* \bigl(-\frac12\Delta_\x\bigr) a_\x + \frac12\int\int\d
			\x\d \y\; a_\x^*a_\y^*v(\x-\y)a_\y a_\x,\\
			N&:=\mathop{\oplus}\limits_{N=0}^\infty N=\int \d\x\;a_\x^*a_\x,\\
			\mathbf{P}
			&:=\mathop{\oplus}\limits_{N=0}^\infty \mathbf{P}_N=-\i\int\d\x\; a_\x^*\nabla_\x a_\x
			,\end{align}
		acting on the Fock space
                $\Gamma_\s(L^2(\rr^d))=\mathop\oplus\limits_{
                  N=0}^\infty L^{2}_{\s}\big((\rr^d)^N\big)$.  (Note that we abuse the notation using the same
                  letter $N$ for the number operator and its
                  eigenvalue.) Then, it
                will always be possible to recover the $N$--body
                Hamiltonian by projecting on $\otimes_\s^{ N} L^2(\rr^d)$.
		One can rewrite  these expressions in the momentum representation:
		\begin{align}
			H
			&=\int\d\p\;\frac{1}{2}|\p|^2a^*_{\p}
			a_{\p}\label{proiu}
			+\frac{1}{2(2\pi)^d}\int\d\p\d\q\d\k\; {\hat v}(\k)
			a^*_{\p+\k}a^*_{\q-\k}a_{\q}a_{\p},
			\\
			N&=\int\d\p\; a^*_{\p}a_\p,\nonumber\\
			\mathbf{P}&=\int\d\p\;\p a^*_{\p}a_\p.\nonumber
		\end{align}

		Unfortunately, if we are interested in a positive density of
		particles, these Hamiltonians are not very useful.

		Some  quantum systems   in infinite volume can be elegantly described
		with the help of the $C^*$-algebraic approach. This works well for
		spin systems, sometimes also for fermionic systems: One introduces a 
		$C^*$-algebra describing quasilocal observables equipped
		with a dynamics given by a family of automorphisms. Then one
		tries to consider a state on this $C^*$-algebra which describes a positive
		density of particles.
		Unfortunately,  for interacting bosons this is problematic, because
		physically motivated  dynamics often do not preserve our $C^*$-algebra. Maybe,
		this approach can be implemented using $C^*$-algebras introduced by
		Buchholz \cite{B_2020,BY_2025}.
		Anyway,  we will proceed with the
		traditional method: We will  approximate the infinite volume system with a
		sequence of finite volume systems with periodic boundary conditions.

		\subsection{The first-principles Hamiltonian in finite volume}
		\label{The first-principles Hamiltonian in finite volume}

		Let us  consider a family of systems of finite size
		$L$.
		The  space $\rr^d$ is replaced with
		the  the $d$-dimensional cubic box of side length $L$, that is, $$\Lambda_L=\Bigl]-\frac{L}{2},\frac{L}{2}\Bigr]^{d}.$$
		In order to preserve the translation invariance, 
		$\Delta$ and $\nabla$
		are replaced with $\Delta^L$ and $\nabla^L$ with periodic boundary
		conditions acting on $L^2(\Lambda_L)$.

		The potential  is replaced by its 
		{\em periodized} version
		\[v^{L}(\x):=\frac{1}{L^d}\sum_{\p\in \Xi_L}\e^{\i\p\x}\hat{v}(\p),\]
		where
		$\Xi_L= \frac{2\pi }{L}\zz^d$ is the momentum lattice 
		corresponding to $\Lambda_L$.
		Note that $v^{L}$ is periodic with respect to the domain $\Lambda_L$
		and, for sufficiently decaying $v$,  $v^{L}(\x)\rightarrow v(\x)$ as $L\to \infty$.
		
		Using $\Delta^L$, $\nabla^L$ and $v^L$ we can define the finite volume
		analogs of $H$,$N$, $\mathbf{P}$, denoted $H^L$,
                $N^L$, $\mathbf{P}^L$ acting on
                $\cH^L:=\Gamma_\s(L^2(\Lambda_L))$, which will be the
                main objects of our study.

		Using the identification  
		$L^2(\Lambda_L)\simeq l^2(\Xi_L)$, we pass to the momentum
		representation.
		The Hamiltonian $H^L$, the number operator $N^L$
		and the momentum $\mathbf{P}^L$ are given
                by \begin{align}
                     \label{2BB}
			H^L 
			&=\sum_{\p\in\Xi_L}\frac{1}{2}|\p|^2a^*_{\p}
			a_{\p}+\frac{1}{2L^d}\sum_{\kk,\pp,\qq\in\Xi_L}\hat v(\qq)
			a^*_{\kk-\qq}a^*_{\pp+\qq}a_{\pp}a_{\kk},\\	
			N^L& = \sum_{\p\in\Xi_L}a_\p^*a_\p, \\
			{\mathbf{P}}^L& = \sum_{\p\in\Xi_L}\p\; a_\p^* a_\p.
		\end{align}

 $n>0$ is a fixed number
                 called the \textit{density of the
          gas}. For $L>0$ we consider a family of systems with the total number of
particles $N(L)$ such that \begin{equation}\label{ass: thermodynamic 
                    limit.}
		n(L):=	\frac{N(L)}{L^d} = n(1+o(L^{0})) ,\qquad \text{ as $L\to+\infty$.}
	\end{equation}Here $N(L)$ is the eigenvalue of the number operator $N^L$.
         Thus we assume that  $N^L=N(L)$ throughout
        our paper, except in the section devoted to the grand-canonical approach.

Actually, instead of the density $n(L)$ and $n$ we will usually
prefer to use the parameter $\nu(L)$, resp. $\nu$:
\beq \nu(L):= n(L)\hat{v}(\0)=\frac{N(L)}{L^d}\hat v(\0),\quad
\nu=n\hat v(\0),\label{nunu}\eeq 
whose meaning is roughly the chemical potential; see Section \ref{Grand canonical derivation}.

	In what follows, we often suppress the label $L$ and use $\nu$, $n$
        for both the finite-volume quantities $\nu(L),\, n(L)$ and their
        thermodynamic limit $\nu,\, n$.
		
		\subsection{Diagonal interaction}
		\label{Separating the condensate}
		

		Let us introduce the operators
		\begin{align}
			N_\pp&:=a_\pp^*a_\pp.\end{align}
		We begin by decomposing the interaction term into a \emph{diagonal} and an \emph{off-diagonal} component:\begin{equation}
			\frac{1}{2L^d}\sum_{\p,\q,\k}\hat{v}(\k) a_{\p+\k}^*a_{\q-\k}^* a_\q a_\p = V_{\text{diag}} + V_{\text{off}},
		\end{equation} where  the diagonal component is given by \begin{align}
			V_{\text{diag}} = \frac{1}{2L^d}\sum_{\p,\q} N_\p N_\q
			\big (\hat{v}(\0) +
			\hat{v}(\q-\p)\big)- \frac{\hat{v}(\0)}{2L^d}
                                                                           \sum_{\p} N_\p(N_\p{ +}1) .
		\end{align} We proceed now to separate the condensate components,
		i.e. the zero momentum mode, { in the diagonal component}.  Introducing the momentum lattice
		outside of the condensate
     		 $\Xi_L^>:=\Xi_L\setminus\{\0\}$,
                 we have  \begin{align}
			V_{\text{diag}} =& \frac{1}{2L^d} \left( \hat{v}(\0)N_\0(N_\0-1) + 2 N_\0 \sum_{\p\in\Xi_L^>} N_\p(\hat{v}(\0)+\hat{v}(\p))\right. \notag \\
			& \left. + \sum_{\p,\q\in\Xi_{L}^>} N_\p N_\q
			\big(\hat{v}(\0)+\hat{v}(\p-\q)\big)-
                   \hat{v}(\0) \sum_{\p\in\Xi_L^>} N_\p(N_\p{ +}1)                                                                                                                                           \right).
		\end{align} In the canonical formalism we can directly substitute the equality
		\begin{equation}\label{eq: no as operator}
			N_\0 = N - \sum_{\k\in\Xi^>_L} N_\k 
		\end{equation} 
		inside this expression, obtaing
		\begin{align}\notag
			V_{\text{diag}} =& \frac{1}{2L^d} \left( \hat{v}(\0)N(N-1)
			-(2N-1)\hat{v}(\0)\sum_{\p\in\Xi_L^>}N_\p
			+ \hat{v}(\0)\sum_{\p,\q\in\Xi_L^>}N_\q N_\p\right.\\
			&+ 2 N \sum_{\p\in\Xi_L^>} N_\p(\hat{v}(\0)+\hat{v}(\p))\notag 
			-2 \sum_{\p,\q\in\Xi^>_L}N_\q N_\p(\hat{v}(\0)+\hat{v}(\p)) \notag \\
			& \left. + \sum_{\p,\q\in\Xi_{L}^>} N_\p N_\q
			\big(\hat{v}(\0)+\hat{v}(\p-\q)\big)-
                   \hat{v}(\0) \sum_{\p\in\Xi_L^>} N_\p(N_\p{ +}1)                                                                                                                                           \right)\\
			=  \frac{1}{2L^d} &\left( \hat{v}(\0)N(N-1) + 2
			N \sum_{\p\in\Xi_L^>} N_\p\hat{v}(\p) \right.\notag \\ & -\left.
			\sum_{\p,\q\in\Xi_L^>}N_\q N_\p\big(2\hat{v}(\p)-\hat{v}(\p-\q)\big)\right)
			-\frac{\hat v(\0)}{2L^d}\sum_{\p\in\Xi_L^>}
                                                                               { N_\p^2
                                                                                 }.\label{notta1}
		\end{align}

		In thermodynamic limit $L\to+\infty$, we expect that the most relevant terms are  the ones containing either a product of $N$ with a
		series over momenta, or the product of two such series.
		Thus the  last termin in line \eqref{notta1} is expected to "vanish" in thermodynamic limit.

		\subsection{Extended space}\label{sec: extended space}
		Recall that
           	 $\Xi_L^>:=\Xi_L\setminus\{\0\}$.  
 By the exponential property of Fock spaces we have the identification
		\begin{equation}
			\Gamma_\s(l^2(\Xi_L))\simeq
			\Gamma_\s(\cc)\otimes\Gamma_\s\Big(l^2\bigl(\Xi_L^>\bigr)\Bigr),
			\label{physicalspace}
		\end{equation}
		where the first factor describes the {\em zeroth mode} treated as the
		{\em condensate}
		and the second
		``excitations outside of the condensate''. 
		Let us embed the space of   zero-modes
		$\Gamma_\s(\cc)=l^2(\{0,1,\dots\})$ in a larger space $l^2(\zz)$. Thus we obtain the {\em extended Hilbert space}
		\begin{equation}
			\mathcal{H}^{\text{ext}}:= l^2(\zz)\otimes \Gamma_\s\Big(l^2\bigl(\Xi_L^>\bigr)\Bigr),
			\label{fullspace}
		\end{equation}
		where in the notation we suppress the dependence on $L$.

                The physical space (\ref{physicalspace}) is spanned by vectors of the
		form $|N_\0)\otimes \Psi^>$, where
		$|N_\0)$ represents $N_\0$  zero-modes  ($N_\0\geq0$)
		and $ \Psi^>$ represents a vector outside the zero-mode.
		The space (\ref{fullspace}) is also spanned by vectors  of this  form, where now 
		the relation $N_\0\geq0$ is not imposed. 
		
		On $\cH^\ext$ we have a self-adjoint operator $N_\0^\ext$ such that
		\begin{align}
		N_\0^\ext|N_\0)\otimes \Psi^>&=N_\0|N_\0)\otimes
                                              \Psi^>.\end{align}
Its spectrum equals $\zz$. We also define
 \begin{align}                                           
   N^>&:=\sum_{\p\in\Xi_L^>}N_\p,\\
N^\ext&:=N^>+N_\0^\ext.
 \end{align}
		If $N\in\zz$, we will write $\cH_N^\ext$ for the subspace of
		$\cH^\ext$ corresponding to $N^\ext=N$.
                
{	For $t\in\rr$ we will use the standard notation $
                |t|_+:=\max(0,t)$. We will write}
		\begin{align} N_\0^+&:=\max(0,N_\0^\ext)=|N_\0^\ext|_+.\end{align}
		Clearly\[N_\0^+\restriction{\Gamma_\s(l^2(\Xi_L))}
		=N_\0^\ext\restriction_{\Gamma_\s(l^2(\Xi_L))}=N_\0,\ \ \ \Gamma_\s(l^2(\Xi_L))=\Ran
		\one_{[0,\infty[}(N_\0^\ext).
		\]

		We have also a unitary operator
		\beq
		U|N_\0)\otimes\Psi^>=|N_\0-1)\otimes\Psi^>.
		\eeq Notice that both $U$ and $U^{\dagger}$ commute with both $a_{\p}$ and $a^{*}_{\p}$ with $\p\neq\0$.  We now define for $\p\neq\0$ the following operator on $\mathcal{H}^{\text{ext}}$:
		\beq \tilde a_{\p}:=a_{\p}U^{*}. 
		\eeq
		Operators $\tilde a_\p$ and $\tilde a_\q^*$ satisfy the same CCR as $a_\p$ and
		$a_\q^*$. However, they do not commute with $N_\0^\ext$:
		\beq (N_\0^\ext-1) \tilde a_\p=\tilde a_\p N_\0^\ext.\eeq
		
		We extend $a_\0$ to $\cH^\ext$ by setting $a_\0|N_\0)=0,$
                $N_\0\leq0$, {and consequently $a_0^*|N_\0)=0$, $N_\0<0$}. Note
		that
		\begin{align} U^* a_\0&=\sqrt{N_\0^+},\\
			U^* N_\0^\ext U&=N_\0^\ext-1,\quad   U^* N^\ext 
			U=N^\ext-1. 
		\end{align}
		Let $H^{\ext}$ be the operator on $\cH^\ext$ given by the same expression as
		\eqref{2BB}, where now the operators $a_\0$,
                $a_\0^*$ have the new
		meaning.

		$H^\ext$ preserves
		the direct sum decomposition
		\beq\cH^\ext=\cdots\oplus \Ran\one_{\{-2\}}(N_\0^\ext)\oplus
		\Ran\one_{\{-1\}}(N_\0^\ext)\oplus \Gamma_\s(l^2(\Xi_L)).\label{summa}\eeq
		On the last
		summand $H^{\ext}$ coincides with the physical Hamiltonian $H^L$.

		It is easy to see that all terms in $H^\ext$ commute with $N^\ext$.

		\subsection{Off-diagonal interaction}\label{sec: off-diagonal}
		
		The off-diagonal part of the interaction  is \begin{align}
			V_{\text{off}} =
                                                               \frac{1}{2L^d}\sum_{\k,\q-\p-\k\in
                                                               \Xi_L^>}{
                                                               \hat v(\k)}a_{\p+\k}^*a_{\q-\k}^* a_\q a_\p.
		\end{align}
		Using the shorthand
		\begin{align}
			A_\pp=A_{-\pp}&=a_\pp a_{-\pp},\\
			A_\pp^*=A_{-\pp}^*&=a_{-\pp}^* a_{\pp}^*,
		\end{align}
		the off diagonal part can be written as \begin{align}\allowdisplaybreaks
			V_{\text{off}} =& \frac{1}{2L^d}\sum_{\p\in\Xi_L^>}\hat{v}(\p)(A_0^* A_\p + \text{h.c.}) \label{sumus0} \\
			& + \frac{1}{L^d}\sum_{\p,\q,\p+\q\in\Xi_L^>} \hat{v}(\p)(a_\0^*a_{\p+\q}^*a_\p a_\q + \text{h.c.})\label{sumus1} \\
			&+\frac{1}{2L^d} \sum_{\p,\q ,\p\pm \q\in \Xi_L^>}\hat{v}(\p-\q)A_{\p}^*  A_{\q} \label{sumus2}\\ 
			&+\frac{1}{2L^d}\sum_{\substack{\p,\q,\k,\p{
                   +}\q,\\ \p+\k, \q-\k,\p+\k-\q \in \Xi_L^>}}\hat{v}(\k)a_{\p+\k}^*a_{\q-\k}^* a_\q a_\p .\label{sumus}
		\end{align} 
		Note new conditions under the sum in \eqref{sumus}:
          $\p+\k$, $\q-\k$, $\p$, $\q$ should be nonzero because the
          zero components are included in \eqref{sumus0} and
          \eqref{sumus1}; the $\p+\q=0$
          terms are included          in \eqref{sumus0} and \eqref{sumus2}.
         The conditions $\p,\q\neq0$ under \eqref{sumus2} exclude terms
         which
         were in \eqref{sumus0} and the $N_\0$ terms of the diagonal
         part. 
		
		Let us use the extended space, setting
		\begin{align}\allowdisplaybreaks
			\tilde A_\p&:=\tilde a_\p\tilde a_{-\p}= U^{*2}A_\p,\quad
			\p\in\Xi_L^>;\\
			\text{and using }\quad                      \sqrt{N_\0^+(N_\0^+ -1)}&=U^{*2}a_0^2=U^{*2}A_0.
		\end{align}
		In the extended space the off-diagonal interaction becomes
		\begin{align}\allowdisplaybreaks
			V_{\text{off}} =&
			\frac{1}{2L^d}\sum_{\p\in\Xi_L^>}\hat{v}(\p)\sqrt{N_\0^+(N_\0^+ -1)} \tilde A_\p + \text{h.c.} \label{notag1} \\
			& + \frac{1}{L^d}\sum_{\p,\q,\p+\q\in\Xi^>_L}
			\hat{v}(\p)\sqrt{N_\0^+}\tilde a_{\p+\q}^*\tilde a_\p \tilde a_\q + \text{h.c.}\label{notag2} \\
			&+\frac{1}{2L^d} \sum_{\p,\q ,\p\pm \q\in
				\Xi_L^>}\hat{v}(\p-\q)\tilde A_{\p}^*  \tilde A_{\q} \notag\\ 
			&+\frac{1}{2L^d}\sum_{\substack{\p,\q,\k,\p{
                   +}\q,\\ \p+\k,
					\q-\k,\p+\k-\q \in
					\Xi_L^>}}\hat{v}(\k)\tilde
			a_{\p+\k}^*\tilde
			a_{\q-\k}^* \tilde
			a_\q \tilde a_\p .
		\end{align}

We		rewrite  the first two lines of the off-diagonal potential  as\begin{align}\allowdisplaybreaks
                                                                                \eqref{notag1} =
	& \frac{1}{2L^d}
   \Big(\sqrt{N_\0^+(N_\0^+-1)}-\sqrt{N_\0^\ext(N_\0^\ext-1)}\Big)
			\sum_{\p\in\Xi_L^>}\hat{v}(\p)\tilde A_\p  + \text{h.c.}\label{w1}\\
+&
\frac{1}{2L^d}
   \Big(\sqrt{N_\0^\ext(N_\0^\ext-1)}-N_\0^\ext+\frac12\Big) 
			\sum_{\p\in\Xi_L^>}\hat{v}(\p)\tilde A_\p  + \text{h.c.}\label{w2}\\
+&\frac{1}{2L^d}\Big(
   N_\0^\ext-\frac12\Big) 
			\sum_{\p\in\Xi_L^>}\hat{v}(\p)\tilde A_\p  +
   \text{h.c.}\label{w3}
                                                                              \end{align} and

			\begin{align}\allowdisplaybreaks
			\eqref{notag2}=                                                                         
			& \frac{1}{L^d}\Big(\sqrt{N_\0^+}-\sqrt{N}\Big)\sum_{\p,\q,\p+\q\in\Xi_L^>}
			\hat{v}(\p)\tilde a_{\p+\q}^*\tilde a_\p
                   \tilde a_\q + \text{h.c.}\label{w4}
                          \\
                                  			&+ \frac{1}{L^d}\sqrt{N}\sum_{\p,\q,\p+\q\in\Xi_L^>}
			\hat{v}(\p)\tilde a_{\p+\q}^*\tilde a_\p \tilde a_\q + \text{h.c.}\label{w6}   		
		\end{align}
                We will see that \eqref{w1}, \eqref{w2}
                are negligible {\ in thermodynamic limit}.
Introducing  the parameter $
			\nu$ \eqref{nunu} we further
                transform
                \begin{align}\allowdisplaybreaks
                                    \eqref{w3}&=
\frac{\nu}{2\hat{v}(\0)}\sum_{\p\in\Xi_L^>}\hat{v}(\p)\tilde A_\p^*  +
                              \text{h.c.}\\
                                    -&\frac{1}{4L^d}
                     \sum_{\p\in\Xi_L^>}\hat{v}(\p)\tilde 
                     A_\p^*  +\text{h.c.}\notag \\
                  -&\frac{1}{2L^d}
                     \sum_{\q\in\Xi^>_L}N_\q\sum_{\p\in\Xi_L^>}\hat{v}(\p)\tilde 
                     A_\p^*  + \text{h.c.}\notag \\
                  \eqref{w4}&=\frac{\sqrt\nu}{L^{\frac{d}{2}}\sqrt{\hat{v}(\0)}}
\Big(\Big|1-\frac{\hat{v}(\0)}{\nu L^d}\sum_{\p\in\Xi_L^>}N_\p\Big|_+^{\frac12}-1\Big)
                              \sum_{\p,\q,\p+\q\in\Xi_L^>}
\hat{v}(\p)\tilde a_{\p+\q}^*\tilde a_\p \tilde a_\q + \text{h.c.}\notag\\
                   \eqref{w6}&=\frac{\sqrt\nu}{L^{\frac{d}{2}}\sqrt{\hat{v}(\0)}}\sum_{\p,\q,\p+\q\in\Xi_L^>}
\hat{v}(\p)\tilde a_{\p+\q}^*\tilde a_\p \tilde a_\q + \text{h.c.}
                  \end{align}

From now on we will omit the tildes from the notation.

		\subsection{Full Hamiltonian}\label{sec: full hamiltonian}
	Let us include also a dimensionless coupling constant $\kappa$. We will rescale
		the potential as $\hat{v}\to \kappa \hat{v}$, with $\kappa$  assumed
		\textit{small but finite} throughout the computation.  So,
                  taking into account of this rescaling, 
		$
			\nu(L)=\kappa n(L)\hat{v}(\0)$.
		
		The full interacting Hamiltonian in the extended space becomes
		\begin{align}
			H^L =& H^L_{\Bog} + \sqrt{\kappa}H^L_{3} +
\kappa H_2^L+			\kappa H^L_{4,2} +\kappa H_{4,1}^L+\kappa
                               H_{4,3}^L+\kappa\sqrt\kappa H_{3,x}^L\\
                             &              
                                                      +\kappa R_{4,1}^L+\kappa R_{4,2}^L, \label{eq: effective Hamiltonian}
\end{align} where we have introduced the  Hamiltonian operators
\begin{align}
			H^L_{\Bog} :=&
                                       \sum_{\p\in\Xi_L^>}\left( \frac{|\p|^2}{2}+ \frac{\nu\hat{v}(\p)}{\hat{v}(\0)}\right)N_\p +\sum_{\p\in\Xi_L^>} \frac{\nu\hat{v}(\p)}{2\hat{v}(\0)}\big(  A_\p^*+ \text{h.c.}\big)\notag\\
                  H_2^L:=&-\frac{1}{4L^d}
                     \sum_{\p\in\Xi_L^>}\hat{v}(\p)  
                     A_\p^*  +\text{h.c.}\notag \\                  
			H^{L}_{3}
                  :=&\frac{1}{L^{\frac{d}{2}}}\sum_{\p,\q,\p+\q\in\Xi_L^>}
                      \sqrt{\frac{\nu}{\hat{v}(\0)}}\hat{v}(\p)\big(  a_{\p+\q}^*  a_\p
                        a_\q + \text{h.c.}\big)\notag\\
                  	H^{L}_{3,x} :=&\frac{\sqrt\nu}{L^{\frac{d}{2}}\sqrt{\hat 
                        v(0)}\kappa}\left(\Big|1-\frac{\kappa\hat{v}(\0)}{\nu L^d}\sum_{\p\in\Xi^>_L}N_\p\Big|_+^{\frac{1}{2}}-1\right) \sum_{\p,\q,\p+\q\in\Xi_L^>} \hat{v}(\p)  a_{\p+\q}^*  a_\p   a_\q + \text{h.c.}\notag\\
			H^L_{4,2} :=&
                                      -\frac{1}{2L^d}\sum_{\q\in\Xi_L^>}N_\q\sum_{\p\in\Xi_L^>} N_\p (\hat{v}(\p)+\hat v(\q)-\hat{v}(\p-\q))\notag \\ 
			& - \frac{1}{2L^d}\sum_{\q\in\Xi_L^>}N_\q\sum_{\p\in\Xi_L^>} \hat{v}(\p)  A_\p^*+\text{h.c.}\notag\\
			&+ \frac{1}{2L^d}\sum_{\p,\q,\p\pm\q\in \Xi^>_L}
			\hat{v}(\p-\q)   A_\p^*  A_\q,\notag\\
			H_{4,1}^L:=&  -\frac{\hat v(\0)}{2L^d}\sum_{\p\in\Xi_L^>}
			N_\p^2\notag\\H_{4,3}^L:=
			&\frac{1}{2L^d}\sum_{\substack{\p,\q,\k,\p+\q,\\ \p+\k,
					\q-\k,\p+\k-\q \in
					\Xi_L^>}}\hat{v}(\k)  a_{\p+\k}^*  a_{\q-\k}^*
                    a_\q   a_\p,\notag
                  \end{align}
                  and the "tail terms"
                  \begin{align}
		R_{4,1}^L:=& \frac{1}{2L^d}
   \Big(\sqrt{N_\0^+(N_\0^+-1)}-\sqrt{N_\0^\ext(N_\0^\ext-1)}\Big)
			\sum_{\p\in\Xi_L^>}\hat{v}(\p)A_\p  + \text{h.c.}\notag\\
R_{4,2}^L:=&
\frac{1}{2L^d}
   \Big(\sqrt{N_\0^\ext(N_\0^\ext-1)}-N_\0^\ext+\frac12\Big) 
			\sum_{\p\in\Xi_L^>}\hat{v}(\p) A_\p  + \text{h.c.}\notag
                \end{align} 
		
			\subsection{Bogoliubov rotation}\label{sec: implementing Bog}
			The Hamiltonian $H_{\Bog}^L$, as well as its thermodynamic limit $H_{\Bog}$,
			will be called the {\em Bogoliubov Hamiltonian} and they will be treated
			as the main part of the full Hamiltonian. It  is purely quadratic,
			therefore it can be diagonalized using an appropriate {\em Bogoliubov
				transformation}.
		
			In fact, consider  $\theta=(\theta_\kk)_{\kk\in \Xi_L^>}$, a sequence of real numbers. Set
			\begin{equation}
				c_\kk:=\cosh\theta_\kk,\ \  s_\kk:=\sinh\theta_\kk.\label{hyper}\end{equation}
			For $\k\in\Xi_L^>$ we 
			make the substitution 
			\begin{eqnarray}
				& a_\kk^*=c_\kk b_\kk^*- s_\kk b_{-\kk},&
				a_\kk=c_\kk b_\kk-s_\kk b_{-\kk}^*.\label{rota}\end{eqnarray}

			We choose the Bogoliubov rotation that  kills double creators and
			annihilators, which amounts to choosing $s_\k$  as 
			\begin{eqnarray}\label{eq: bogo roto}
				s_\kk&=&\operatorname{sign}(\hat{v}(\k))\frac{1}{\sqrt2}\left(\left(1-\left(\frac{\hat
					v(\kk)\frac{\nu}{\hat
						v(\0)}}{\frac12\kk^2+\hat
					v(\kk)\frac{\nu}{\hat
						v(\0)}}\right)^2\right)^{-1/2}-1\right)^{1/2}, \label{sk4}\end{eqnarray} and $c_\k= \sqrt{1+s_\k^2}$.
						Note  that we have
					$s_\kk=s_{-\kk}$ and $c_\kk =c_{-\kk}$.	Another useful relation between $c_\k$ and $s_\k$ is \begin{equation}
						s_\k = \frac{\hat{v}(\k)\nu}{2\omega_\Bog(\k)\hat{v}(\0)}\frac{1}{c_\k}.
					\end{equation}	
								
			Equation \eqref{eq: bogo roto} is well defined on account of Assumption \eqref{eq: positivity of Fourier1}. Morever, Assumption \eqref{eq: continuity of Fourier1} ensures that $\hat{v}\in C_0(\mathbb{R}^d)$. Then, we see that \begin{align}
				&s_\k^2 = O\left(\frac{\hat{v}(\k)^2}{|\k|^4}\right) \text{ as }|\k|\to +\infty,\\
				&s_\k^2 = O\left(\frac{1}{|\k|}\right) \text{ as } |\k|\to 0,
			\end{align} and thus, $s_\k\in \ell^2(\Xi_L^>)$ for $d= 2,3$. This last property continues to hold also in the flat Fourier limit $\hat{v}(\k)=\hat{v}(\0)$. Moreover, $s_\k$ can be also defined for $\k\in\mathbb{R}^d$, and as a consequence of its decay properties at infinity we see that \begin{equation}
			\lim_{L\to+\infty} \frac{1}{L^d}\sum_{\k\in\Xi^>_L}s_\k^2 = \int \frac{\d\k}{(2\pi)^d} \s_\k^2 <+\infty.
			\end{equation}

					After the diagonalization, we obtain
					\begin{align}
			H_{\Bog}^{L}&=E_{\Bog}^{L}
			+\sum_{\kk\in\Xi_L^>} \omega_{\Bog}(\kk) b_\kk^*b_\kk,\label{elem}\\
			\mathbf{P}^{L}&=\sum_{\kk\in\Xi_L^>} \kk b_\kk^*b_\kk,
			\end{align}
			where
			the quasiparticle excitation spectrum is defined as \begin{equation}
				\omega_\Bog(\k) = \sqrt{\frac{|\k|^4}{4}+ \frac{\hat{v}(\k)}{\hat{v}(\0)}\nu |\k|^2}.
			\end{equation} 
		
		The ground state energy is
	\begin{eqnarray}
	E_{\Bog}^{L}&=
	&  -\sum_{\k\in\Xi_L^>}\frac12
	\left(
	\frac12\kk^2+\nu \frac{\hat{v}(\k)}{\hat{v}(\0)}-\omega_{\Bog}(\kk)\right).\label{elem1}\end{eqnarray}
	In the following instead of $H^L_{\Bog}$ in Equation \eqref{elem} we will use the centered version $H^L_{\Bog}-E^L_\Bog$. With some abuse of notation, this will be again denoted by $H^L_{\Bog}$.

		It is convenient to introduce the operators
		\begin{align}
			 M_{\p}&:=b_\pp^*b_\pp,\\
			B_\pp=B_{-\pp}&:=b_\pp b_{-\pp},\\
			B_\pp^*=B_{-\pp}^*&:=b_{-\pp}^* b_{\pp}^*.
		\end{align}
When we express the full Hamiltonian $H^L$ in the Bogoliubov
		operators, the following identities are  useful:
{		\begin{align}
			&N_\p= s_\p^2 + c_\p^2M_\p+s_\p^2M_{-\p} -s_\p c_\p(B_\p + B_\p^*),\label{eq: identity bog number}\\
			&\label{eq: identity bog quadratic } A_\p=
                   -s_\p c_\p -s_\p c_\p (M_\p+M_{-\p})
                   + c_\p^2B_\p +s_\p^2 B_\p^*;
		\end{align}
                \begin{align}
                  [N_{\p},b_\k^*]&=c_\p^2 
                                 b_\k^*\delta_{\p\k}+s_\p^2b_\k^*\delta_{-\p\k}
                                 -s_\p c_\p 
                                    b_{-\k}(\delta_{\p\k}+\delta_{-\p\k}),\label{comm1}\\
                  [A_{\p},b_\k^*]&=-s_\p c_\p 
                                 b_\k^*\delta_{\p\k}-s_\p c_\p b_\k^*\delta_{-\p\k}
                                 +c_\p ^2
                                    b_{-\k}(\delta_{\p\k}+\delta_{-\p\k});\label{comm2}\end{align}
                \begin{align}
[b_\k,                  [N_{\p},b_\k^*]]&=c_\p^2 
  \delta_{\p\k}+s_\p^2\delta_{-\p\k},\label{commu1}\\
       [b_\k,           [A_{\p},b_\k^*]]&=-s_\p c_\p 
(                               \delta_{\p\k}+\delta_{-\p\k})
.\label{commu2}\end{align}

              }

The ground state of the Bogoliubov Hamiltonian  $H_\Bog^L$ in the extended
                    space in the $N$-particle sector is
                    \begin{equation}
		\Omega_\Bog = U_\theta |N) \otimes \Omega^> = \prod_{\p\in\Xi^>_L}\frac{1}{(\cosh\theta_\p)^{\frac12}}\e^{-\frac{1}{2}\tanh\theta_\p a_\p^* a_{-\p}^*}|N)\otimes \Omega^>,
		\end{equation} where $\Omega^>$ is the Fock vacuum of $\Gamma_\s(l^2(\Xi_L^>))$. It is important to remark that the operators $a_\k^*a_{-\k}^*$ act also on $|N)$, by lowering its value by two. In particular, the state $\Omega_\Bog$ sticks out of the physical space by its tail. We can further exploit the symmetry under reflection to write\begin{equation}
		\Omega_\Bog = \prod_{\p\in\Xi^{>,+}_L}\frac1{\cosh\theta_\p}\e^{-\tanh\theta_\p a_\p^* a_{-\p}^*}|N)\otimes \Omega^>,
		\end{equation} where we have added an additional constrain
		\begin{align}
			\Xi^{>,+}_L =& \{\p\in\Xi^>_L | p_d >0\}\cup \{\p\in\Xi^>_L| p_d = 0, p_{d-1}>0\}\cup \dots\notag \\
			 &\dots \cup\{\p\in\Xi^>_L | p_d = \dots = p_2 = 0,p_1>0\}.
		\end{align}

{
\subsection{Weak coupling limit at fixed $\nu$}\label{sec: weak at fixed nu}

All our considerations are founded on the expectation that various
quantities related to the Bose gas simplify in  {\bf thermodynamic limit.} That is, $L\to\infty$ with
 $n(L)=\frac{N(L)}{L^d}\to n$.
Note that we do not have a limiting object at $L=\infty$: all our
statements go through the limit of a sequence of systems with finite $L$

On top of  thermodynamic limit  in this paper we 
consider the {\bf weak coupling limit at fixed $\nu$.}
This means replacing $v$ with $\kappa v$ and computing the 
lowest terms in perturbation theory in terms of $\kappa$, keeping
$\nu =\kappa\hat{v}(\0) n$ fixed.

Note that when the profile of the potential $v$ and the parameter
$\nu$ are
kept fixed, then
	$\omega_\Bog(\k) = \sqrt{\frac{|\k|^4}{4}+
          \frac{\nu\hat{v}(\k)}{\hat v(\0)}|\k|^2}$ does not change with varying
        $\kappa$, and has an interesting shape, with
          phonons, and often also with maxons, rotons, etc.
Therefore, in the limit we consider it makes then sense to
        assume that the Bogoliubov Hamiltonian $H_\Bog^L$ yields a good
approximation of our system at low energies. In particular, we will
treat $\Omega_\Bog $ as close to the ground state of our system.

 It is possible to compute the mean value of the extended zero-mode number operator on $\Omega_\Bog$ (for $d= 2,3$ ) \begin{align}
			n_{0,\Bog}(L) &:= \frac{1}{L^d}(\Omega_\Bog|
                                                                                                                       N^\ext_\0
                                                                                                                       \Omega_\Bog)
                                                                                                                       =
                                                                                                                       n(L)-n_\Bog^>(L),\\
                                                                                                                       n_\Bog^>(L)&
:= \frac{1}{L^d}(\Omega_\Bog|
                                                                                                                       N^>
                                                                                                                       \Omega_\Bog)                                                                             =                                                                                                \frac{1}{L^d}\sum_{\p\in\Xi^>_L}s_\p^2.
                                                                                                                     \end{align}
                                                                                                          
 We will also need the following quantity \begin{align}\label{eq: exponential computation}
	&(\Omega_\Bog| \e^{\mu N^>}\Omega_\Bog) \notag \\
	=&\prod_{\p\in\Xi^{>,+}_L}\frac{1}{\cosh^2\theta_\p}\sum_{n,m \geq 0}\frac{(-\e^{\mu }\tanh\theta_\p)^{n+m}}{n!m!}((a_\p^*a_{-\p}^*)^n|N)\otimes\Omega^>|(a_\p^*a_{-\p}^*)^m|N)\otimes\Omega^>)\notag \\
	=& \prod_{\p\in\Xi^{>,+}_L}\frac{1-\tanh^2\theta_\p}{1-\e^{2\mu}\tanh^2\theta_\p},
\end{align} which is well defined only for \begin{equation}	\label{eq: important inequality}
e^{\mu}< \frac{1}{\sup_{\p\in\Xi^{>,+}_L}|\tanh\theta_\p|}.
\end{equation}

We remark that $n_{0,\Bog}(L)$  could be possibly vanishing or negative. Indeed, the operator $N_\0^\ext$ is not the physical zero--mode number operator, and in particular its spectrum is $\mathbb{Z}$. 
We can compute its thermodynamic limit \begin{align}
			n_{0,\Bog}&:= \lim_{L\to+\infty} n_{0,\Bog}(L)
                                                                                     = n -n_\Bog^>,\\
                                                                                     n_\Bog^>&:=\int \frac{\d \p}{(2\pi)^d}s_\p^2.
		\end{align}We will refer interchangeably to
                $n_{0,\Bog}(L)$ and $n_{0,\Bog}$  as the \textit{extended Bogoliubov condensate density},
		omitting the qualifier ``finite--volume'' when referring to $n_{0,\Bog}(L)$.

		We now state another important assumption for the subsequent computation. We shall refer	to it as the \textit{condensate assumption}, since, as will become clear below, it implies	that the tail of the extended Bogoliubov state $\Omega_{\Bog}$ is exponentially suppressed
		in $L^{d-1}$, equivalently in the number of particles.
		\begin{assump}\label{ass: condensate hypotheses}\textbf{(Condensate assumption)}
			Define  $n_{0,\Bog}$  as above. We assume that $$n_{0,\Bog}>0.$$ 
		\end{assump}

We have
                $n_{0,\Bog}=\frac{\nu}{\kappa\hat v(\0)}-n_\Bog^>$,
              where $n_\Bog^>$ depends only on $v$ and
                  $\nu$. 
                  Therefore,
Assumption \ref{ass: condensate hypotheses} essentially says that
$\kappa$ is  sufficiently small.

\subsection{Limit $\nu\to0$}

Note that keeping $\nu$ fixed and $\kappa\to0$ means that the
density $n\to+\infty$.  So this limit involves the high density
regime.
One can also consider the limit $\kappa\to0$ at the constant density
$n$. This implies $\nu\to0$. It is then natural to rescale the
momenta, obtaining a universal dispersion relation, which
unfortunately does not have
 maxons or rotons any more:
               \begin{align}\label{ngeq1}
\lim_{\nu\to0}           \nu^{-1}
                 \omega_\Bog(\sqrt\nu\p,\nu)=&\sqrt{\frac14|\p|^4+|\p|^2} ,\\\label{ngeq2}
                 \lim_{\nu\to0} s_{\sqrt\nu\p}=&\frac{1}{\sqrt2}\left(\left(1-\left(\frac{1}{\frac12|\p|^2+1}\right)^2\right)^{-1/2}-1\right)^{1/2},\\
			\lim_{\nu\to0}\nu^{-\frac{d}{2}}	n_\Bog^>
  = & \int \frac{\d \p}{(2\pi)^d}\frac{ \frac{|\p|^2}{2} +
      1-\sqrt{\frac{|\p|^4}{4}+|\p|^2} }{2\sqrt{\frac{|\p|^{4}}{4}+|\p|^{2}
      }}. \label{ngeq}
			\end{align} Thus, 
                        also in this limit $n_{\Bog}^>$ converges to zero, so that the condensate
                        assumption holds.
                        For $d=2,3$, \eqref{ngeq}  remains
                        finite and can be computed explicitly.
                        For $d=3$ it equals $\frac{1}{3\pi^2}$.

                     The regime of  small $\nu$ is closely related to
                     the dilute limit, which will be discussed in
                                            Subsection \ref{Dilute limit}.

  \subsection{Dilute limit}
  \label{Dilute limit}
The material of this subsection will not be used in this paper,
 except for Subsect. \ref{sec: dilute gas limit}.
  
  As we mentioned above, the limit considered in this paper involves
  the high density regime. It is also interesting, and often
  physically more relevant, to consider dilute Bose gas. One can
  introduce the {\em dilute limit } as follows:
 we keep the potential $v$ fixed  (thus $\kappa=1$), we take
   thermodynamic limit, and then we consider
  $n\to0$. It
  is expected that in this limit in dimension $d=3$  the natural
  parameter is not
  $\nu=\hat{v}(\0)n$, it is $\mu=n4\pi
                \mathfrak{a}$, where $\mathfrak{a}$ is the scattering
                length of the potential $v$. Moreover, in the
                Bogoliubov approximation one should replace the
                formulas for the dispersion relation and related quantities
by
 \begin{align}
                 \omega_\Bog(\p,\mu)=&\sqrt{\frac14|\p|^4+\mu|\p|^2} ,\\
                  s_{\p}=&\frac{1}{\sqrt2}\left(\left(1-\left(\frac{\mu}{\frac12|\p|^2+\mu}\right)^2\right)^{-1/2}-1\right)^{1/2},\\
			n_\Bog^>
  = &  \int \frac{\d
  	\p}{(2\pi)^3} \frac{ \frac{|\p|^2}{2} +
      \mu-\sqrt{\frac{|\p|^4}{4}+|\p|^2\mu}
      }{2\sqrt{\frac{|\p|^{4}}{4}+|\p|^{2}\mu }}= \frac{\mu^{\frac32}}{3\pi^2}. 
			\end{align}
               Note that these formulas have essentially the
                 same form as in
               the limit $\nu\to0$ described in
               \eqref{ngeq1}--\eqref{ngeq}.

                The two approaches (the weak coupling limit with $\nu$ fixed
                and the dilute limit) are consistent with one
                another. Indeed, in the first order of the Born
                approximation the scattering length 
satisfies $ 4\pi
\mathfrak{a}=\kappa\hat{v}(\0)+O(\kappa^2)$
Thus $\mu=\nu(1+O(\kappa))$.
Note  also that\begin{equation}
				n_{0,\Bog} = n - \frac{1}{3\pi^2}\mu^{\frac{3}{2}} = n\left(1- \frac{1}{3\pi^2}\sqrt{n(4\pi\mathfrak{a})^{3}}\right),
			\end{equation} so that the condition $n_{0,\Bog}>0$ corresponds to \begin{equation}
				\sqrt{n \mathfrak{a}^3} < \frac{3\sqrt{\pi}}{8 }.
			\end{equation}
Thus for sufficiently small scattering length $\mathfrak{a}$ the
condensate assumption is satisfied.
                        
In the literature on dimension
$3$,
often, instead of the dilute limit, 
a seemingly different limit is considered, which is usually
called the {\em Gross-Pitaevskii limit}.  The size of the box
$L$ is kept fixed,  the number of particles $N$ goes to
infinity, and the potential is rescaled as
$v_N(x):=N^2v(Nx)$. Note that the scattering
length depends on $N$ as $\mathfrak{a}^{(N)}:=\frac{\mathfrak{a}}{N}$.
Thus $\mu=\frac{N}{L^3}4\pi\mathfrak{a}^{(N)}$ does not change. 

The Gross–Pitaevskii limit has also been referred to as the {\em ultra-dilute limit} \cite{BBCS_2019}. Indeed, following the discussion after Theorem 1.1 in \cite{BBCS_2019}, one may rescale the side length of the box by a factor of $N$, namely $L\to NL$. The system is then equivalent to an $N$-particle problem with a fixed interaction potential $v(x)$ in a box of volume $O(N^3)$. Under this scaling, the vanishing-density limit $n\to0$ is taken simultaneously with the infinite-volume and large-particle-number limits. In particular,
$$
n=O\left(N^{-2}\right)
\qquad\text{as }N\to+\infty.
$$
This differs from both our approach and the dilute limit: in
  both approaches
  the  limits $\nu\to0$, resp. $n\to0$ are taken only after the thermodynamic limit at fixed density; see \ref{ass: thermodynamic limit.}. The two procedures need not to be equivalent. Indeed, the excitation spectrum obtained in \cite{BBCS_2019} retains a discrete structure, with a fixed step, whereas the perturbative expansion developed in the present paper suggests that the spectrum becomes purely continuous in the thermodynamic limit. Similarly, the ground-state energy obtained in the Gross–Pitaevskii regime in \cite[Equation 1.10]{BBCS_2019} differs from the prediction of the Lee–Huang–Yang formula \cite[Equation 1.15] {BBCS_2019}.

\subsection{Shift of dispersion relation}
Let us now describe the main quantity computed in our paper. We
will choose a sequence 
$\k_L\in\Xi_L^>$ converging to $\k\in\rr^d\backslash\{\0\}$ and
consider the perturbation theory around the sequence of vectors
$b_{\k_L}^*\Omega_\Bog$. With a slight abuse of notation, we will drop
$L$ from $\k_L$.

The difference between the physical and Bogoliubov Hamiltonians, that
is $H^L-H_\Bog^L$ is quite complicated and consists of several terms
with different roles and different orders in $\kappa$, see \eqref{eq: effective
    Hamiltonian}.
  
Let us first consider
the operators $R_{4,1}$ and $R_{4,2}$, which will be called the {\em
  tail terms}. They 
depend on the condensate degree
of freedom in 
complicated way. In particular, they cannot be represented as good
second quantized operators of the out-of-condensate degrees of
freedom.
In Theorem \ref{thm: effective hamiltonian} we will show
that these terms are irrelevant in thermodynamic
  limit for all  low-lying eigenstates.

All the remaining terms in \eqref{eq: effective Hamiltonian}
$H_{\Bog}^L$, $H_2^L$, $H_3^L$, $H_{4,1}^L$, $H_{4,2}^L$, $H_{4,3}^L$, 
$H_{3,x}^L$ 
do not depend on the condensate degrees of freedom. With respect to
the factorization \eqref{fullspace}, they can be
written  as
\beq \one\otimes \tilde H_j^L\eeq
where $\tilde H_j^L$ is the corresponding ``reduced'' operator, acting on the
modes in $ \Xi_L^>$. In what
follows, we will drop these tildes, and we will denote the reduced and
unreduced operators the same way. We will treat 
$ \Gamma_\s\Big(l^2\bigl(\Xi_L^>\bigr)\Bigr)$ as our basic Hilbert
space.

All of these operators but 
$H_{3,x}^L$ do not depend 
on $\kappa$.
The leading behavior of $H_{3,x}^L$
in terms of $\kappa $ is  of order $0$, as we show later on.
In a perturbative setting it can be expanded in a power series in
$\kappa$.
           Thus we expect that to study Bose gas in the weak coupling
           limit at fixed $\nu$ we can use the perturbation theory for
                \beq
                 H^L_{\Bog} +  \kappa H_2^L+\sqrt{\kappa}H^L_{3} +
 \kappa H^L_{4,1}+ \kappa H^L_{4,2}+ \kappa H^L_{4,3}+\kappa\sqrt{\kappa}H_{3,x}^L \eeq
 acting on the space $ \Gamma_\s\Big(l^2\bigl(\Xi_L^>\bigr)\Bigr)$. 

 The term 
		$\kappa\sqrt\kappa H_{3,x}^L$ yields contributions of 
                a higher order in $\kappa$ than the order considered here, therefore 
                  we will ignore it as well. 

           Thus we can use perturbation theory for
                \beq
                 H^L_{\Bog} +  \sqrt{\kappa}H^L_{3} +
 \kappa H^L_{4}. \eeq
 where
 \beq
 H_4^L:=H_2^L+H_{4,1}^L+H_{4,2}^L+H_{4,3}^L.\eeq

The shift of the dispersion relation at order $\kappa^1$ will be
computed from the formula
\begin{align}\label{formi1}
  \delta(\kk)
  :=		\lim_{\epsilon\searrow
  0}\lim_{L\rightarrow\infty}&
\bigg{\{} \left(b_\k^*\Omega_\Bog |
		H_4^Lb_\k^*\Omega_\Bog \right)
                             \\&- \left(H_{3}^Lb_\k^*\Omega_\Bog |
		\big(H_\Bog^L-\omega_\Bog(\k)-\i\epsilon\big)^{-1}H_3^Lb_\k^*\Omega_\Bog \right)
                           \label{formi2}     \\
  -& \left(\Omega_\Bog \right.\left|H_4^L\Omega_\Bog \right) +\left(H_3^L \Omega_\Bog \right.\left| (H_\Bog^L  -\i\epsilon)^{-1}H_3^L\Omega_\Bog \right) \bigg{\}} 
\label{eq: physical correction braket I.}\end{align}
Note that  it is the difference of the shift of the energy level of $b_\k^*\Omega_\Bog$
 \eqref{formi1}+\eqref{formi2}  and  the  shift for the ground state energy
\eqref{eq: physical correction braket I.}.

    
The formula for      $\delta(\kk)$ can be transformed as follows:
{\begin{align}\label{forri}
        \delta(\kk)&=\delta_4(\k)+\delta_3(\k),\\
\delta_4(\k)  &:=		\lim_{L\rightarrow\infty}
(\Omega_\Bog |[b_\k,
                [H_4^L, b_\k^*]]\Omega_\Bog \big)
                \label{hellfeyn}
           \\\notag
  \delta_3(\k)&:=
                \lim_{\epsilon\searrow
  0}\lim_{L\rightarrow\infty}\bigg\{
                                    -\big(\Omega_\Bog |[b_\k,
                                                  H_{3}^L](H_\Bog^L-\omega_\Bog(\k)-\i\epsilon)^{-1}[H_3^L,  
                                                  b_\k^*]\Omega_\Bog \big)  \\
  &-\big(\Omega_\Bog |[b_\k^*, H_{3}^L](H_\Bog^L+\omega_\Bog(\k)-\i\epsilon)^{-1}[H_3^L, 
                       b_\k]\Omega_\Bog \big)\notag\\
&  -\big(\Omega_\Bog | H_{3}^L(H_\Bog^L-\i\epsilon)^{-1}[b_\k ,
                                                           [H_3^L, b_\k^*]]\Omega_\Bog \big)
                                                           \notag \\ &-\big(\Omega_\Bog |[b_\k^*,
  [H_3^L, b_\k]](H_\Bog^L-\i\epsilon)^{-1}H_3^L\Omega_\Bog \big)\bigg\}
      .\notag\end{align}
   } To obtain this we used
    \begin{align}
    &  [b_\k, b_\k^*]=1,\\&b_\k^*(H_\Bog^L+\omega_\Bog(\k)-\i\epsilon)^{-1}=
      (H_\Bog^L-\i\epsilon)^{-1}b_\k^*,\\
&      b_\k(H_\Bog^L-\omega_\Bog(\k)-\i\epsilon)^{-1}=
      (H_\Bog^L-\i\epsilon)^{-1}b_\k.
            \end{align}

                 		We will argue that the Hellmann-Feynman
                                terms  involving	$\kappa H_2^L$,	$\kappa H_{4,1}^L$,  $\kappa H_{4,3}^L$
vanish in thermodynamic limit.
                So eventually, we will use 
                \beq 
                H_\eff^L:=
                 H^L_{\Bog} + \sqrt{\kappa}H^L_{3} +
 \kappa H^L_{4,2} .\eeq 
	on the space $\Gamma_\s\big(l^2(\Xi_L^>)\big)$ as an effective
Hamiltonian up to the order $\kappa^1$.

            \subsection{Irrelevant terms for thermodynamic limit}\label{sec: irrelevant terms}
     In this subsection we discuss various terms which vanish in
thermodynamic limit, and hence do not contribute to the shift of the
dispersion relation.

 The tail terms $R_{4,1}$, $R_{4,2}$ formally do
              not depend on $\kappa$ and are multiplied by
              $\kappa$. So, formally, they should contribute to the
              Hellmann-Feynman correction. However, they vanish very
              fast in thermodynamic limit. This is described in the
              theorem below, which will be shown in Sect. \ref{Tail estimates}, where we will
describe the mechanism which causes unphysical degrees of freedom to
be invisible for large volumes.

\begin{thm}\label{thm: effective hamiltonian} Fix $d=2,3$ and assume the validity of Assumptions \ref{ass: primary assumption on the potential} and \ref{ass: condensate hypotheses}. Consider the normalized $k$--quasiparticle vector \begin{equation}
		\psi_k
		=
		\mathcal{N} b_{\k_1}^*\cdots b_{\k_k}^*\Omega_{\Bog},
		\qquad
		\mathcal{N}
		=
		\big\| b_{\k_1}^*\cdots b_{\k_k}^*\Omega_{\Bog} \big\|^{-1},
	\end{equation} with momenta $\k_1,\dots \k_k\in \Xi^>_L$ satisfying $\liminf_{L\to+\infty}|\k_i|>0$.
	The Hellmann-Feynman corrections at order $\kappa^1$ arising from the remainder terms  $R_{4,1}^L$ $R_{4,2}^L$ are zero in the thermodynamic limit. That is \begin{align}
		&\lim_{L\to+\infty }\left((\psi_k| R_{4,1}^L \psi_k) -(\Omega_\Bog| R_{4,1}^L\Omega_\Bog)\right) = 0,\\
			&\lim_{L\to+\infty }\left((\psi_k| R_{4,2}^L \psi_k) - (\Omega_\Bog| R_{4,2}^L\Omega_\Bog) \right)= 0.
	\end{align} 
\end{thm}

			The following three lemmas show that the
                        Hellmann-Feynman contribution arising from
                        $H^L_2$ and $H^L_{4,1}$ and $H^L_{4,3}$ are
                        vanishing in the thermodynamic limit. We will always assume the potential to satisfy Assumption \ref{ass: primary assumption on the potential} and $d\geq 2$.

                          Recall  that
                          \beq H_2^L:=-\frac{1}{4L^d}
                     \sum_{\p\in\Xi_L^>}\hat{v}(\p)
                     A_\p^*  +\text{h.c.}\eeq

\begin{lemma}
		\label{lemm: thermodynamic for H2.}
		                We have
                \beq
                (b_\k^*\Omega_\Bog|H_2^L b_\k^*\Omega_\Bog)-
                (\Omega_\Bog|H_2^L \Omega_\Bog)=O(L^{-d}).\eeq
              \end{lemma}
              \proof
              We rewrite the left hand side in the commutator form and
              use \eqref{commu2}:
              \begin{align}&
                (\Omega_\Bog|[b_\k,[H_2^L ,b_\k^*]]\Omega_\Bog)
                =-\frac{\hat v(\k)}{4
                   L^d}4s_\k c_\k=O(L^{-d}).\end{align}
               \qed

               Recall that
               \beq
               	H_{4,1}^L:=  -\frac{\hat v(\0)}{2L^d}\sum_{\p\in\Xi_L^>}
			N_\p^2.\eeq
\begin{lemma}
		\label{lemm: thermodynamic for H41.}
                We have
                \beq
                (b_\k^*\Omega_\Bog|H_{4,1}^L b_\k^*\Omega_\Bog)-
                (\Omega_\Bog|H_{4,1}^L \Omega_\Bog)=O(L^{-d}).\eeq
              \end{lemma}
              \proof
We proceed similarly as in the proof of the previous lemma, using
\eqref{comm1} and \eqref{commu1}:
              \begin{align}&
                (\Omega_\Bog|[b_\k,[H_{4,1}^L ,b_\k^*]]\Omega_\Bog)
                =-\frac{\hat{v}(\0)}{2
                   L^d}
(\Omega_\Bog|[b_\k,[N_\k^2+N_{-\k}^2,b_\k^*]]\Omega_\Bog)
                   =O(L^{-d}).\end{align}
               \qed

               Recall that
               \beq H_{4,3}^L:=
			\frac{1}{2L^d}\sum_{\substack{\p,\q,\rrr,\p\pm\q,\\ \p+\rrr, 
					\q-\rrr,\p+\rrr-\q \in 
					\Xi_L^>}}\hat{v}(\rrr)a_{\p+\rrr}^* a_{\q-\rrr}^*
   a_\q a_\p,\eeq
\begin{lemma}
		\label{lemm: thermodynamic for H43.} 
                We have
                \beq
                (b_\k^*\Omega_\Bog|H_{4,3}^L b_\k^*\Omega_\Bog)-
                (\Omega_\Bog|H_{4,3}^L \Omega_\Bog)=0.\eeq
              \end{lemma}
              \proof
We can again rewrite the lhs in the commutator form:
              \begin{align}&
                (\Omega_\Bog|[b_\k, [H_{4,3}^L ,b_\k^*]]\Omega_\Bog).
\end{align}
Using $b_\k^*=c_\k a_\k^*+s_\k a_{-\k}$ we obtain
\begin{align}
[H_{4,3}^L ,b_\k^*]=&
			\frac{2c_\k}{2L^d}\sum_{\substack{\p,\rrr,\p+\k,\\ \p+\rrr, 
					\k-\rrr,\p+\rrr-\k \in 
					\Xi_L^>}}\hat{v}(\rrr)a_{\p+\rrr}^* a_{\k-\rrr}^*
   a_\p \\
  &+	-\frac{2s_\k}{2L^d}\sum_{\substack{\p,\rrr,\p-\k,\\ \p+\rrr, 
					\k+\rrr,\p+\rrr+\k \in 
					\Xi_L^>}}\hat{v}(\rrr)a_\p^* a_{\p+\rrr} a_{-\k-\rrr}
                .
\end{align}
               Now the commutator of this with $b_\k=c_\k a_\k+s_\k
               a_{-\k}^*$ is zero. \qed

}
               
	\begin{remark}(\textbf{Naive thermodynamic limit})
		\label{Thermodynamic  limit}		
			Let us try to describe the thermodynamic limit naively.
		Remember that we consider $L\to\infty$ and we keep
		$\nu$ and $\kk\mapsto \hat v(\kk)$ fixed. 
		Formally, $\Xi_L^>$  converges to $\rr^d$. $\frac{1}{L^d
		}\sum\limits_{\kb\in\Xi_L^>}$ should be replaced with $\frac{1}{(2\pi)^d }\int\d
		\kb$. The operators $ a_\kb$, $ a_\kb^*$ for discrete $\kb$ should
		be replaced by $\frac{(2\pi)^{\frac d 2}}{L^{\frac d 2}} a_\kb$,
		$\frac{(2\pi)^{\frac d 2}}{L^{\frac d 2}}  a_\kb^*$ for
		continuous $\kb$.
		Thus 
		$H_\Bog^L$ formally
		converges to
		\beq\label{therm1}
		H_\Bog:=\int\d\p\;\Big(\frac{\p^2}{2}+\frac{\nu\hat v(\p)}{\hat v(\0)}\Big) a_\p^* 
		a_\p
		+
		\int\d\p\;\Big(\frac{\nu\hat v(\p)}{2\hat v(\0)}
		a_\p 
		a_{-\p}
		+\hc\Big).
		\eeq
		Similarly,  in thermodynamic limit $H_3^L$ and $H_{4,3}^L$ formally converge to
		\begin{align}\label{therm2}
			H_3&:=\frac{1}{(2\pi)^{\frac{d}{2}}}\int \d\k\d \p\;\frac{\sqrt{\nu}\hat 
				v(\k)}{\sqrt{\hat v(\0)}}\Big( a_{\p+\k}^*  a_\k  a_{\p}
			+\hc\Big),\\
			H_{4,3}&:=\frac{1}{2(2\pi)^d}\int\d\p\d\k\d\q\;\hat v(\k) a_{\p+\k}^*  a_{\q-\k}^* 
			a_\q  a_\p.  \notag
		\end{align} 
		However, as we have just noted, 
		$H_{4,3}^L$
		is of little relevance for our analysis.
		Formally,  in thermodynamic limit $H_{4,2}^L$ and $H_{4,1}^L$ converge to
		zero. However, $H_{4,2}^L$ plays an important role in calculations of the
		quasiparticle spectrum.
		
		This illustrates the fact that thermodynamic limit 	in 
		our problem is quite subtle.

	\end{remark}
		
			Finally, let us now consider  $H_{3,x}^L$. This Hamiltonian contains a factor  \begin{equation*}\left|1- \frac{N^>}{N}\right|_+^{\frac{1}{2}} -1 = \left|1- \frac{\kappa\hat{v}(\0)}{\nu L^d}\sum_{\p\in\Xi^>_L}N_\p\right|_+^{\frac{1}{2}} -1. \end{equation*}
			
			As we will now show, the matrix elements of
                        this operator in low energy states contribute at least as an order $\kappa^1$ \begin{lemma}\label{lem: square root remainder }
		 Suppose 
                        that the
                        condensate assumption \ref{ass: condensate
                          hypotheses} is satisfied.
                        Take momenta $\q_1,\dots \q_m; \k_1,\dots \k_n \in \Xi^>_L$. Then, we have \begin{equation}
				\left|\left(b_{\q_1}^*\dots b_{\q_m}^*\Omega_\Bog| \left(\left| 1-\frac{N^>}{N}\right|^{\frac{1}{2}}_+-1\right)b_{\k_1}^*\dots b_{\k_n}^*\Omega_\Bog\right)\right| = O(\kappa).
			\end{equation}
			\end{lemma}
			\begin{proof} 
				We have the operator inequality \begin{equation}
				\left|1-	\left|1-\frac{N^>}{N}\right|^{\frac{1}{2}} _+\right|\leq \frac{N^>}{N} = \kappa \frac{\hat{v}(\0)}{\nu L^d}\sum_{\p\in\Xi^>_L} N_\p\label{eq: explicit expansion of F>}
				\end{equation}
			 If we restrict the analysis to states with a finite number of Bogoliubov excitations, the excited number operator
			 $\sum_{\p\in\Xi^>_L}N_\p$ is bounded on the corresponding finite-particle subspace and thus the matrix elements are well defined and of order $O(\kappa)$ as $\kappa\to 0$.
                       \end{proof}

\section{ Computing perturbative corrections}
\label{sec: computation of the real part 1}

In this section we compute  perturbative corrections of order
$\kappa^1$ to the
dispersion relation. They consist of two terms: the 
Hellmann-Feynman correction and the real part of the Fermi golden rule
correction. We also extract the leading behavior of these corrections
in the low momentum limit, which yield the phonon velocity. In this
section we content ourselves with formal calculations---rigorous
estimates are deferred to Section 
\ref {Estimates on perturbative corrections}.

  In addition to Assumptions \ref{ass: primary assumption on the potential} and \ref{ass: condensate hypotheses} , we impose the following
  regularity conditions, which are needed for the detailed computation
  of the real energy shift.
  
\begin{assump}\label{ass: secondary hypotheses}
	\begin{subequations}\label{eq: assumptions on potential1} 
	Suppose there exists a $\nu_0\in ]0,+\infty[$ such that the potential satisfies the following assumptions for all the possible choices of $\nu \in ] 0, \nu_0 ] $\begin{align} 
	& v \text{ is rotationally invariant}; \label{eq: rot invariace}\\
	&\hat{v} \text{ is } C^5 \text{ in a neighborhood of } \k = \0.\label{eq: regularity at 0}\\
	& \hat{v} \text{ is } C^2 \text{ and } \partial_i\partial_j\hat{v} \text{ is bounded for every }i,j\in\{1,2,\dots,d\};  \label{eq: twice differentiability}\\
	& \partial_i\hat{v} \text{ and } \partial_i\partial_j \hat{v} \in L^2(\mathbb{R}^d),\text{ for every } i,j \in\{1,2,\dots,d\}. \label{eq: integrability}\\
	& \frac{\nu}{\hat{v}(\0)}\frac{\d^2\hat{v}}{\d k^2}(\0) > -\frac{1}{2}.  \label{eq: convexity}
	\end{align}
\end{subequations} 
	\end{assump}

We remark that under these hypotheses  \begin{equation}
	\hat{v}(\k) -\hat{v}(\0) = O(|\k|^2),\text{ as }\k\to \0.
\end{equation}

Since the functions
considered below are rotationally invariant, we sometimes identify a function
$f(\mathbf{p})$ with the corresponding radial function $f(p)$,
where $p=|\mathbf{p}|$; for example, we write
$\omega_{\mathrm{bg}}(\mathbf{p})=\omega_{\mathrm{bg}}(p)$ and
$\widehat{v}(\mathbf{p})=\widehat{v}(p)$. When no confusion can
arise, primes denote derivatives with respect to the scalar argument:
\begin{equation}
	\frac{d}{dx}f(x)\equiv f'(x).
\end{equation}

\subsection{Hellmann-Feynman correction} \label{sec: HF matrix
  elements}

This section is devoted to the Hellmann-Feynman term $\delta_4^\can(\k,\nu,v)$ defined in
\eqref{hellfeyn}. As we argued, in this term we can replace $H_4^L$ with
$H_{4,2}^L$. Thus we need to compute \begin{align}
\delta_4^\can(\k,\nu,\hat v):=&\lim_{L\to\infty}\Big(	\left(b_\k^*\Omega_\Bog\right| H_{4,2}^L \left. b_\k^*\Omega_\Bog\right)-	\left(\Omega_\Bog\right| H_{4,2}^L \left. \Omega_\Bog\right) \Big)\\=& \lim_{L\to\infty}	\left(\Omega_\Bog\right|\left. [b_\k,[H_{4,2}^L , b_\k^*]]\Omega_\Bog\right).
                                                                                                                                       \end{align}
                                                                                                               Recall that
\begin{align}	H^L_{4,2} :=&H_{NN}^L+H_{NA}^L+H_{AA}^L,\\
H_{NN}^L:=  &  -\frac{1}{2L^d}\sum_{\q\in\Xi_L^>}N_\q\sum_{\p\in\Xi_L^>} N_\p (\hat{v}(\p)+\hat v(\q)-\hat{v}(\p-\q)),\notag \\ 
H_{NA}^L:=			& - \frac{1}{2L^d}\sum_{\q\in\Xi_L^>}N_\q\sum_{\p\in\Xi_L^>} \hat{v}(\p) A_\p^*+\text{h.c.},\notag\\
		H_{AA}^L:=	& \frac{1}{2L^d}\sum_{\p,\q,\p\pm\q\in \Xi^>_L}
			\hat{v}(\p-\q)  A_\p^* A_\q.\notag\end{align}

{

To evaluate  the $NN$ term in $H_{4,2}^L$ we compute
\begin{align}
  [b_\k,[N_\q
  N_\p,b_\k^*]]&=N_\q[b_\k,[N_\p,b_\k^*]]+[b_\k,[N_\q,b_\k^*]]N_\p\\
  &+[b_\k,N_\q][N_\p,b_\k^*]+[b_\k^*,N_\q][N_\p,b_\k].\label{hell}
\end{align}
The terms in \eqref{hell} contribute $O(L^{-d})$, so that it will be
dropped. Thus we obtain
\begin{align}
&\left(\Omega_\Bog|
                \Bigl[b_\k,\Bigl[H_{NN}^L , b_\k^*\Bigr]\Bigr]\Omega_\Bog\right) \notag\\
  =&-\frac{1}{L^d}\sum_{\p\in\Xi^>_L}s_\p^2c_\k^2
(\hat v(\p)+\hat v(\k)-\hat v(\p-\k))
     \notag \\
  &-\frac{1}{L^d}\sum_{\p\in\Xi_L^>}s_\p^2s_\k^2\left(\hat{v}(\p)
    +\hat v(\k)-\hat{v}(\p+\k)\right) +O(L^{-d}).
\end{align}}
  Similarly, the $NA$ term gives
	 \begin{align}
		&\left(\Omega_\Bog\right|  [b_\k,[H_{NA}^L, b_\k^*]]\Omega_\Bog) \notag\\
		=&  \frac{1}{L^d}\sum_{\p\in\Xi_L^>}2s_\p^2 c_\k s_\k \hat{v}(\k)  +\frac{1}{L^d}\sum_{\p\in\Xi_L^>} (c_\k^2+s_\k^2) c_\p s_\p \hat{v}(\p) + O(L^{-d}).
\end{align} Finally, the $AA$ term gives
\begin{align}
	&\left(\Omega_\Bog\right|  [b_\k,[H_{AA}^L , b_\k^*]]\Omega_\Bog) \notag\\
	=& \frac{1}{L^d}\sum_{\p,\p\pm\k\in\Xi^>_L}  c_\k s_\k c_\p s_\p (\hat{v}(\p+\k)+\hat{v}(\p-\k)) + O(L^{-d})
\end{align}

  Combining these contributions and using $s_\k=s_{-\k}$,
  $c_\k=c_{-\k}$,
we obtain
\begin{align}
  &	\left(\Omega_\Bog\right|\left. [b_\k,[H_{4,2}^L,
    b_\k^*]]\Omega_\Bog\right) = \notag \\
=	&
   (c_\k+s_\k)^2\frac{1}{L^d}\sum_{\p\in\Xi_L^>}c_\p s_\p
    \hat{v}(\p)\\
  	&+	(c_\k^2+s_\k^2)\frac{1}{L^d}\sum_{\p\in\Xi_L^>}s_\p^2\left(-\hat{v}(\p) +\frac{\hat{v}(\p+\k)+\hat{v}(\p-\k)}{2}\right)  \\
  &   + 2c_\k s_\k \frac{1}{L^d}\sum_{\p,\p\pm\k\in\Xi^>_L} c_\p s_\p\left (-\hat v(\p)+\frac{\hat{v}(\p+\k)+\hat{v}(\p-\k)}{2}\right)\\
	&	-(c_\k-s_\k)^2 \hat{v}(\k)\frac{1}{L^d}\sum_{\p\in\Xi_L^>}s_\p^2 + O(L^{-d}). \label{eq: finite volume HF}
\end{align} 
These expressions have a thermodynamic limit:
\begin{align}
	&\delta^\can_4(\k,\nu,\hat{v})=\lim_{L\to+\infty}
                                                                                                             \left(\Omega\right|\left. [b_\k,[H_{4,2}^L, b_\k^*]]\Omega\right)\label{eq: HF thermodynamic limit} \\
=	&
   (c_\k+s_\k)^2\frac{1}{(2\pi)^d}\int \d\p c_\p s_\p
    \hat{v}(\p)\\
  	&	+(c_\k^2+s_\k^2)\frac{1}{(2\pi)^d}\int \d\p s_\p^2\left(-\hat{v}(\p) +\frac{\hat{v}(\p+\k)+\hat{v}(\p-\k)}{2}\right)  \\
  &   +2 c_\k s_\k \frac{1}{(2\pi)^d}\int\d\p c_\p s_\p\left (-\hat v(\p)+\frac{\hat{v}(\p+\k)+\hat{v}(\p-\k)}{2}\right)\\
	&	-(c_\k-s_\k)^2 \hat{v}(\k)\frac{1}{(2\pi)^d}\int\d\p s_\p^2. \label{eq: infinite volume HF}
                                                                                                \end{align} 
                                                                                                We can then express $c_\k$, $s_\k$ in simpler quantities using
                                                                                                \begin{align}
                                                                                                  c_\k
                                                                                                  s_\k=\frac{\hat
                                                                                                  v(\k)\nu}{\hat
                                                                                                  v(\0)2\omega_\Bog(\k)},&\quad
                                                                                                                           (c_\k-s_\k)^2=\frac{\k^2}{2\omega_\Bog(\k)},\quad
                                                                                                                           (c_\k+s_\k)^2=\frac{2\omega_\Bog(\k)}{
                                                                                                                           \k^2},\\c_\k^2+s_\k^2=\frac{\frac{\k^2}{2}+\frac{\hat 
                                                                                                  v(\k)}{\hat 
                                                                                                  v(\0)}\nu}{\omega_\Bog(\k)},&\quad s_\k^2=                             \frac{\frac{\k^2}{2}+\frac{\hat 
                                                                                                  v(\k)}{\hat 
                                                                                                  v(\0)}\nu-\omega_\Bog(\k)}{2\omega_\Bog(\k)}. \label{eq: simple form for cksk}
                                                                                 \end{align}

From this point 
onward, we fix the spatial dimension $d=3$.  
                                                                         The
                                                                         above
                                                                         computations                                                                        lead to the following infrared asymptotics of the
Hellman--Feynman correction, which we will prove rigorously in Section
\ref{Estimates on perturbative corrections}.
\begin{prop}\label{prop: HF term}
	Suppose that the potential satisfies Assumptions \ref{ass: primary assumption on the potential} and \ref{ass: secondary hypotheses}.
Then	
\begin{align}
		\delta_4^\can(\k,\nu,\hat{v})
 = &c_\ir(k,\nu,\hat v)  + k c_4\left(\nu, \hat{v}\right) + o\left(\frac{k}{\sqrt{\nu}}\right),\label{eq: HF final form.}
	\end{align} as $\frac{k}{\sqrt{\nu}}\to 0$, where the  terms
        $c_\ir$ and  $c_4$ are equal
        to \begin{align}\label{cir}
c_\ir(k,\nu,\hat v)&:=
\frac{\omega_\Bog(k)}{k^2}\int_0^{+\infty}\frac{\d p}{2\pi^2}\; p^2 \frac{\hat{v}(p)^2\nu}{\hat{v}(0)\omega_\Bog(p)} , 
             \\\label{c4a}
		c_4\left(\nu,\hat{v}\right)                                                                                                                               ={}&
                                                                                                                                    \frac{\sqrt\nu}{12\pi^2}\int_0^\infty\d
                                                                                                                                                                               p\frac{                                                                                                                      \left(\frac{p^2}{2}+\frac{2\hat
                                                      v(p)\nu}{\hat
                                                                                                                                    v(0)}-\omega_\Bog(p)\right)}{2\omega_\Bog(p)}\left(p^2\hat
                                                                                                                               v''(p)+2p\hat
                                                                                                                                                                               v'(p)\right)\\\label{c4b}
                                                                                                                               &-\frac{\hat{v}(0)}{4\pi^2\sqrt\nu}\int_0^\infty
                                                                                                                                 \d
                                                                                                                                 p
                                                                                                                                \frac{\left(\frac{p^2}{2}+\frac{\hat
                                                                                                                                 v(p)}{\hat v(0)}\nu-\omega_\Bog(p)\right)p^2}{2\omega_\Bog(p)}.
	\end{align}
\end{prop}

Note that the term $c_\ir$ behaves as $O(k^{-1})$. It will be
cancelled by an identical term coming from the Fermi golden rule.

\subsection{Fermi golden rule correction}

 In this section we will
analyze
 the real part of the Fermi golden rule correction to the dispersion
 relation (we take $\delta_3^\can \equiv \Re \delta_3^\can$ ):
\begin{align}\label{eq: real divergence fermi1}
\delta^\can_3(\k,\nu,\hat{v})
= &	\lim_{\epsilon\searrow 0}\lim_{L\to+\infty}\Re\Bigl[ -(b_\k^*\Omega_\Bog | H_{3}^L(H^L_\Bog -\omega_\Bog(\k)-\i\epsilon)^{-1}H^L_{3}b_\k^*\Omega_\Bog)  \\ &+ (\Omega_\Bog | H^L_{3}(H^L_\Bog -\i\epsilon)^{-1}H^L_{3}\Omega_\Bog)\Bigr]
\end{align} 

   In particular, we will 
 estimate its behavior  for small $\frac{|\k|}{\sqrt{\nu}}\to 0$.
The integral representation for $\delta_3^\can$ is derived in \cite{DP_2026}: \begin{align}\label{eq: real divergence fermi}
\delta^\can_3(\k,\nu,\hat{v})=&-\frac{1}{2}\Re\int \frac{\d \p}{(2\pi)^3}\frac{j(\k;\p,\p-\k)^2}{\omega_\Bog(\p)+\omega_\Bog(\p-\k)-\omega_\Bog(\k)-\i 0} \\  &-\frac{1}{2}\int \frac{\d \p}{(2\pi)^3}\frac{\kappa(\k,\p,\p-\k)^2}{\omega_\Bog(\k)+\omega_\Bog(\p)+\omega_\Bog(\p-\k)} .\label{eq: real divergence fermi2}
\end{align}
The function  $j(\k;\p,\q)$ contained 
in \eqref{eq: real divergence fermi} is defined as
  \begin{align}
	j(\k;\p,\q) = &\sqrt{\frac{\nu}{\hat{v}(\0)}}(-\hat{v}(\k)(c_\k-s_\k)(c_\p s_\q + s_\p c_\q) \notag \\
	&+ \hat{v}(\p)(c_\p-s_\p)(c_\q c_\k +s_\q s_\k)+ \hat{v}(\q)(c_\q -s_\q)(c_\p c_\k + s_\p s_\k ))\label{eq: def of j}.
                                                 \end{align}
                                                 $j$ 
is symmetric under interchange of its last two arguments, i.e.  $j(\k;\p,\q) = j(\k;\q,\p)$. 
The second function,  $\kappa(\k,\p,\q)$, which  appears inside the
second integral in Equation \eqref{eq: real divergence fermi2}, is  \begin{align}
	\kappa(\k,\p,\q) = &\sqrt{\frac{\nu}{\hat{v}(\0)}}(-\hat{v}(\k)(c_\k-s_\k)(c_\p s_\q + s_\p c_\q) \notag \\
	&- \hat{v}(\p)(c_\p-s_\p)(c_\q s_\k +s_\q c_\k)- \hat{v}(\q)(c_\q -s_\q)(s_\p c_\k + c_\p s_\k ))\label{eq: def of kappa}.
\end{align} $\kappa$  is symmetric under permutations of its three
arguments;  $\kappa(\k,\p,\q) = \kappa(\k,\q,\p) = \kappa(\q,\k,\p)$.

A rigorous analysis of the integrals \eqref{eq: real divergence fermi} and
\eqref{eq: real divergence fermi2} requires lengthy estimates of the relevant
contributions and their remainders. We therefore begin with a formal derivation
of their asymptotic structure as
$\frac{|\k|}{\sqrt{\nu}}\to0$. In particular, in the following formal
derivation we forget about the principal value in $\delta^\can_3$,
assuming that  it does not cause any issue. This is justified in Proposition \ref{prop: FG zero}.

We use the notation
\begin{equation}
	k:=|\k|,
	\qquad
	p:=|\p|,
	\qquad
	q:=|\p-\k|.
\end{equation}
The triangle inequality gives
\begin{equation}
	-k\leq p-q\leq k,
\end{equation}
which motivates the following change of variables.

\begin{defn}\textbf{($t$--$s$ coordinates)}
	\label{def: ts coordinates}
	We define
	\begin{equation}\label{eq: change in t s}
		t=\frac{p+q}{2},
		\qquad
		s=\frac{p-q}{2k},
	\end{equation}
	or, equivalently,
	\begin{equation}
		p=t+ks,
		\qquad
		q=t-ks,
	\end{equation}
	with
	\begin{equation}
		t\in\Bigl[\frac{k}{2},+\infty\Bigr[,
		\qquad
		s\in\Bigl[-\frac{1}{2},\frac{1}{2}\Bigr].
	\end{equation}
\end{defn}

Under this change of variables, the symmetry $p\leftrightarrow q$ becomes
$s\leftrightarrow-s$, while the measure transforms as
\begin{equation}\label{eq: ts coordinate-}
	\d\p
	=
	\frac{pq}{k}\,\d p\,\d q\,\d\varphi
	=
	2pq\,\d\varphi\,\d t\,\d s.
\end{equation}
Consequently,
\begin{align}
\delta^\can_3(\k,\nu,\hat{v})
	=&
	-\int_{\frac{k}{2}}^{+\infty}
	\int_{-\frac{1}{2}}^{\frac{1}{2}}
	\frac{\d t\,\d s}{4\pi^2}\,
	\frac{j(k;p,q)^2pq}
	{\omega_\Bog(p)+\omega_\Bog(q)-\omega_\Bog(k)}
	\label{eq: real divergence fermi-}
	\\
	&-
	\int_{\frac{k}{2}}^{+\infty}
	\int_{-\frac{1}{2}}^{\frac{1}{2}}
	\frac{\d t\,\d s}{4\pi^2}\,
	\frac{\kappa(k,p,q)^2pq}
	{\omega_\Bog(p)+\omega_\Bog(q)+\omega_\Bog(k)}.
	\label{eq: real divergence fermi2-}
\end{align}

The two interaction kernels can be decomposed as
\begin{align}
	j(k;p,q)
	&=
	\frac{1}{2}\sqrt{\frac{\nu}{\hat v(0)}}
	\Bigl[
	(c_k+s_k)j_+(p,q)
	+(c_k-s_k)j_-(k,p,q)
	\Bigr],
	\\
	\kappa(k,p,q)
	&=
	\frac{1}{2}\sqrt{\frac{\nu}{\hat v(0)}}
	\Bigl[
	-(c_k+s_k)j_+(p,q)
	+(c_k-s_k)j_-(k,p,q)
	\Bigr],
\end{align}
where
\begin{align}
	j_+(p,q)
	={}&
	(c_pc_q-s_ps_q)\bigl(\hat v(p)+\hat v(q)\bigr)
	\nonumber\\
	&+
	(c_ps_q-s_pc_q)\bigl(\hat v(p)-\hat v(q)\bigr),
	\\
	j_-(k,p,q)
	={}&
	(c_pc_q+s_ps_q)\bigl(\hat v(p)+\hat v(q)\bigr)
	\nonumber\\
	&-
	\bigl(2\hat v(k)+\hat v(p)+\hat v(q)\bigr)
	(c_ps_q+s_pc_q).
\end{align}

Using $(c_k+s_k)(c_k-s_k)=1$, we can therefore combine the two
Fermi golden rule channels as
\begin{align}
	&\delta^\can_3(\k,\nu,\hat{v})
	=
	-\frac{\nu}{16\pi^2\hat v(0)}
	\int_{\frac{k}{2}}^{+\infty}
	\int_{-\frac{1}{2}}^{\frac{1}{2}}
	\d t\,\d s\,pq
	\Bigg[ \notag \\ 
	&(c_k+s_k)^2j_+(p,q)^2	\frac{2(\omega_\Bog(p)+\omega_\Bog(q))}
	{(\omega_\Bog(p)+\omega_\Bog(q))^2-\omega_\Bog(k)^2}
	\nonumber\\
	&+
	2j_+(p,q)j_-(k,p,q)	\frac{2\omega_\Bog(k)}
	{(\omega_\Bog(p)+\omega_\Bog(q))^2-\omega_\Bog(k)^2}
	\nonumber\\
	&+
	(c_k-s_k)^2j_-(k,p,q)^2
	\frac{2(\omega_\Bog(p)+\omega_\Bog(q))}
	{(\omega_\Bog(p)+\omega_\Bog(q))^2-\omega_\Bog(k)^2}
	\Bigg].
	\label{eq: combined FGR channels}
\end{align}
We have
\begin{align}\label{prefa}
	(c_k+s_k)^2
	&=
	\frac{2\sqrt{\nu}}{k}
	+O\left(\frac{k}{\sqrt{\nu}}\right),
	\\\omega_\Bog(k)&=\sqrt\nu k+O\left(\frac{k^2}{\nu}\right),\\
	(c_k-s_k)^2
	&=
	\frac{k}{2\sqrt{\nu}}+o\left(\frac{k}{\sqrt{\nu}}\right).
\end{align}
The dominant contributions to the integrals in $\delta_3^\can$ will come from the 
diagonal $p=q=t$. For $j_-$ it will suffice to use
\begin{align}
  	j_-(k,p,q)
	&=
   j_-(0,t,t)+O\left(\frac{k}{\sqrt{\nu}}\right),\\
	j_-(0,t,t)&
 	=
 	\frac{\hat{v}(t)\bigl(t^2-2\nu\bigr)}
 	{\omega_\Bog(t)}.  
\end{align}
For $j_+$, because of the more singular prefactor \eqref{prefa}, we will need a more precise asymptotics around the diagonal:
 \begin{align}
 	j_+(p,q)
 	={}
 	j_+(t,t)&
 	+\frac{k^2s^2}{2}
 	\left.
 	\bigl(\partial_p-\partial_q\bigr)^2j_+(p,q)
 	\right|_{p=q=t}
 	+o\left(\frac{k^2}{\nu}\right),
   \\	j_+(t,t)&=2\hat v(t),\\
\label{eq: derivative values I.}
 	\left.
 	\bigl(\partial_p-\partial_q\bigr)^2j_+(p,q)
 	\right|_{p=q=t}
 	&=
 	2\hat{v}''(t) -
 	8\hat{v}'(t)
 	\left(
 	\frac{\omega_\Bog'(t)}
 	{2\omega_\Bog(t)}
 	-\frac{1}{t}
    \right)
 	\nonumber\\
 	&
+8\hat{v}(t)
 	\left(
 	\frac{\omega_\Bog'(t)}
 	{2\omega_\Bog(t)}
 	-\frac{1}{t}
 	\right)^2
 	.
 \end{align}
The identity \eqref{eq: derivative values I.}  in Lemma in Subsection \ref{app: proofs estimates} below.

The remaining factors are symmetric under $p\leftrightarrow q$. Hence, in
the $t$--$s$ coordinates, their formal expansions contain only even powers
of $k$:
\begin{align}
	pq
	&=
	t^2-k^2s^2,
\\
 	\omega_\Bog(p)+\omega_\Bog(q)
 	=&
 	2\omega_\Bog(t)
 	+k^2s^2\omega_\Bog''(t)
 	+o(k^2),
 	\\
 	\frac{2\bigl(\omega_\Bog(p)+\omega_\Bog(q)\bigr)}
 	{\bigl(\omega_\Bog(p)+\omega_\Bog(q)\bigr)^2-\omega_\Bog(k)^2}
 	=&
 	\frac{1}{\omega_\Bog(t)}
 	+\frac{\nu k^2}{4\omega_\Bog(t)^3}
 	\nonumber\\
 	&-
 	\frac{k^2s^2\omega_\Bog''(t)}
 	{2\omega_\Bog(t)^2}
 	+o\left(\frac{k^2}{\nu}\right),
 	\\
 	\frac{1}
 	{\bigl(\omega_\Bog(p)+\omega_\Bog(q)\bigr)^2-\omega_\Bog(k)^2}
 	=&
 	\frac{1}
 	{4\omega_\Bog(t)^2}
 	+O\left(\frac{k^2}{\nu}\right).
 \end{align}

Keeping the terms of the order $k^{-1}$ and $k$ in \eqref{eq: combined
  FGR channels}, we expect the following asymptotics for small $k$:
\begin{align}
\delta^\can_3(\k,\nu,\hat{v})
	=&-c_\ir(k,\nu,v)
	+k\,c_3(\nu,\hat 
           v)+o\left(\frac{k}{\sqrt{\nu}}\right),
\end{align}
where
\begin{align}c_\ir(k,\nu,\hat v)&=
	\frac{\nu}{16\pi^2\hat v(0)}
	(c_k+s_k)^2\int_{0}^{+\infty}
	\d t   \,       t^2 \frac{j_+(t,t)^2 	}   {\omega_\Bog(t)}
\\
	&=
	\frac{\nu\omega_\Bog(k)}{2k^2\pi^2\hat v(0)}
	\int_{0}^{+\infty}
	\d t\,\frac{ t^2\hat v(t)^2}  {\omega_\Bog(t)}
	\label{eq: formal FGR asymptotics}
\end{align}
is the term of order $O(k^{-1})$, the same as in \eqref{cir}, except
that it has  the opposite
sign as in the Hellman--Feynman correction, and
\begin{align}\label{expli}
 c_3(\nu,\hat 
           v)
  =&\frac{\nu \hat{v}(0)}{16\pi^2}-\frac{\nu}{16\pi^2\hat v(0)}
	\int_{0}^{+\infty}
     \d t 
     \\&\Bigg[ 2\sqrt\nu j_+(t,t)^2	\left(-\frac{1}
    {12\omega_\Bog(t)}
    +\frac{\nu t^2}{4\omega_\Bog(t)^3}- \frac{t^2\omega_\Bog''(t)}{24\omega_\Bog(t)^2}
  \right)\\&+
\sqrt\nu\frac{t^2j_+(t,t)}{6\omega_\Bog(t)} 	\left.
 	\bigl(\partial_p-\partial_q\bigr)^2j_+(p,q)
 	\right|_{p=q=t}\\
	&+
	\sqrt\nu \frac{j_+(t,t)j_-(0,t,t)	t^2}{\omega_\Bog(t)^2}
+
	\frac{1}{\sqrt\nu}\frac{j_-(0,t,t)^2t^2}
	{2\omega_\Bog(t)}
	\Bigg]
	\label{eq: combined FGR channels.}
\end{align}
is the Fermi golden rule contribution to the phonon velocity correction. In the above,
we used $\int_{-\frac12}^{\frac12}s^2\d s=\frac1{12}$.  The first term
on the right of \eqref{expli} (without  integration) comes from the
integral of the infrared term \eqref{eq: formal FGR asymptotics}
in the region $\Bigl[0,\frac{k}{2}\Bigr]$: \begin{equation}\frac{\nu\omega_\Bog(k)}{2k^2 \pi^2 \hat{v}(0)} \int^{\frac{k}{2}}_0 \d t \, \frac{t^2 \hat{v}(t)^2}{\omega_\Bog(t)}=
k\frac{\nu \hat{v}(0)}{16\pi^2} + o\left(\frac{k}{\sqrt{\nu}}\right) .
\end{equation}

Next we  observe that
\begin{align}
	&
	\frac{\nu^{3/2}t^2j_+(t,t)^2}{2\omega_\Bog(t)^3}
	+
	\sqrt{\nu}\,
	\frac{t^2j_+(t,t)j_-(0,t,t)}{\omega_\Bog(t)^2}
	+
	\frac{t^2j_-(0,t,t)^2}
	{2\sqrt{\nu}\,\omega_\Bog(t)}
	\notag \\
	=&
	\frac{t^6\hat v(t)^2}
	{2\sqrt{\nu}\,\omega_\Bog(t)^3}.\label{idon}
\end{align}
Using \eqref{idon}, inserting
\eqref{eq: derivative values I.} we obtain
\begin{align}
 c_3(\nu,\hat 
           v)
  =&\frac{\nu\hat{v}(0)}{16\pi^2}-\frac{\nu}{16\pi^2\hat v(0)}
	\int_{0}^{+\infty}
     \d t \label{eq: combined FGR channels..1}
  \\&\Bigg[
-\frac{\sqrt{\nu}t^2\hat{v}(t)^2\omega_\Bog''(t)}{3\omega_\Bog(t)^2}+\frac{2\sqrt\nu t^2\hat
  v(t)}{3\omega_\Bog(t)}\left(\hat v''(t)-4\hat
  v'(t)\Big(\frac{\omega_\Bog'(t)}{2\omega_\Bog(t)}-\frac1t\Big)\right)\\
&+\frac{2\sqrt{\nu}\hat
                                                                            v(t)^2}{3\omega_\Bog(t)}\left(-1+4t^2\Big(\frac{\omega_\Bog'(t)}{2\omega_\Bog(t)}-\frac1t\Big)^2\right)+\frac{t^6\hat v(t)^2}{2\nu^{\frac12}\omega_\Bog(t)^3}
	\Bigg].
	\label{eq: combined FGR channels..}
\end{align}

The integral in the preceding expression is absolutely convergent.
The second term in \eqref{eq: combined FGR channels..} (which comes from
\eqref{idon}) is regular at origin.

Let us look at the first term of
\eqref{eq: combined FGR channels..}.
By the regularity and
radial symmetry of the potential,
\begin{equation}
\hat v(t)
=
\hat v(0)+\frac{\hat v''(0)}{2}t^2+O(t^4),
\end{equation}
while
\begin{equation}
\omega_\Bog(t)
=
\sqrt{\nu}\,t
+
\left(
\frac{1}{8\sqrt{\nu}}
+
\frac{\sqrt{\nu}\,\hat v''(0)}
{4\hat v(0)}
\right)t^3
+O(t^5).
\end{equation}
It follows that
\begin{equation}
\frac{\omega_\Bog'(t)}{2\omega_\Bog(t)}
-\frac{1}{t}
=
-\frac{1}{2t}+O(t),
\qquad
\omega_\Bog''(t)=O(t),
\end{equation}
and therefore,
\begin{equation}
  -1+4t^2\Big(\frac{\omega_\Bog'(t)}{2\omega_\Bog(t)}-\frac1t\Big)^2=O(t^2).\end{equation}
Thus the integrand in
\eqref{eq: combined FGR channels..} is $O(t)$ at origin.

We next consider the behavior at infinity $t\to+\infty$. In this regime,
\begin{equation}
\omega_\Bog(t)= O( t^2),
\qquad
\omega_\Bog''(t)=O(1).
\end{equation}
Moreover, direct differentiation of the Bogoliubov dispersion relation
gives
\begin{equation}
\frac{\omega_\Bog'(t)}{2\omega_\Bog(t)}
-\frac{1}{t}
=
\frac{
	\nu\bigl(t\hat v'(t)-2\hat v(t)\bigr)
}{
	t\bigl(\hat v(0)t^2+4\nu\hat v(t)\bigr)
}.
\end{equation}
Consequently,
\begin{equation}
\left|
\frac{\omega_\Bog'(t)}{2\omega_\Bog(t)}
-\frac{1}{t}
\right|
\leq
C\left(
\frac{|\hat v'(t)|}{t^2}
+
\frac{|\hat v(t)|}{t^3}
\right).
\end{equation}
Since $\hat v$ and $\hat v'$ are bounded under our assumptions, the
explicit formula for the derivative of $j_+$ yields
\begin{equation}
\left|
\left.
(\partial_p-\partial_q)^2j_+(p,q)
\right|_{p=q=t}
\right|
\leq
C\left(
|\hat v''(t)|
+
\frac{|\hat v'(t)|}{t^2}
+
\frac{|\hat v(t)|}{t^4}
\right).
\end{equation}
It follows that integral \eqref{eq: combined FGR channels} is absolutely integrable at infinity.

In Section \ref{Estimates on perturbative corrections} we justify these expansions
  proving the following proposition:
  \begin{prop}\label{prop: FG term}
    Under Assumptions \ref{ass: primary assumption on the potential} and \ref{ass: secondary hypotheses}, we have
    \begin{align}
	\delta_3^\can(\k,\nu,\hat{v})
	=&-c_\ir(k,\nu,\hat{v})
	+k\,c_3(\nu,\hat 
           v)+o\left(\frac{k}{\sqrt{\nu}}\right),
\end{align}
where $c_\ir$ is given in \eqref{eq: formal FGR asymptotics} and $c_3$
in
\eqref{eq: combined FGR channels..1}-\eqref{eq: combined FGR channels..}.
    \end{prop}

\subsection{Real energy shift}\label{sec: real energy shift} In this subsection we present the general second-order expression
for the real energy shift. This is obtained by summing the contributions $\delta_3^\can$ and $\delta_4^\can$ from Propositions \ref{prop: FG term} and \ref{prop: HF term}.

After summing the two contributions, the infrared-divergent terms $c_\ir(\k,\nu,\hat{v})$ and $-c_\ir(\k,\nu,\hat{v})$ cancel, leaving a  correction whose leading term is linear in $\k$ as $\frac{|\k|}{\sqrt{\nu}}\to0$. To simplify the derivation, throughout this subsection we will fix $\nu= \hat{v}(\0) = 1$. The dependence on these parameters will be restored at the end. 

Writing \begin{equation}
s_p^2 = \frac{\frac{p^2}{2}+\hat{v}(p)-\omega_\Bog(p)}{2\omega_\Bog(p)},
\end{equation} we can rewrite the Hellmann-Feynman correction
\eqref{c4a}-\eqref{c4b} as 
\begin{align}
  c_4(1,\hat v)=&\frac{1}{12\pi^2}\int_0^\infty\d p s_p^2\frac{\d}{\d 
                  p}\big(p^2\hat v'(p)-p^3\big)\label{c4.1}\\
  +&\frac{1}{24\pi^2}\int_0^\infty\d p \frac{\hat v(p)}{\omega_\Bog(p)}\frac{\d}{\d 
                  p}\big(p^2\hat v'(p)\big).\label{c4.2}
\end{align}

According to
\eqref{eq: combined FGR channels..1}--\eqref{eq: combined FGR channels..} the Fermi Golden
Rule contribution to the phonon velocity is
\begin{align}
	& c_3(1,\hat{v})
	=-\frac{1}{32\pi^2}\int_0^\infty\d p\,
	\frac{p^6\hat v(p)^2}{\omega_\Bog(p)^3}
	\label{eq: relevant term ccan}
		+\frac{1}{16\pi^2}\\
	&-\frac{1}{16\pi^2}\int_0^\infty\d p\,
	\Biggl[-\frac{p^2\hat v(p)^2\omega_\Bog''(p)}
	{3\omega_\Bog(p)^2}
	+\frac{2p^2\hat v(p)}{3\omega_\Bog(p)}
	\Biggl(\hat v''(p)\notag \\ 
	&\left. -4\hat v'(p)
	\left(\frac{\omega_\Bog'(p)}{2\omega_\Bog(p)}-\frac1p\right)
	\right)
	+\frac{2\hat v(p)^2}{3\omega_\Bog(p)}
	\left(-1+4p^2
	\left(\frac{\omega_\Bog'(p)}{2\omega_\Bog(p)}-\frac1p\right)^2
	\right)\Biggr] .
\end{align}

While the above formula might look extremely overwhelming, there are
some hidden simplifications that will render it considerably
simpler. The integrand in the last two lines can be rewritten as
\begin{align}
  &+\frac{p^2\omega_\Bog''(p)\hat{v}(p)^2}{48\omega_\Bog(p)^2}-\frac{p^2
    \hat
    v(p)\hat{v}''(p)}{24\omega_\Bog(p)}+\frac{p^2\omega_\Bog'(p)\hat{v}(p)\hat{v}'(p)}{12\omega_\Bog(p)^2}-\frac{p\hat{v}'(p)\hat{v}(p)}{6\omega_\Bog(p)}
  \\&
  -\frac{\omega_\Bog'(p)^2\hat{v}(p)^2p^2}{24\omega_\Bog(p)^3}+
\frac{\omega_\Bog'(p)\hat{v}(p)^2p}{6\omega_\Bog(p)^2}-\frac{\hat{v}(p)^2}{8\omega_\Bog(p)}\\
= 
 &-\frac{\omega_\Bog'(p)^2\hat{v}(p)^2p^2}{24\omega_\Bog(p)^3}+\frac{p^2\omega_\Bog''(p)\hat{v}(p)^2}{48\omega_\Bog(p)^2}+\frac{p^2\omega_\Bog'(p)\hat{v}(p)\hat{v}'(p)}{24\omega_\Bog(p)^2}\label{line1}\\ &+\frac{\hat{v}(p)^2}{24\omega_\Bog(p)} +\frac{p\hat{v}'(p)\hat{v}(p)}{12\omega_\Bog(p)}
\label{line2}\\&
-\frac{\hat{v}(p)}{24\omega_\Bog(p)}\left(4\hat{v}(p)+4p\hat{v}'(p)-\frac{\omega_\Bog'(p)}{\omega_\Bog(p)}[p^2\hat{v}'(p)+4\hat{v}(p)p]\right)\label{line3}
\\ &-\frac{p\hat{v}(p)}{24\omega_\Bog(p)}(p\hat{v}''(p)+2\hat{v}'(p)).\label{line4}\end{align}
Let us analyze separately the expressions in the four lines \eqref{line1}--\eqref{line4}:
\begin{align} \eqref{line1}
  +\frac{p\omega_\Bog'(p)\hat{v}(p)^2}{24\omega_\Bog(p)^2}
	=&\frac{\d}{\d p}\frac{p^2\omega_\Bog'(p)\hat{v}(p)^2}{48\omega_\Bog(p)^2}
,\\
\eqref{line2}-\frac{p\omega_\Bog'(p)\hat{v}(p)^2}{24\omega_\Bog(p)^2}
=&\frac{\d}{\d p}\frac{p\hat{v}(p)^2}{24\omega_\Bog(p)}. 
  \label{eq: last two line integrand1}
\end{align}
Next, using
      \begin{equation}\omega_\Bog(p)^2=\frac{1}{4}p^4+p^2\hat v(p),\quad
\omega_\Bog(p)        \omega_\Bog'(p) = \frac12p^3+p\hat{v}(p)+\frac12p^2\hat{v}'(p),
\end{equation}
we compute
\begin{align}
\eqref{line3}&= \frac{\hat{v}(p)}{24\omega_\Bog(p)^3}\Big( -
   4\omega_\Bog(p)^2(\hat{v}(p)+p\hat{v}'(p))+\left. \omega_\Bog(p)\omega_\Bog'(p)(p^2\hat
   v'(p)+4p\hat v(p)\right)\Big)\notag \\
=& \frac{p^2}{12}(\hat{v}'(p)-p)\frac{p\hat{v}(p)(\hat{v}'(p)p-2\hat{v}(p))}{4\omega_\Bog(p)^3}\notag \\
=& \frac{p^2}{12}(\hat{v}'(p)-p)\frac{\d}{\d p}s_p^2\label{eq: second final}
      \end{align}
Finally,
 \begin{equation}\label{eq: third term}
\eqref{line4}=
  -\frac{\hat{v}(p)}{24\omega_\Bog(p)}\frac{\d}{\d p} p^2\hat{v}'(p).
\end{equation}

Thus
\begin{align}
	& c_3(1,\hat{v})
	=-\frac{1}{32\pi^2}\int_0^\infty\d p\,
	\frac{p^6\hat v(p)^2}{\omega_\Bog(p)^3}
		+\frac{1}{16\pi^2}\label{c3.1}\\
	&+\frac{1}{12\pi^2}\int_0^\infty\d p\,
(p^2\hat v'(p)-p^3)\frac{\d}{\d p}s_p^2\\
&-\frac{1}{24\pi^2}\int_0^\infty\d p \frac{\hat v(p)}{\omega_\Bog(p)}\frac{\d}{\d 
                                            p}\big(p^2\hat v'(p)\big)\\
        &+\frac1{\pi^2}
          \int_0^\infty\d p
          \Big(\frac{\d}{\d p}\frac{p^2\omega_\Bog'(p)\hat{v}(p)^2}{48\omega_\Bog(p)^2}
+
\frac{\d}{\d p}\frac{p\hat{v}(p)^2}{24\omega_\Bog(p)}\Big).  \label{c3.4}
\end{align}
Adding \eqref{c4.1}-\eqref{c4.2} and \eqref{c3.1}--\eqref{c3.4}, and performing the integration of the
derivatives we obtain
\begin{align}
	& c_3(1,\hat{v})+c_4(1,\hat{v})
	=-\frac{1}{32\pi^2}\int_0^\infty\d p\,
   \frac{p^6\hat v(p)^2}{\omega_\Bog(p)^3}+\frac{1}{16\pi^2}\\+&
                                               \frac{1}{8\pi^2}\Big(
	\frac23p^2s_p^2\bigl(p-\hat v'(p)\bigr)
	+\frac{p^2\hat v(p)^2\omega_\Bog'(p)}{6\omega_\Bog(p)^2}
	+\frac{p\hat v(p)^2}{3\omega_\Bog(p)}\Big)\Bigg|_0^{+\infty}.  
\end{align}

At the origin $$s_p = O(\sqrt{p}^{-1}),\quad\hat{v}(p) = 1 +O(p),\quad \hat{v}'(p) = O(p),\quad \omega_\Bog(p) =O(p),\quad \omega_\Bog'(p) = 1 + O(p).$$ Therefore,  we have 

\begin{equation}
	\lim_{p\to 0}\frac{1}{8\pi^2}\Bigl[
	\frac23p^2s_p^2\bigl(p-\hat v'(p)\bigr)
	+\frac{p^2\hat v(p)^2\omega_\Bog'(p)}{6\omega_\Bog(p)^2}
	+\frac{p\hat v(p)^2}{3\omega_\Bog(p)}\Bigr] = \frac{1}{16\pi^2},
\end{equation} which cancels with the positive $\frac{1}{16\pi^2}$ contribution in $c_3+c_4$.
At infinity instead, by  $s_p = O(p^{-2})$,  $\omega_\Bog(p) = \frac{p^2}{2}+O(p)$ and $\omega_\Bog'(p) = O(p)$  we find \begin{equation}
	\lim_{p\to +\infty}-\frac{1}{8\pi^2}\Bigl[
	\frac23p^2s_p^2\bigl(p-\hat v'(p)\bigr)
	-\frac{p^2\hat v(p)^2\omega_\Bog'(p)}{6\omega_\Bog(p)^2}
	-\frac{p\hat v(p)^2}{3\omega_\Bog(p)}\Bigr] = 0.
\end{equation}

We conclude that
\begin{equation*}
	c_3(1,\hat v)+c_4(1,\hat v)
	=-\frac{1}{32\pi^2}\int_0^\infty\d p\,
	\frac{p^6\hat v(p)^2}{\omega_\Bog(p)^3}.
\end{equation*}

Finally, rescaling momenta by $\sqrt\nu$ and replacing the normalized
potential by $\frac{\hat v(\sqrt\nu\,\cdot)}{\hat v(0)}$ gives
\begin{align}
	&c_3(\nu,\hat v)+c_4(\nu,\hat v)=\nu\hat v(0)
	\Bigl[c_3\left(1,\frac{\hat v(\sqrt\nu\,\cdot)}{\hat v(0)}\right)
	+c_4\left(1,\frac{\hat v(\sqrt\nu\,\cdot)}{\hat v(0)}\right)\Bigr]\notag \\
	&=-\frac{\sqrt\nu}{32\pi^2\hat v(0)}
	\int_0^\infty\d p\,
	\frac{p^6\hat v(p)^2}{\omega_\Bog(p)^3}.
\end{align}
Restoring the perturbative
factor $\kappa$ and using $\omega_\Bog(k)=\sqrt\nu\,k+O(k^3)$ for
fixed $\nu>0$, we have obtained:
\begin{thm}\label{thm: canonical dispersion}
	Suppose that the potential satisfies Assumptions \ref{ass: primary assumption on the potential} and \ref{ass: secondary hypotheses}. Then, for fixed $\nu\in ]0,\nu_0]$, the Bogoliubov dispersion relation, including the second-order perturbative correction, satisfies \begin{align}
		\omega_\Bog(k)+\kappa\left(\delta^\can_3(\k,\nu,\hat{v})+\delta_4^\can(\k,\nu,\hat{v})\right)= kc_\mathrm{can}(\nu, \hat v) + o\left(\frac{k}{\sqrt{\nu}}\right),\notag \\ \text{ as } \frac{k}{\sqrt{\nu}} \to 0
		\label{eq: physical correction braket I}\end{align} where the coefficient $c_\mathrm{can}$ is given by 
	\begin{align}
		&	c_\mathrm{can}(\nu,\hat v) =
		\sqrt{\nu} - \kappa \hat{v}(0)\sqrt{\nu}\int^{+\infty}_{0}\frac{\d p}{32\pi^2}\frac{\hat v(p)^2}{\hat{v}(0)^2}\frac{p^6}{\omega_\Bog(p)^3}  \label{eq: cbel} 
	\end{align} 
\end{thm}

      \subsection{Proof of identity \eqref{eq: derivative values I.}}
\label{app: proofs estimates}
Recall that we are fixing $\nu = 1$ and $\hat{v}(0)=1$

\begin{lemma} The identity \eqref{eq: derivative values I.} holds.
	\label{identi}  \end{lemma}
\begin{proof}
	We first compute
	\begin{align}\label{eq: derivative values}
		\left.
		\bigl(\partial_p-\partial_q\bigr)^2j_+(p,q)
		\right|_{p=q=t}
		=&
		2\hat{v}''(t)
		+ 8\hat{v}'(t)(c_t's_t -s_t'c_t) -8((c_t'^2)-(s_t')^2)\hat{v}(t)
	\end{align} This can be expressed explicitly in terms of $\hat{v}(t),\; \hat{v}'(t)$ and $\omega_\Bog(t),\omega_\Bog'(t)$. We write first $(c_t')^2-(s_t')^2$. We have  \begin{equation}
		c_t'-s_t' = \frac{\d}{\d t}\left( \frac{t}{\sqrt{2\omega_\Bog(t)}}\right) =(c_t-s_t) \left( \frac{1}{t} -\frac{\omega_\Bog'(t)}{2\omega_\Bog(t)}\right),\label{eq:ratio and derivative}
	\end{equation} and \begin{equation}
		c_t'+s_t' = -\frac{c_t'-s_t'}{(c_t-s_t)^2},
	\end{equation} which imply \begin{align}
		(c_t')^2-(s_t')^2 = & (c_t'-s_t')(c_t'+s_t') \notag \\
		=& -\left(\frac{c_t'-s_t'}{c_t-s_t}\right)^2  =  -\left( \frac{1}{t} -\frac{\omega_\Bog'(t)}{2\omega_\Bog(t)}\right)^2.
	\end{align} In the last equality above we used Equation \eqref{eq:ratio and derivative}. To derive a useful expression for $c_t's_t -s_t'c_t$  we write \begin{equation}
		\frac{\d}{\d t}(c_t^2-s_t^2) = 2c_tc_t' -2s_ts_t' = 0, 
	\end{equation} which gives us  \begin{equation}
		c_t's_t - s_t'c_t = c_t's_t-s_t'c_t +c_t'c_t-s_t's_t = \frac{c_t+s_t}{c_t+s_t}\frac{c_t'-s_t'}{c_t-s_t}.
	\end{equation} Using again Equation \eqref{eq:ratio and derivative} we arrive at \begin{equation}
		c_t's_t -s_t'c_t = \left( \frac{1}{t} -\frac{\omega_\Bog'(t)}{2\omega_\Bog(t)}\right).
	\end{equation} Thus, we finally arrive at the expression
	\eqref{eq: derivative values I.}.
\end{proof}

	\subsection{Comparison of weak coupling and dilute  limits}\label{sec: dilute gas limit}
        Equation \eqref{eq: cbel} provides the first-order correction
        in $\kappa$ to the Bogoliubov dispersion relation. We
        emphasize that this formula is derived within perturbation
        theory, in the weak-coupling regime $\hat{v}\to \kappa
        \hat{v}$. Our aim is to relate this result, at least at a
        formal level, to the better known corrections obtained by Beliaev in the dilute regime.
	In that setting, the relevant parameters are the gas density
        $n$ and the scattering length $\mathfrak{a}$, with the product
        $n\mathfrak{a}^3$ assumed to be small.   It is well known
        that, under suitable assumptions, the scattering length can be
        expanded as a series in the interaction potential, see
        \cite{BBCS_2019,ErdosSchleinYau2008} { and Appendix \ref{app: born series}:\begin{align}
&		4\pi\mathfrak{a} = \kappa\hat{v}(\0) \notag \\ &+\sum_{k=1}^{+\infty}(-1)^k\kappa^{k+1}\int \frac{\d\p_1\dots \d\p_k}{((2\pi)^{3})^k}\frac{\hat{v}(\p_1)\hat{v}(\p_1-\p_2)\dots \hat{v}(\p_{k-1}-\p_k)\hat{v}(\p_k)}{|\p_1|^2\dots |\p_k|^2},
                                                                  \notag	\\
                                                                       &=4\pi\sum_{k=0}^\infty\kappa^{k+1}\mathfrak{a}_k,\end{align}}which is convergent for sufficiently regular and small potentials.

	The first two terms of this series are \begin{equation}
		\mathfrak{a}_0 = \frac{\hat{v}(\0)}{4\pi},\qquad \mathfrak{a}_1 = -\frac{1}{4\pi}\int \frac{\d \p}{(2\pi)^3}\frac{\hat{v}(\p)^2}{|\p|^2}=-\int^{+\infty}_0 \frac{\d p}{(2\pi)^3}\hat{v}(p)^2.
	\end{equation}
	
	All terms in the Born series beyond $\mathfrak{a}_0$ diverge in the formal
	flat-Fourier limit $\hat v(k)=\hat v(0)$.

Recall that $\nu=\kappa\hat v(\0)n$,
$\omega_\Bog(p)=\sqrt{\frac{p^2}{4}+\frac{\hat v(p)}{\hat v(0)}\nu
  p^2}$.    
		The second term  in $c_\mathrm{can}$ can be written
                as \begin{align}
&-\kappa\hat v(0)\nu
			\int_0^\infty\frac{\d p}{32\pi^2}\,
			\frac{\hat v(\sqrt\nu p)^2}{\hat
                                   v(0)^2}\frac{\nu^3 p^6}{\omega_\Bog(\sqrt\nu
                                   p)^3}\\
=&-\frac{\kappa\sqrt\nu}{\hat v(0)}\int_0^{+\infty} \frac{\d p}{4\pi^2}\hat{v}(p)^2.
                  \label{term1}   \\
		&-\kappa\hat v(0)\nu \int_0^{+\infty} \frac{\d p}{32\pi^2}\left(\frac{p^6}{(\frac{p^4}{4}+p^2)^{\frac32}}- 8\right)  \label{term2}\\
&-\kappa\hat v(0)\nu
			\int_0^\infty\frac{\d p}{32\pi^2}\,
			\Big(\frac{\hat v(\sqrt\nu p)^2}{\hat
                                   v(0)^2}-1\Big)\Big(\frac{\nu^3p^6}{(\frac{p^4}{4}+p^2)^{\frac32}}-8\Big)\label{term3}\\
&-\kappa\hat v(0)\nu
			\int_0^\infty\frac{\d p}{32\pi^2}\,
			\frac{\hat v(\sqrt\nu p)^2}{\hat
                                   v(0)^2}\Big(
                                                    \frac{\nu^3p^6}{\omega_\Bog(\sqrt\nu
                                   p)^3}-\frac{p^6}{(\frac{p^4}{4}+p^2)^{\frac32}}\Big).\label{term4}
                   \end{align}
                   Now $c_\can$ consists of the following terms:
                   \begin{align}\sqrt\nu&=\sqrt{ 4\pi \kappa\mathfrak{a}_0n},\label{term0}\\
                     \eqref{term1}&=\frac{\sqrt{4\pi
                                    n}\kappa^2\mathfrak{a_1}}{2\sqrt{\kappa\mathfrak{a}_0}},\\
                     \eqref{term2}&=\frac{\kappa\hat
                                    v(0)\nu}{\pi^2}=\frac{\sqrt{4\pi 
                                    n\kappa \mathfrak{a}_0}\sqrt{n(4\pi 
                                  \kappa  \mathfrak{a}_0)^3}}{\pi^2}\\
                     &=\frac{\sqrt{4\pi 
                                    n\mathfrak{a}}\sqrt{n(4\pi 
                       \mathfrak{a})^3}}{\pi^2}+O(\kappa^{\frac52}),\\
                     \eqref{term3}&=O(\kappa^{\frac52}),\\
                                          \eqref{term4}&=O(\kappa^{\frac52}).
                                    \end{align}

                   Now the first two terms of $c_\mathrm{can}$ 
                   give \begin{align}\sqrt\nu+\eqref{term1}&=\sqrt{4\pi 
                                                                  n}\Big(\sqrt{\kappa\mathfrak{a}_0}+\frac{\kappa^2\mathfrak{a}_1}{2\sqrt{\kappa 
                                                                  \mathfrak{a}_0}}\Big)\\
                          &=\sqrt{4\pi 
                            n}\sqrt{\kappa\mathfrak{a}_0+\kappa^2\mathfrak{a}_1}+O(\kappa^{\frac52})
\,=\, \sqrt{4\pi 
                            n\mathfrak{a}}+O(\kappa^{\frac52}).
         	\end{align} 
	Adding the contribution from \eqref{term2}, we obtain \begin{align}
	c_\mathrm{can}(\nu,\hat{v}) 
	=&\sqrt{4\pi n\mathfrak{a}}\left(1+ \frac{1}{\pi^2}\sqrt{n(4\pi\mathfrak{a})^3}\right)+O(\kappa^{\frac52}).
	\end{align}
	
Next we express the gas density in terms of the density of the condensate: \begin{align}
          n =& n_{0,\Bog} + \frac{\nu^{\frac{3}{2}}}{3\pi^2}
= n_{0,\Bog}\left(1 +\frac{\sqrt{n_{0,\Bog}(4\pi \kappa \mathfrak{a}_0)^3}}{3\pi^2}\right)\\
          =& n_{0,\Bog}\left(1 +\frac{\sqrt{n_{0,\Bog}(4\pi
             \mathfrak{a})^3}}{3\pi^2}\right)+O(\kappa^{\frac52}),\\
\sqrt{          n} 
          =&\sqrt{ n_{0,\Bog}}\left(1 +\frac{\sqrt{n_{0,\Bog}(4\pi
             \mathfrak{a})^3}}{6\pi^2}\right)+O(\kappa^{\frac52}).                                            
	\end{align}Thus we can recast the correction as \begin{equation}
	c_\mathrm{can}(\nu,\hat{v}) =  \sqrt{4\pi n_{0,\Bog}\mathfrak{a}}\left(1+ \frac{7}{6\pi^2}\sqrt{n_{0,\Bog}(4\pi\mathfrak{a})^3}\right)+O(\kappa^{\frac52}),
	\end{equation} which is consistent with Beliaev's original result \cite{B_58}. This essentially proves Corollary \ref{cor: dilute limit}

				\section{Grand-canonical derivation}
	\label{Grand canonical derivation}
	
	Let us describe an alternative approach to the Bose gas based on
        the grand-canonical formalism.
We will see that it yields equivalent description of the shift of the
quasiparticle spectrum as the canonical approach, which we applied so
far in our paper.
        
	Let $H^L$ be the ``first-principles Hamiltonian'' \eqref{2BB},
        where we make the coupling constant $\kappa$ explicit. It acts
        on the full Fock space $\Gamma_\s(l^2(\Xi_L))$, which includes the
        zero-mode.
	Introduce the  grand-canonical Hamiltonian
	\beq
	H_\mu^L:= H^L-\mu N^L,
	\eeq
	where $\mu$ is the \textit{grand canonical chemical
          potential}.
       We expect that at zero temperature
          the physical system is described by the
  ground state of $H_\mu^L$ and its low-lying states.

        Again, we will view $H_\mu^L$ as a perturbation of an appropriate quadratic
        Hamiltonian, whose choice we are now going to describe.
	First for $\alpha\geq0$ we introduce the Weyl operator of the mode $\kb=0$
	\beq \label{def:weylop}
	W_\alpha=\exp(-\alpha a_0^*+\bar{\alpha}a_0).
	\eeq
We set $\Omega_\alpha =W^*_\alpha \Omega$.
  Instead of
   $\alpha$ in the sequel we will often
  prefer the parameter $\mu_0 \in \mathbb{R}_{+}$ so that \begin{equation}
		\alpha = \sqrt{\frac{\mu_0 L^d}{\kappa \hat{v}(\0)}}.
	\end{equation}

        	We compute
	\begin{align}
		(\Omega_\alpha|a_0\Omega_\alpha)= (\Omega_\alpha|a_0^*\Omega_\alpha)&=\frac{\sqrt{\mu_0 
				L^d}}{\sqrt{\kappa\hat 
				v(\0)}},\label{denso0}\\
	   (\Omega_\alpha|a_0a_0\Omega_\alpha)=   (\Omega_\alpha|a_0^*a_0^*\Omega_\alpha)&=	(\Omega_\alpha|a_0^*a_0\Omega_\alpha)=\frac{\mu_0 
			L^d}{\kappa\hat 
                                                                                                                                         v(\0)},\label{denso}\\
\label{minimi}
		\big( \Omega_\alpha|(H^L-\mu N^L)\Omega_\alpha\big)=& -\mu |\alpha|^2+ \frac{ \kappa \hat{v}(\0)}{2L^d}|\alpha|^4\notag\\
		&= \frac{L^d}{\kappa\hat{v}(\0)}\left( -\mu\mu_0 +\frac{\mu_0^2}{2} \right).
        \end{align}
By \eqref{denso},  $\mu_0$ is responsible for the density of the condensate.        

One possibility, see e.g. \cite{DLN_2023}, consists in choosing the
quadratic Hamiltonian determined by $\alpha =\sqrt{\frac{\mu
    L^d}{\kappa\hat{v}(\0)}}$,  so that 
$\Omega_\alpha$  minimizes \eqref{minimi}.
 This would mean  $\mu = \mu_0$.
However,
we prefer to make a better choice of $\mu_0$, which will be different
from $\mu$. For a while, we keep $\mu_0$ as an additional free
parameter, which will be fixed soon.

        It is convenient to introduce new
        annihilation operators
	\beq \label{trafo}
	W_\alpha^* a_\kb W_\alpha = a_\kb-\alpha \delta_{\kb,0}
	=: \tilde{a}_{\kb}.
	\eeq 
They kill  the new vacuum 
	$\Omega_\alpha $.
	Omitting the tildes, we obtain the Hamiltonian translated by the transformation \eqref{trafo}:
	\begin{align}
		\label{eq:WHWalpha}
		H^L-\mu N^L =&H_{0,1}^L+H_{0,2}^L + H_{0,3}^L+ H_{0,4}^L + E_0^L+H_{\Bog,\mu_0}^L+\sqrt\kappa H_{3,\mu_0}^L+\kappa H_{4,\mu_0}^L \notag \\& -(\mu-\mu_0)N^>,\\\notag
		E_0^L:=&
		\frac{L^d}{\kappa \hat{v}(\0)}\left(\frac{\mu_0^2}{2}- \mu_0\mu\right),\\\notag
		H_{\Bog,\mu_0}^L:=& \sum_{\p\in \Xi^>_L} \left( \frac{\p^2}{2}  +   \frac{\mu_0 \hat{v}(\p)}{\hat{v}(\0)} \right) a_\p^* a_\p + \sum_{\p\in \Xi^>_L }  \frac{\mu_0 \hat{v}(\p)}{2\hat{v}(\0)} \left(   a_\p^*a_{-\p}^*+ 
		a_\p a_{-\p} \right),\\\notag
		H_{3,\mu_0}^L:=& \frac{1}{L^{\frac{d}{2}}
		}\sum_{\p,\q,\p+\q\in\Xi^>_L}\frac{\hat{v}(\p)\sqrt{\mu_0}}{\sqrt{\hat{v}(\0)}} \left(
		a_{\p+\q}^*a_{\p}a_{\q}+
		a_{\p}^*a_{\q}^*a_{\p+\q}\right), \\\notag
		H_{4,\mu_0}^L:=&\frac{1 }{2L^d}\sum_{\substack{\p,\q, \p+\k,\q-\k\in\Xi^>_L} }\hat{v}(\k)a_{\p+\k}^*a_{\q-\k}^* a_{\q}a_{\p}.\notag
	\end{align} The  operators
	$H^L_{0,1},H^L_{0,2},H^L_{0,3},\text{ and }H^L_{0,4}$ (which we do not
        write explicitly) contain
	respectively products of two, three and four
	annihilation and creation operators, always including
	at least one normally ordered $a_0$ or $a_0^*$.

	Following the intuition in \cite{HP_59},
        we will  start from 
$\mu_0$  
          minimizing approximately the ground state energy of $H_\mu^L$. More precisely, we will tune $\mu_0$  so as to minimize the
          expectation value of $H_\mu^L$ in 
 $\Omega_{\Bog}(\mu_0)$, the ground state of the Bogoliubov
        Hamiltonian associated with the parameter $\mu_0$.  Differently from the canonical approach, the state $\Omega_{\Bog}(\mu_0)$ depends now on $\mu_0$, whereas the canonical state was constructed starting from $\nu$. This state can  be  written in terms of the shifted vacuum $\Omega_\alpha$ as \begin{equation}
        \Omega_{\Bog}(\mu_0) = \prod_{\p\in\Xi^>_L}\frac{1}{(\cosh\theta_\p)^{\frac{1}{2}}}\e^{-\frac{1}{2}\tanh\theta_\p a_\p^* a_{-\p}^*}\Omega_\alpha.
        \end{equation} Now,
        \begin{align}
          		(\Omega_{\Bog}(\mu_0)|H^L_\mu\Omega_{\Bog}(\mu_0))
=			\frac{L^d}{\kappa\hat{v}(\0)}
			\left(
			\frac{\mu_0^2}{2}-\mu\mu_0
			\right)
			+
			(\Omega_{\Bog}(\mu_0)|
   H^L_{\Bog,\mu_0}\Omega_{\Bog}(\mu_0))\\
          -(\mu-\mu_0)(\Omega_{\Bog}(\mu_0)|N^>
   \Omega_{\Bog}(\mu_0))
   + \kappa(\Omega_{\Bog}(\mu_0)|H_{4,\mu_0}^L \Omega_{\Bog}(\mu_0)).
		\end{align}
		Indeed, the expectation values of the terms containing zero-mode
		operators and of the cubic term vanish in
                $\Omega_{\Bog}(\mu_0)$.
                
Recall from \eqref{elem1} that
	\begin{align}
(\Omega_{\Bog}(\mu_0)|H_{\Bog,\mu_0}^L\Omega_{\Bog}(\mu_0))&=
                                                           E_{\Bog}^{L}(\mu_0)\\
        &  =
	  -\sum\limits_{\p\in\Xi_L^>}\frac12
	\left(
	\frac12\p^2+\mu_0
   \frac{\hat{v}(\p)}{\hat{v}(\0)}-\omega_{\Bog}(\p,\mu_0)\right).\label{elem1a}
        \end{align}
    Therefore,
    \begin{align}
			\frac{\partial}{\partial\mu_0}
			(\Omega_{\Bog}(\mu_0)|H^L_{\Bog,\mu_0}\Omega_{\Bog}(\mu_0)) 
&=			
				\sum_{\p\in\Xi^>_L}
				\frac{\hat{v}(\p)(\frac{\p^2}{2}-\omega_{\Bog}(\p,\mu_0))}{\hat{v}(\0)2\omega_{\Bog}(\p,\mu_0)}					\\	&=
				\sum_{\p\in\Xi^>_L}
				\frac{\hat{v}(\p)}{\hat{v}(\0)}
				\left(s_\p^2-c_\p s_\p\right).
      \end{align}
      Moreover,
      \beq           	(\Omega_{\Bog}(\mu_0)|N^>\Omega_{\Bog}(\mu_0)) 
				=
				\sum_{\p\in\Xi^>_L}s_\p^2. \eeq
                                Thus
                                \begin{align}
	0=&		\frac{\partial}{\partial\mu_0}
			(\Omega_{\Bog}(\mu_0)|H^L_{\mu}\Omega_{\Bog}(\mu_0)) 
                                  \\=&
                                       			\frac{L^d}{\kappa\hat{v}(\0)}(\mu_0-\mu)
				+
				\sum_{\p\in\Xi^>_L}
				\Bigl[
				\left(
				1+\frac{\hat{v}(\p)}{\hat{v}(\0)}
				\right)s_\p^2
				-
				\frac{\hat{v}(\p)}{\hat{v}(\0)}c_\p s_\p
				\Bigr]\\
				&+
				(\mu_0-\mu)
				\frac{\partial}{\partial\mu_0}
				\sum_{\p\in\Xi^>_L}s_\p^2+		\kappa		\frac{\partial}{\partial\mu_0}(\Omega_{\Bog}(\mu_0)|H_{4,\mu_0}^L\Omega_{\Bog}(\mu_0)).
	\end{align}
 Multiplying the stationarity equation by
		$\frac{\kappa\hat{v}(\0)}{L^d}$, we obtain
		\begin{align}\label{eq: mu-mugc}
							\mu_0-\mu
				=&
				-\frac{\kappa\hat{v}(\0)}{L^d}
				\sum_{\p\in\Xi^>_L}
				\Bigl[
				\left(
				1+\frac{\hat{v}(\p)}{\hat{v}(\0)}
				\right)s_\p^2
				-
				\frac{\hat{v}(\p)}{\hat{v}(\0)}c_\p s_\p
				\Bigr]\\
				&-
				\frac{\kappa\hat{v}(\0)(\mu_0-\mu)}{L^d}
				\frac{\partial}{\partial\mu_0}
				\sum_{\p\in\Xi^>_L}s_\p^2-
                    		\frac{\kappa^2\hat{v}(\0)}{L^d}		\frac{\partial}{\partial\mu_0}(\Omega_{\Bog}(\mu_0)|H_{4,\mu_0}^L\Omega_{\Bog}(\mu_0)).
			\label{pere}                      \end{align}
      The
		 term involving $H_4^L$ carries an additional factor
                 of $\kappa^2$.
                 Therefore, we will drop it.
The first term in the line \eqref{pere} carries $\kappa$ and
$\mu-\mu_0$, which we know to be of the order $\kappa$. Therefore,
we will drop it as well.

 This justifies setting
		\begin{equation}
			\begin{aligned}
				\mu_0
				&=
				\mu-\kappa\mu_1
                                ,\label{eq: mu 0 implicit}\\
				\mu_1
				&:=
				\frac{1}{L^d}
				\sum_{\p\in\Xi^>_L}
				\Bigl[
				\left(
				\hat v(\0)+\hat{v}(\p)
				\right)s_\p^2
				-
				\hat{v}(\p)c_\p s_\p
				\Bigr],
			\end{aligned}
                      \end{equation}
                      where $s_\p$, $c_\p$ are evaluated for
                      $\mu_0$. (Actually, since $\mu_0-\mu=O(\kappa)$, 
$c_\p$ and $s_\p$ in the definition of $\mu_1$ can be evaluated at
		$\mu$ up to errors $O(\kappa^2)$).

	\subsection{Grand-canonical energy shift}\label{sec: grand canonical estimate}                
	
	We can now compute the dispersion-relation shift in the grand-canonical formalism. To this end, in the formula derived for the canonical shift, we replace the canonical Hamiltonian by the corresponding grand-canonical Hamiltonian, with $\mu_0$ chosen as above.
	\begin{align}\label{forri-grand}
		\delta^\gcan(\k,\mu,\hat{v})&=		\delta_{4}^\gcan(\k,\mu,\hat{v})+	\delta_{3}^\gcan(\k,\mu,\hat{v}),\end{align}
		with
		\begin{align}
		\delta_{4}^\gcan(\k,\mu,\hat{v})	&:=
                                                   \lim_{L\rightarrow\infty}
		(\Omega_{\Bog}(\mu_0) |[b_\k,
		[H_{4,\mu_0}^L - \mu_1N^>,
                                                   b_\k^*]]\Omega_{\Bog}(\mu_0) \big),\end{align} and
                                                   \begin{align}
&\delta_{3}^\gcan(\k,\mu,\hat{v})\notag\\
                                            &\notag:=\lim_{\epsilon\searrow
			0}\lim_{L\rightarrow\infty}\bigg\{
		-\big(\Omega_{\Bog}(\mu_0) |[b_\k,
		H_{3,\mu_0}^L](H_{\Bog,\mu_0}^L-\omega_{\Bog}(\k,\mu_0)-\i\epsilon)^{-1}[H_{3,\mu_0}^L,  
		b_\k^*]\Omega_{\Bog}(\mu_0) \big)  \\
		&-\big(\Omega_{\Bog}(\mu_0) |[b_\k^*, H_{3,\mu_0}^L](H_{\Bog,\mu_0}^L+\omega_{\Bog}(\k,\mu_0)-\i\epsilon)^{-1}[H_{3,\mu_0}^L, 
		b_\k]\Omega_{\Bog}(\mu_0) \big)\notag\\
		&  -\big(\Omega_{\Bog}(\mu_0) | H_{3,\mu_0}^L(H_{\Bog,\mu_0}^L-\i\epsilon)^{-1}[b_\k ,
		[H_{3,\mu_0}^L, b_\k^*]]\Omega_{\Bog}(\mu_0) \big)
		\notag \\ &-\big(\Omega_{\Bog}(\mu_0) |[b_\k^*,
		[H_{3,\mu_0}^L, b_\k]](H_{\Bog,\mu_0}^L-\i\epsilon)^{-1}H_{3,\mu_0}^L\Omega_{\Bog}(\mu_0) \big)\bigg\}.
                \end{align}

 The Fermi golden rule contribution is exactly the same in the two formalisms after setting $\mu_0 = \mu= \nu$ (which is true after discarding terms of order $\kappa$).
 
 Let us compute the Hellman--Feynman correction.
          An easy computation gives \begin{align}
                                      &  (\Omega_{\Bog}(\mu_0)|
                                        [b_\k,[H_{4,\mu_0}^L,b_\k^*]]\Omega_{\Bog}(\mu_0))\\
                                        =&
                                        (c_\k^2+s_\k^2)\frac{1}{L^d}\sum_{\p\in\Xi^>_L}s_\p^2\left(\hat
                                        v(\0)+\frac{\hat v(\p-\k)+\hat
                                        v(\p+\k)}{2}\right)\notag \\
		&+2c_\k s_\k \frac{1}{L^d}\sum_{\p\in\Xi^>_L}c_\p s_\p\left(\frac{\hat{v}(\p-\k)+\hat{v}(\p+\k)}{2}\right) + O(L^{-d}).
                                    \end{align}
                           Even after including the
                                    contribution of the Fermi Golden
                                    Rule, this  expression remains
                                    infrared divergent (in the
                                    thermodynamic limit) as $\k\to
                                    \0$. However, we still have to add
                                    the counterterm involving
              \begin{align}\mu_1&=
                 \frac{1}{L^d}\sum_{\p\in\Xi^>_L}\left(\left(\hat
                v(\0)+\hat{v}(\p)\right)s_\p^2-\hat{v}(\p) c_\p
                s_\p\right),\\
&                		(\Omega_{\Bog}(\mu_0)|[b_\k,[N^>,b_\k^*]]\Omega_{\Bog}(\mu_0)) =
               (c_\k^2+s_\k^2).
		\end{align}
             The full expression for the grand-canonical Hellmann-Feynman correction is \begin{align}\label{eq: delta4 grand canonical}
		&\delta_{4}^{\gcan,L}(\k,\mu,\hat{v})=(\Omega_{\Bog}(\mu_0) |[b_\k,
		[H_{4,\mu_0}^L -\kappa \mu_1N^>, b_\k^*]]\Omega_{\Bog}(\mu_0) \big) \restriction_{\mu_0 = \nu} \\ 
		=&(c_\k+s_\k)^2\frac{1}{L^d}\sum_{\p\in\Xi^>_L}c_\p s_\p \hat{v}(\p) \\
		&+
                   (c_\k^2+s_\k^2)\frac{1}{L^d}\sum_{\p\in\Xi^>_L}s_\p^2\left(-\hat{v}(\p)+\frac{\hat{v}(\p+\k)+\hat{v}(\p-\k)}{2}\right)
                    \\&+2 c_\k s_\k
                                         \frac{1}{L^d}\sum_{\p\in\Xi^>_L}c_\p s_\p \left(-\hat{v}(\p)+\frac{\hat{v}(\p+\k)+\hat{v}(\p-\k)}{2}\right) +O(L^{-d})     .
	\end{align}

\subsection{Comparison of the grand-canonical and canonical approaches}

	Let us compare \eqref{eq: delta4 grand canonical} with the canonical expression for the Hellmann-Feynman term \begin{align}
		&\delta _{4}^{\can,L}(\k,\nu,\hat{v})	:=\left(\Omega_{\Bog}(\nu)\right|\left. [b_\k,[H_{4,2}^L,
		b_\k^*]]\Omega_{\Bog}(\nu)\right) = \notag \\
		=	&
		(c_\k+s_\k)^2\frac{1}{L^d}\sum_{\p\in\Xi_L^>}c_\p s_\p
		\hat{v}(\p)\\
		&+	(c_\k^2+s_\k^2)\frac{1}{L^d}\sum_{\p\in\Xi_L^>}s_\p^2\left(-\hat{v}(\p) +\frac{\hat{v}(\p+\k)+\hat{v}(\p-\k)}{2}\right)  \\
		&   + 2c_\k s_\k \frac{1}{L^d}\sum_{\p,\p\pm\k\in\Xi^>_L} c_\p s_\p\left (-\hat v(\p)+\frac{\hat{v}(\p+\k)+\hat{v}(\p-\k)}{2}\right)\\
		&	-(c_\k-s_\k)^2 \hat{v}(\k)\frac{1}{L^d}\sum_{\p\in\Xi_L^>}s_\p^2 + O(L^{-d}), \label{eq: finite volume HF2}
                                                                                        \end{align}
where the above quantities depend on the parameter $\nu$. If we
naively set $\nu=\mu_0$, we see that the two expressions differ by the
last line in Equation \eqref{eq: finite volume HF2}, which is of the order $O(\kappa)$. Their difference is \begin{align}                                                                                                                                                                                   
&\delta_{4}^{\can,L}(\k,\nu,\hat{v})-\delta_{4}^{\gcan,L}(\k,\mu,\hat{v})
                                \\                                                                                                          =& -\hat{v}(\k)(c_\k-s_\k)^2\frac{1}{L^d}\sum_{\p\in\Xi^>_L}s_\p^2 + O(L^{-d})+O(\kappa).\label{eq: HF difference}
	\end{align}  

Let us explain this discrepancy.
As we saw, the chemical potential $\mu$ and the parameter of the
corresponding Bogoliubov Hamiltonian $\mu_0$ differ by a term of order
$\kappa$. They both differ by a quantity of order $\kappa$ from the
canonical parameter $\nu$. To determine
the relationship between the grand-canonical and canonical parameters,
let us use the grand-canonical Bogoliubov ground state $\Omega_{\Bog}(\mu_0)$:
\begin{align}
  \nu=\kappa\hat v(\0) \frac{N}{L^d}= &\frac{\kappa\hat v(\0)}{L^d}\Big((\Omega_{\Bog}(\mu_0)|N_\0
  \Omega_{\Bog}(\mu_0))+(\Omega_{\Bog}(\mu_0)|N^>\Omega_{\Bog}(\mu_0))\Big)\\
  =&\mu_0+\frac{\kappa\hat v(\0)}{L^d}\sum_{\p\in\Xi^>_L}s_\p^2\end{align}
(Here $N_\0$ is the unshifted number operator). Equivalently, 
        \begin{align}\label{eq: relation nu and mu}
	 \mu 
	= &\nu +\kappa	\frac{1}{L^d}
	 \sum_{\p\in\Xi^>_L}
	 \Bigl[
	\hat{v}(\p)
	s_\p^2
	 -
	\hat{v}(\p)c_\p s_\p
	 \Bigr].
			\end{align}

Therefore the unperturbed  dispersion
relation in the canonical formalism differs from the relation in
the grand-canonical
formalism by a term of order $O(\kappa)$: both are given by the same functional form as in $\omega_{\Bog}(\k,\mu_0)$, but  the
former  involves $\nu$ and the latter 
depends on $\mu_0$. Expanding in the Taylor series we have
\beq\omega_{\Bog}(\k,\nu)=\omega_{\Bog}(\k,\mu_0)+(\mu_0-\nu)\frac\partial{\partial\mu_0}\omega_{\Bog}(\k,\mu_0) \restriction_{\mu_0 = \nu}+O(\kappa^2).\eeq
Now
 \begin{align}
   \mu_0-\nu&=-\frac{\kappa\hat
                v(\0)}{L^d}\sum_{\p\in\Xi_L}s_\p^2,\\
      \frac{\partial}{\partial \mu_0}\omega_{\Bog}(\k,\mu_0) 
   \restriction_{\mu_0 = \nu} = &\frac{\hat{v}(\k)|\k|^2}{2\hat 
                                  v(\0)\omega_\Bog(\k)}\notag \\
   c_\k-s_\k = &\frac{|\k|}{\sqrt{2\omega_\Bog(\k)}}. \label{eq: taylor
                 correction}
	\end{align}
        Hence,
        \begin{align}   (\mu_0-\nu)
                \frac{\partial}{\partial \mu_0}\omega_{\Bog}(\k,\mu_0) 
   \restriction_{\mu_0 = \nu}                  
			 =& -\kappa \hat{v}(\k)(c_\k-s_\k)^2 \frac{1}{L^d}\sum_{\p\in\Xi^>_L}s_\p^2.\label{eq: miracolous correction} \end{align}

        We see that the correction in Equation \eqref{eq: miracolous correction} exactly coincides with the difference between the Hellmann-Feynman corrections computed in the canonical and grand-canonical formalisms \eqref{eq: HF difference}.
	Thus,  one has
\begin{align}
\kappa\delta_{4}^\gcan(\k,\mu,\hat{v})+  	(\mu_0-\nu)\frac{\partial}{\partial \mu_0}\omega_{\Bog}(\k,\mu_0)\restriction_{\mu_0 = \nu} = 	\kappa\delta_{4}^\can(\k,\nu,\hat{v}) +O(\kappa^2)\end{align}
        Since the Fermi golden rule corrections already essentially coincide, we
  conclude that
\begin{align}
\kappa\delta^\gcan(\k,\mu,\hat{v})+  	(\mu_0-\nu)\frac{\partial}{\partial \mu_0}\omega_{\Bog}(\k,\mu_0)\restriction_{\mu_0 = \nu} = 	\kappa \delta^\can(\k,\nu,\hat{v})
+O(\kappa^2)\end{align}
 which shows the consistency of the grand-canonical and canonical approaches.
	
	If we compare with the canonical approach, the present
	derivation is, in a sense, cleaner
	because we do not
	have to deal with
	dirty ``tail terms''. However, the parameter $\mu$  is perhaps
  sometimes less transparent. Indeed, one needs to renormalize the "effective chemical potential" $\mu_0$ at every step of the perturbative expansion.

\section{Tail estimates}
\label{Tail estimates}

\subsection{ Localization of the Bogoliubov vacuum}

{ The terms $R_{4,1}$ and $R_{4,2}$ arise when passing from the
physical Hilbert space to the extended Hilbert space. We call them
\emph{tail terms} and show that their contributions vanish in the
thermodynamic limit. The main ingredient is the exponential
concentration of $N_\0^\ext$ around $n_{0,\Bog}(L)L^d$ in the
Bogoliubov ground state.

This section contains the technical estimates needed to prove
Theorem \ref{thm: effective hamiltonian}. Readers primarily interested
in the perturbative computation may take that theorem for granted
and skip this section.

We begin with an estimate on the fourth powers of $s_\p$.

\begin{lemma}\label{lem: sp4}
	Fix $d=2,3$ and assume Assumption
	\ref{ass: primary assumption on the potential}. Then
	\begin{equation}\label{eq: estimate for sp4}
		\sum_{\p\in\Xi_L^{>,+}}s_\p^4
		=
		\begin{cases}
			O(L^2\log L),&d=2,\\
			O(L^3),&d=3,
		\end{cases}
		\qquad L\longrightarrow+\infty.
	\end{equation}
\end{lemma}
\begin{proof}
	Under Assumption \ref{ass: primary assumption on the potential},
	the small-momentum asymptotics, continuity, strict stability
	and large-momentum behaviour imply
	\begin{align*}
		s_\p^4&\leq\frac{C}{|\p|^2}
		&&\text{for }0<|\p|\leq2\pi,\\
		s_\p^4&\leq Cs_\p^2
		&&\text{for }|\p|>2\pi,
	\end{align*}
	with $C>0$ independent of $L$. Therefore,
	\begin{align}
		\sum_{\p\in\Xi_L^{>,+}}s_\p^4
		&\leq
		C\sum_{\substack{\p\in\Xi_L^>\\|\p|>2\pi}}s_\p^2
		+C\sum_{\substack{\p\in\Xi_L^>\\0<|\p|\leq2\pi}}
		\frac{1}{|\p|^2}\leq
		CL^dn_\Bog^>(L)
		+C'L^2\sum_{n=1}^{\lceil L\rceil}n^{d-3},
	\end{align}
	where $C'>0$ is also independent of $L$. 
	The remaining sum is $O(\log L)$ for $d=2$ and $O(L)$ for $d=3$,
	proving \eqref{eq: estimate for sp4}.
\end{proof}

Next, we derive an exponential Markov bound.

\begin{lemma}\label{lem: exponential estimates}
	Let $A$ be a self-adjoint operator and $\Omega$ a vector of norm $1$
	such that the quadratic-form expectations
	\begin{align}
		\e^{c_1}&:=(\Omega|\e^A\Omega),\\
		\e^{c_2}&:=(\Omega|\e^{-A}\Omega)
	\end{align}
	are finite. If $I_\lambda:=[a-\lambda,a+\lambda]$, with
	$a\in\rr$ and $\lambda>0$, then
	\begin{equation}
		(\Omega|\one_{\rr\setminus I_\lambda}(A)\Omega)
		\leq \e^{-\lambda-a+c_1}+\e^{-\lambda+a+c_2}.
		\label{markov1}
	\end{equation}
\end{lemma}
\begin{proof}
	Integrate the pointwise inequality
	\begin{equation}
		\one_{\rr\setminus I_\lambda}(x)
		\leq \e^{-\lambda}\big(\e^{x-a}+\e^{-(x-a)}\big)
	\end{equation}
	against the spectral measure of $A$ in $\Omega$.
\end{proof}

We apply this bound to $N^>$.

\begin{lemma}\label{lemma: estimate for N>}
	Fix $d=2,3$ and assume Assumption
	\ref{ass: primary assumption on the potential}. Let $m>0$ and
	\begin{equation}
		I_m=[n_\Bog^>(L)L^d-mL^d,n_\Bog^>(L)L^d+mL^d].
	\end{equation}
	For every fixed $\delta\in]0,1[$ and all sufficiently large $L$,
	\begin{equation}
		(\Omega_\Bog|\one_{\rr\setminus I_m}(N^>)\Omega_\Bog)
		\leq
		\exp\left\{-\frac{\pi(1+\delta)m}{\sqrt\nu}L^{d-1}\right\}.
	\end{equation}
\end{lemma}
\begin{proof}
	Fix $\delta'\in]\delta,1[$.
	We apply Lemma \ref{lem: exponential estimates} with
	$A=\mu N^>$, where $\mu>0$ is chosen so that
	\begin{equation}
		\e^{2\mu}-1
		=\frac{1+\delta'}{2}
		\left(\sup_{\p\in\Xi_L^>}s_\p^2\right)^{-1}.
	\end{equation}
	The uniform small-momentum expansion
	\begin{equation}
		s_\p^2=\frac{\sqrt\nu}{2|\p|}+o(|\p|^{-1}),
		\qquad \p\longrightarrow0,
	\end{equation}
	together with boundedness of $s_\p^2$ outside any fixed
	neighbourhood of the origin, gives
	\begin{equation}
		\sup_{\p\in\Xi_L^>}s_\p^2
		=\frac{\sqrt\nu}{4\pi}L+o(L),
	\end{equation}
	Consequently,
	\begin{equation}
		\mu=\frac{\pi(1+\delta')}{\sqrt\nu L}+o(L^{-1}),
		\qquad
		0\leq|(\e^{\pm2\mu}-1)s_\p^2|
		\leq\frac{1+\delta'}{2}<1.
	\end{equation}
	For the negative sign, we also used
	$1-\e^{-2\mu}\leq\e^{2\mu}-1$.
	
	By \eqref{eq: exponential computation},
	\begin{align}
		&\log\big(
		\Omega_\Bog\big|
		\e^{\pm\mu(N^>-n_\Bog^>(L)L^d)}
		\Omega_\Bog
		\big)\notag\\
		&\qquad=
		-\sum_{\p\in\Xi_L^{>,+}}
		\log\left(1-(\e^{\pm2\mu}-1)s_\p^2\right)
		\mp\mu n_\Bog^>(L)L^d.
		\label{eq: log}
	\end{align}

	For $|x|\leq\frac{1+\delta'}{2}$, we have
	\begin{equation*}
		0\leq-\log(1-x)-x
		=x^2\int_0^1\frac{v}{1-vx}\,\mathrm dv
		\leq\frac{x^2}{1-\delta'}.
	\end{equation*}
	Applying this inequality term by term in \eqref{eq: log},
	and using
	$2\sum_{\p\in\Xi_L^{>,+}}s_\p^2=L^dn_\Bog^>(L)$,
	yields
	\begin{align*}
		\log\big(
		\Omega_\Bog\big|
		\e^{\pm\mu(N^>-n_\Bog^>(L)L^d)}
		\Omega_\Bog
		\big)\leq
		\frac{\e^{\pm2\mu}-1\mp2\mu}{2}L^dn_\Bog^>(L)
		+
		\frac{(\e^{\pm2\mu}-1)^2}{1-\delta'}
		\sum_{\p\in\Xi_L^{>,+}}s_\p^4.
	\end{align*}

	Lemma \ref{lem: sp4} gives
	\begin{equation}
		\log\big(
		\Omega_\Bog\big|
		\e^{\pm\mu(N^>-n_\Bog^>(L)L^d)}
		\Omega_\Bog
		\big)
		=
		\begin{cases}
			O(\log L),&d=2,\\
			O(L),&d=3,
		\end{cases}
		=o(L^{d-1}).
	\end{equation}
	
	In \eqref{markov1}, take $A=\mu N^>$,
	$a=\mu n_\Bog^>(L)L^d$ and $\lambda=\mu mL^d$.
	The preceding estimate gives
	$c_1-a=o(L^{d-1})$ and $c_2+a=o(L^{d-1})$.
	Hence
	\begin{align*}
		(\Omega_\Bog|\one_{\rr\setminus I_m}(N^>)\Omega_\Bog)
		&\leq
		2\exp\left\{
		-\frac{\pi(1+\delta')m}{\sqrt\nu}L^{d-1}
		+o(L^{d-1})
		\right\}
	\end{align*}
	for all sufficiently large $L$. 
	After replacing $\delta'$ with $\delta\in ]0,1[$, $\delta'>\delta$ we can absorb both the
	$o(L^{d-1})$ remainder and the factor $2$.
\end{proof}

The identity $N_\0^\ext=N-N^>$ yields the corresponding estimates
for the zero-mode.

\begin{cor}\label{cor: N0 estimate<}
	Fix $d=2,3$ and assume Assumption
	\ref{ass: primary assumption on the potential}.
	\begin{itemize}
		\item[a)]
		Suppose that Assumption \ref{ass: condensate hypotheses} holds.
		Let $m_0>0$ and
		\begin{equation}
			I_{m_0}
			=[n_{0,\Bog}(L)L^d-m_0L^d,n_{0,\Bog}(L)L^d+m_0L^d].
		\end{equation}
		For every fixed $\delta\in]0,1[$ and all sufficiently large $L$,
		\begin{equation}
			(\Omega_\Bog|
			\one_{\rr\setminus I_{m_0}}(N_\0^\ext)\Omega_\Bog)
			\leq
			\exp\left\{
			-\frac{\pi(1+\delta)m_0}{\sqrt\nu}L^{d-1}
			\right\}.
		\end{equation}
		
		\item[b)]
		If $n_{0,\Bog}<0$, then
		\begin{equation}
			\lim_{L\to+\infty}
			(\Omega_\Bog|
			\one_{]-\infty,0[}(N_\0^\ext)\Omega_\Bog)=1.
		\end{equation}
	\end{itemize}
\end{cor}
\begin{proof}
	\textit{a)}
	Since $n_{0,\Bog}(L)L^d=N-n_\Bog^>(L)L^d$,
	functional calculus gives
	\begin{equation*}
		(\Omega_\Bog|
		\one_{\rr\setminus I_{m_0}}(N_\0^\ext)\Omega_\Bog)
		=
		(\Omega_\Bog|
		\one_{\rr\setminus I_{m_0}^>}(N^>)\Omega_\Bog),
	\end{equation*}
	where
	\begin{equation*}
		I_{m_0}^>
		=[n_\Bog^>(L)L^d-m_0L^d,n_\Bog^>(L)L^d+m_0L^d].
	\end{equation*}
	Lemma \ref{lemma: estimate for N>} therefore gives the claimed bound.
	
	\textit{b)}
	By the Cauchy--Schwarz inequality,
	\begin{align*}
		n_{0,\Bog}(L)
		&\geq
		-L^{-d}(\Omega_\Bog|
		\one_{]-\infty,0[}(N_\0^\ext)|N_\0^\ext|\Omega_\Bog)\\
		&\geq
		-\sqrt{(\Omega_\Bog|
			\one_{]-\infty,0[}(N_\0^\ext)\Omega_\Bog)}
		\,L^{-d}
		\sqrt{(\Omega_\Bog|(N_\0^\ext)^2\Omega_\Bog)}.
	\end{align*}
	The pair distribution of the Bogoliubov ground state gives
	\begin{align*}
		L^{-2d}(\Omega_\Bog|(N_\0^\ext)^2\Omega_\Bog)
		&=
		n_{0,\Bog}(L)^2
		+2L^{-2d}\sum_{\p\in\Xi_L^>}s_\p^2(1+s_\p^2)\\
		&=
		n_{0,\Bog}(L)^2+
		\begin{cases}
			O(L^{-2}\log L),&d=2,\\
			O(L^{-3}),&d=3.
		\end{cases}
	\end{align*}
	Here we used Lemma \ref{lem: sp4} and
	$\sum_{\p\in\Xi_L^>}s_\p^2=L^dn_\Bog^>(L)=O(L^d)$.
	Thus
	\begin{equation*}
		L^{-d}\sqrt{(\Omega_\Bog|(N_\0^\ext)^2\Omega_\Bog)}
		\longrightarrow |n_{0,\Bog}|.
	\end{equation*}
	Since $n_{0,\Bog}(L)\to n_{0,\Bog}<0$, we have
	$n_{0,\Bog}(L)<0$ for all sufficiently large $L$.
	Rearranging the Cauchy--Schwarz bound yields
	\begin{equation*}
		\sqrt{(\Omega_\Bog|
			\one_{]-\infty,0[}(N_\0^\ext)\Omega_\Bog)}
		\geq
		\frac{-n_{0,\Bog}(L)}
		{L^{-d}\sqrt{(\Omega_\Bog|(N_\0^\ext)^2\Omega_\Bog)}}
		\longrightarrow1.
	\end{equation*}
	As the left-hand side is at most $1$, this proves \textit{b)}.
\end{proof}

Part \textit{a)} shows that, for every fixed $m_0>0$, the Bogoliubov
ground state is concentrated in the spectral subspace
\begin{equation*}
	\one_{I_{m_0}}(N_\0^\ext)\mathcal H^\ext.
\end{equation*}
This justifies interpreting $n_{0,\Bog}$ as the condensate density:
when $n_{0,\Bog}>0$, the zero-mode occupation is concentrated at
macroscopically large values. Choosing $0<m_0<n_{0,\Bog}$ ensures
that $I_{m_0}\subset]0,+\infty[$ for all sufficiently large $L$.
Consequently, the unphysical tail
$\one_{]-\infty,0[}(N_\0^\ext)\Omega_\Bog$ is exponentially
suppressed in $L^{d-1}$.

\begin{remark}
It is clear from the proof of Lemma \ref{lemma: estimate for N>} and Corollary \ref{cor: N0 estimate<} that one gets similar exponential decay even for vanishing condensate densities \begin{equation}\lim_{L\to+\infty }n_{\Bog,0}(L)\end{equation} and for spectral intervals of the form \begin{equation}
	I_{m_0,\epsilon} 
	=[n_{0,\Bog}(L)L^d-m_0L^{d-1+\epsilon},n_{0,\Bog}(L)L^d+m_0L^{d-1+\epsilon}],\qquad \epsilon>0.
\end{equation}
\end{remark}

A complementary lower bound shows that, for $d=2,3$, the power
$L^{d-1}$ in the exponential cannot be improved.

\begin{prop}\label{lemm: lower bound on I}
	Fix $d\geq1$ and assume Assumption
	\ref{ass: primary assumption on the potential}. Assume that the Bogoliubov vacuum
	$\Omega_\Bog$ is well defined. Then, for every $\lambda>1$ and
	all sufficiently large $N$,
	\begin{equation*}
		(\Omega_\Bog|
		\one_{]-\infty,0[}(N_\0^\ext)\Omega_\Bog)
		\geq
		\exp\left\{-\lambda n\frac{2\pi}{\sqrt\nu}
		L^{d-1}\right\}.
	\end{equation*}
\end{prop}
\begin{proof}
	Let $(\k_1)_i=\frac{2\pi}{L}\delta_{i,1}$. The marginal distribtution of $N_{\k_1}$ is geometric. Thus
	\begin{align*}
		(\Omega_\Bog|
		\one_{]-\infty,0[}(N_\0^\ext)\Omega_\Bog)\geq
		(\Omega_\Bog|
		\one_{[\lfloor \frac{N}{2}\rfloor+1,+\infty[}(N_{\k_1})
		\Omega_\Bog)
		=\bigl(\tanh^2\theta_{\k_1}\bigr)^{\lfloor \frac{N}{2}\rfloor+1}.
	\end{align*}
	As $L\to+\infty$, $\k_1\to 0$. Therefore, we can expand	\begin{equation*}
		\tanh^2\theta_{\k_1}
		=1-\frac{4\pi}{\sqrt\nu L}+o(L^{-1}).
	\end{equation*}
	Consequently,
	\begin{align*}
		\left(\lfloor  \tfrac{N}{2}\rfloor+1\right)\log(\tanh^2\theta_{\k_1})
		&=-\frac{2\pi}{\sqrt\nu}nL^{d-1}(1+o(1))
	\end{align*}
	
	For each fixed $\lambda>1$, the claimed lower bound follows for
	all sufficiently large $N$.
\end{proof}
For $d=1$, this proves
\begin{equation*}
	\liminf_{N\to\infty}
	(\Omega_\Bog|
	\one_{]-\infty,0[}(N_\0^\ext)\Omega_\Bog)
	\geq e^{-\frac{2\pi n}{\sqrt\nu}}>0.
\end{equation*}
Thus the unphysical tail does not tend to zero in dimension one.

The estimate in Corollary \ref{cor: N0 estimate<} remains valid after applying a fixed
number of creation and annihilation operators and inserting a polynomial
weight in the extended zero-mode occupation number.

\begin{prop}\label{cor: decay for particles}
	Fix $d=2,3$ and suppose that Assumptions
	\ref{ass: primary assumption on the potential} and \ref{ass: condensate hypotheses} hold.
	Let $m_0>0$ and $\alpha\geq0$ be fixed, let $m_0\geq0$ be a fixed integer,
	and set
	\begin{equation}
		I_{m_0}=[n_{0,\Bog}L^d-m_0L^d,n_{0,\Bog}L^d+m_0L^d].
	\end{equation}
	Then, for all $\delta \in ]0,1[$ 
	and, for all sufficiently large $L$,
	\begin{multline}
		\bigl(a_{\k_1}^{?}\cdots a_{\k_m}^{?}\Omega_\Bog\big|
		|N_\0^\ext|^\alpha\one_{\rr\setminus I_{m_0}}(N_\0^\ext)
		a_{\k_1}^{?}\cdots a_{\k_m}^{?}\Omega_\Bog\bigr)\\ \leq
			\exp\left\{
		-\frac{\pi(1+\delta)m_0}{\sqrt\nu}L^{d-1}
		\right\}.
	\end{multline}
	The estimate is uniform in $\k_1,\ldots,\k_m\in\Xi_L^>$, including
	repeated momenta, and in the choices
	$a_\k^{?}\in\{a_\k,a_\k^*\}$.
\end{prop}

\begin{proof}
	Let $s$ be the number of creation operators minus the number of
	annihilation operators, so that $|s|\leq m$. Since
	$N_\0^\ext=N-N^>$, spectral calculus gives
	\begin{align*}
		|N_\0^\ext|^\alpha\one_{\rr\setminus I_{m_0}}(N_\0^\ext)
		a_{\k_1}^{?}\cdots a_{\k_m}^{?}=
		a_{\k_1}^{?}\cdots a_{\k_m}^{?}
		|N_\0^\ext-s|^\alpha
		\one_{(\rr\setminus I_{m_0})+s}(N_\0^\ext).
	\end{align*}
	Successively applying the usual creation and annihilation bounds on
	each eigenspace of $N^>$ yields
	\begin{equation*}
		0\leq
		(a_{\k_m}^{?})^*\cdots(a_{\k_1}^{?})^*
		a_{\k_1}^{?}\cdots a_{\k_m}^{?}
		\leq(N^>+m+1)^m.
	\end{equation*}
	This product commutes with $N^>$ and $N_\0^\ext$.
	
	We first bound the polynomial moments needed below. 
Equation 
	\eqref{eq: exponential computation} implies, for sufficiently large $L$,
	\begin{align*}
		\log(\Omega_\Bog|\e^{\frac{N^>}{L^d}}\Omega_\Bog)
		&=-\frac12\sum_{\p\in\Xi_L^>}
		\log\bigl(1-(\e^{\frac{2}{L^d}}-1)s_\p^2\bigr)\leq CL^{-d}\sum_{\p\in\Xi_L^>}s_\p^2\leq C.
	\end{align*}
	Since $N=O(L^d)$ and for any $r\geq 0$ \begin{equation}
		\left(\frac{N^>}{L^d}\right)^r \leq D_r e^{\frac{N^>}{L^d}}
	\end{equation}
For some positive constant $D_r>0$ independent from $L$, we obtain
	\begin{equation*}
		(\Omega_\Bog|(N+N^>+m+1)^r\Omega_\Bog)\leq C_rL^{dr},
		\qquad r\geq0,
	\end{equation*} where $C_r>0$ is again independent from the volume.
	In particular, using the preceding operator bound and then
	Cauchy--Schwarz, the matrix element in the statement is at most
	\begin{align*}
		&\left\|
		|N_\0^\ext-s|^\alpha
		(a_{\k_m}^{?})^*\cdots(a_{\k_1}^{?})^*
		a_{\k_1}^{?}\cdots a_{\k_m}^{?}\Omega_\Bog
		\right\|
		\left\|\one_{(\rr\setminus I_{m_0})+s}(N_\0^\ext)
		\Omega_\Bog\right\|\\
		&\quad\leq
		\bigl(\Omega_\Bog|(N+N^>+m+1)^{2(m+\alpha)}
		\Omega_\Bog\bigr)^{\frac{1}{2}}
		\left\|\one_{(\rr\setminus I_{m_0})+s}(N_\0^\ext)
		\Omega_\Bog\right\|\\
		&\quad\leq CL^{d(m+\alpha)}
		\left\|\one_{(\rr\setminus I_{m_0})+s}(N_\0^\ext)
		\Omega_\Bog\right\|.
	\end{align*}
	The polynomial factor in $L$ can be absorbed into the stated exponential rates by applying  Corollary \ref{cor: N0 estimate<} with a smaller $\delta\in (0,1)$. This proves the claim.
\end{proof}
}
\renewcommand{\l}{\mathbf{l}}

\subsection{Estimates for the tail remainders}\label{sec: weak coupling}

Throughout this subsection, the external momenta are bounded away from zero,
and limits are taken along admissible lattice momenta. In addition to the preceding assumptions, we
assume the uniform summability bound
\begin{equation*}
	\limsup_{L\to\infty}\frac{1}{L^d}\sum_{\p\in\Xi_L^>}
	|\hat v(\p)|\bigl(|c_\p s_\p|+s_\p^2\bigr)<\infty.
\end{equation*}
It holds for $d=2,3$ under the additional Assumption
\ref{ass: secondary hypotheses} by standard Sobolev inequalities. 

The following lemma shows that the Hellmann--Feynman contribution of
$R_{4,2}^L$ vanishes in the thermodynamic limit. We retain the operator
ordering in its definition in Section \ref{sec: full hamiltonian}.

\begin{lemma}\label{lemm: thermodynamic of R4}
	Fix $d=2,3$ and assume Assumptions
	\ref{ass: primary assumption on the potential} and \ref{ass: condensate hypotheses},
	together with the uniform summability bound stated above. Let
	\begin{align*}
		R_{4,2}^L=&\frac{1}{2L^d}
		\left(\sqrt{N_\0^\ext(N_\0^\ext-1)}-N_\0^\ext+\frac12\right)
		\sum_{\p\in\Xi_L^>}\hat v(\p)A_\p^*+\mathrm{h.c.}
	\end{align*}
	Then
	\begin{equation*}
		\lim_{L\to\infty}\Bigl[
		(b_\k^*\Omega_\Bog|R_{4,2}^Lb_\k^*\Omega_\Bog)
		-(\Omega_\Bog|R_{4,2}^L\Omega_\Bog)\Bigr]=0.
	\end{equation*}
\end{lemma}

\begin{proof}
	Let
	\begin{equation*}
		f(x):=\sqrt{|x(x-1)|}-x+\frac12.
	\end{equation*}
	As $x\to+\infty$, this function satisfies$
		f(x)=O(x^{-1}),\,f(x+1)-f(x)=O(x^{-2}),\,
		f(x+1)-2f(x)+f(x-1)=O(x^{-3})$.  Since
	$[N_\0^\ext,A_\p]=2A_\p$, we have $A_\p f(N_\0^\ext)=f(N_\0^\ext-2)A_\p$.
	Consequently, the double-commutator identity gives
	\begin{align*}
		&(b_\k^*\Omega_\Bog|R_{4,2}^Lb_\k^*\Omega_\Bog)
		-(\Omega_\Bog|R_{4,2}^L\Omega_\Bog)=\frac{\kappa}{L^d}\operatorname{Re}
		\sum_{\p\in\Xi_L^>}\hat v(\p)
		(\Omega_\Bog|[b_\k,[f(N_\0^\ext-2)A_\p,b_\k^*]]\Omega_\Bog).
	\end{align*}
	We expand the commutators as follows:
	\begin{align*}
		&[b_\k,[f(N_\0^\ext-2)A_\p,b_\k^*]]=f(N_\0^\ext-2)[b_\k,[A_\p,b_\k^*]]
		+[b_\k,f(N_\0^\ext-2)][A_\p,b_\k^*]\\
		&+[f(N_\0^\ext-2),b_\k^*][b_\k,A_\p]
		+[b_\k,[f(N_\0^\ext-2),b_\k^*]]A_\p.
	\end{align*}
	Using the expansion of $A_\p$ in $M_\p$ and $B_\p$, together with
	\eqref{comm2} and \eqref{commu2}, we obtain the four identities
	\begin{align}
		&(\Omega_\Bog|f(N_\0^\ext-2)[b_\k,[A_\p,b_\k^*]]\Omega_\Bog)
		\notag\\
		=&-s_\k c_\k(\delta_{\p,\k}+\delta_{-\p,\k})
		(\Omega_\Bog|f(N_\0^\ext-2)\Omega_\Bog),
		\label{eq: no commutator}\\
		&(\Omega_\Bog|[b_\k,f(N_\0^\ext-2)][A_\p,b_\k^*]\Omega_\Bog)
		\notag\\
=&-s_\k c_\k(\delta_{\p,\k}+\delta_{-\p,\k})
		(\Omega_\Bog|[b_\k,f(N_\0^\ext-2)]b_\k^*\Omega_\Bog),\\
		&(\Omega_\Bog|[f(N_\0^\ext-2),b_\k^*][b_\k,A_\p]\Omega_\Bog)
		\notag\\
	=&s_\k^2(\delta_{\p,\k}+\delta_{-\p,\k})
		(\Omega_\Bog|[f(N_\0^\ext-2),b_\k^*]b_{-\k}^*\Omega_\Bog),\notag\\
		&(\Omega_\Bog|[b_\k,[f(N_\0^\ext-2),b_\k^*]]A_\p\Omega_\Bog)\\
=&s_\p^2
		(\Omega_\Bog|[b_\k,[f(N_\0^\ext-2),b_\k^*]]B_\p^*\Omega_\Bog)-s_\p c_\p
		(\Omega_\Bog|[b_\k,[f(N_\0^\ext-2),b_\k^*]]\Omega_\Bog).
	\end{align}
	Now, we estimate the scalar product in Equation \ref{eq: no commutator}. 
	Choose $0<m_0<n_{0,\Bog}$ and set
	\begin{equation*}
		I_{m_0}:=[n_{0,\Bog}L^d-m_0L^d,n_{0,\Bog}L^d+m_0L^d].
	\end{equation*}
	Using Proposition \ref{cor: decay for particles}, the first expectation in \eqref{eq: no commutator} satisfies
	\begin{align}
		&|(\Omega_\Bog|f(N_\0^\ext-2)\Omega_\Bog)| 
		\leq |(\Omega_\Bog|\one_{I_{m_0}}(N_\0^\ext)
		f(N_\0^\ext-2)\Omega_\Bog)|\notag \\ &+|(\Omega_\Bog|\one_{\mathbb R\setminus I_{m_0}}(N_\0^\ext)
		f(N_\0^\ext-2)\Omega_\Bog)| =O(L^{-d})+O(e^{-cL^{d-1}}).
	\end{align}
	 Thus
	\begin{equation*}
		\frac{1}{L^d}\sum_{\p\in\Xi_L^>}\hat v(\p)
		(\Omega_\Bog|f(N_\0^\ext-2)[b_\k,[A_\p,b_\k^*]]\Omega_\Bog)
		=O(L^{-2d}).
	\end{equation*}
	
	
	Similarly, it follows that
	\begin{align*}
		|(\Omega_\Bog|[b_\k,[f(N_\0^\ext-2),b_\k^*]]\Omega_\Bog)|
		&=O(L^{-2d}),\\
		|(\Omega_\Bog|[f(N_\0^\ext-2),b_\k^*]b_{-\k}^*\Omega_\Bog)|
		&=O(L^{-2d}),\\
		|(\Omega_\Bog|[b_\k,f(N_\0^\ext-2)]b_\k^*\Omega_\Bog)|
		&=O(L^{-2d}).
	\end{align*}
The remaining sum over momenta is controlled by the Kronecker deltas for the first three terms, and by the uniform
	summability assumption for the  fourth one.
	Consequently,
	\begin{equation*}
		\left|\frac{1}{L^d}\sum_{\p\in\Xi_L^>}\hat v(\p)
		(\Omega_\Bog|[b_\k,[f(N_\0^\ext-2)A_\p,b_\k^*]]\Omega_\Bog)\right|
		=O(L^{-2d}),
	\end{equation*}
	which proves the lemma.
\end{proof}
An analogous proof shows 
\begin{lemma}\label{lemm: thermodynamic of R41}
	Under the assumptions of Lemma \ref{lemm: thermodynamic of R4}, let
	\begin{align*}
		R_{4,1}^L=&\frac{1}{2L^d}
		\left(\sqrt{N_\0^+(N_\0^+-1)}
		-\sqrt{N_\0^\ext(N_\0^\ext-1)}\right)
		\sum_{\p\in\Xi_L^>}\hat v(\p)A_\p^*+\mathrm{h.c.}
	\end{align*}
	Then
	\begin{equation*}
		\lim_{L\to\infty}\Bigl[
		(b_\k^*\Omega_\Bog|R_{4,1}^Lb_\k^*\Omega_\Bog)
		-(\Omega_\Bog|R_{4,1}^L\Omega_\Bog)\Bigr]=0.
	\end{equation*}
\end{lemma}

The same arguments apply to normalized states
$\mathcal N b_{\k_1}^*\cdots b_{\k_n}^*\Omega_\Bog$ with a fixed number
of quasiparticles and momenta bounded away from zero. Commuting the finitely many creation and
annihilation operators introduces only finitely many integer shifts and
finitely excited vectors, to which the same tail estimates apply. Thus
the conclusions of Lemmas \ref{lemm: thermodynamic of R4} and
\ref{lemm: thermodynamic of R41}, as well as those of
Lemmas \ref{lemm: thermodynamic for H2.},
\ref{lemm: thermodynamic for H41.} and
\ref{lemm: thermodynamic for H43.}, extend to these states.
In particular, neither $R_{4,1}^L$ nor $R_{4,2}^L$ contributes to the
Hellmann--Feynman correction in the thermodynamic limit.

\section {Estimates on perturbative corrections}
\label{Estimates on perturbative corrections}

In this section we give rigorous proofs of the main results of our
paper presented in  Section \ref{sec: computation of the real part 1}.
First  we analyze rigorously the 
Hellmann--Feynman contribution to the energy shift for low momenta.
In the following subsections we will make rigorous the arguments
leading to Proposition \ref{prop: FG term} about the Fermi golden rule
correction, and thus, to Theorem \ref{thm: canonical dispersion} about
the dispersion relation shift in the canonical setting.
Several auxiliary functions are introduced in Subsections \ref{sec: bog coefficients in momentum rep} and \ref{sec: bog coefficient in energy}. In  Subsection \ref{sec: integration domain}  we partition the integration domain into three
regions adapted to the singularity structure of the integrals.
The corresponding contributions are estimated in Subsection \ref{sec:
  estimates for FG}.

\subsection{Estimates for the Hellmann-Feynman term} \label{sec: HF
  estimates}

\begin{remark}\textbf{($\nu$--dependence)}
	\label{rmk: nu dependence}
	We study the small-momentum asymptotics of
	$\delta_4^\can(\k,\nu,\hat{v})$ and
	$\delta_3^\can(\k,\nu,\hat{v})$ in the regime
	$\frac{|\k|}{\sqrt{\nu}} \to 0$. Since we will eventually consider the dilute
	scaling $\nu\mapsto\kappa\nu$ and the limit $\kappa\nu\to0$, see Sec. \ref{sec: dilute gas limit}, we must
	keep track of the dependence on $\nu$.
	
	After the rescaling $p\mapsto\sqrt{\nu}\,p$, the Bogoliubov dispersion
	relation becomes
	\begin{equation}
		\omega_\Bog(\sqrt{\nu}\,p)
		=
		\nu
		\sqrt{
			\frac{p^4}{4}
			+\frac{\hat{v}(\sqrt{\nu}\,p)}{\hat{v}(0)}p^2
		}.
	\end{equation}
	Thus, apart from explicit powers of $\nu$, the remaining dependence
	on $\nu$ enters through $\hat{v}(\sqrt{\nu}\,\cdot)$. Notice also that
	each derivative of the potential with respect to the rescaled variable produces an
	additional factor of $\sqrt{\nu}$.
	
	Accordingly, we shall perform the computation with $\hat{v}(0)=1$
	and $\nu=1$. The correct dependence is restored by multiplying by
	the global factor $\hat{v}(0)\nu^{3/2}$ and making the replacements
	$$k\mapsto \frac{k}{\sqrt{\nu}}$$ and
	$$\hat{v}(\,\cdot\,)\mapsto
	\frac{\hat{v}(\sqrt{\nu}\,\cdot\,)}{\hat{v}(0)}.$$
\end{remark}


 Before the proof of Proposition \ref{prop: HF term}, we
need the following  simple lemma:
\begin{lemma}\label{lem: L2 convergence}
	Let
	$d=3$
	and
	$f$
	be
	a
	radial
	function in 
        $L^2(\mathbb{R}^3)$.
        Then
	\begin{align}\label{lhs}&
		\frac{1}{(2\pi)^3}\int
		\d\p f(\p)\left(\hat
		v(\p)-\frac{\hat
			v(
			\p+\k)+\hat
			v(\p-\k)}{2}\right)
		\\\label{rhs}
		=&-{\frac{k^2}{12\pi^2}\int_0^\infty\d p
			f(p)\big(\hat
			v''(p)p^2+
			2\hat{v}'(p)p\big)
			+o(k^2).}
	\end{align}
\end{lemma}

\begin{proof}
	\begin{align}
	-	\eqref{lhs}  
		=& \frac12\sum_{j,i=1}^3 k_j k_i \int^1_{-1} \d v (1-|v|)     \frac{1}{(2\pi)^3}\int\d\p f(\p) 
		(\partial_j\partial_i\hat{v})(\p+v\k)\notag\\
		=&\frac12\sum_{j,i=1}^3 k_j k_i \frac{1}{(2\pi)^3}\int\d\p f(\p) 
		(\partial_j\partial_i\hat{v})(\p)+o(k^2). \label{mainterm}
	\end{align}
	Equation \eqref{mainterm} can be justified by an $L^2$-translation continuity argument. We estimate \begin{align}
&\left|	\int\d\p f(\p) 
	(\partial_j\partial_i\hat{v})(\p)- \int\d\p f(\p) \int_{-1}^1 \d v(1-|v|)(\partial_j\partial_i\hat{v})(\p+v\k)\right|\notag \\
	\leq &\|f\|_{L^2(\mathbb{R}^3)} \sup_{|v|\leq 1} \|\partial_i\partial_j\hat{v}(\cdot +v\k)-\partial_i\partial_j\hat{v}\|_{L^2(\mathbb{R}^3)}  \label{eq: L2 estimate I}
	\end{align}
	Now, for every $g\in L^2(\mathbb{R}^3)$ and $\varphi\in C_c^\infty(\mathbb{R}^3)$,
	\begin{equation}
	\sup_{| \q|\leq\k }\|g(\cdot+\q)-g\|_{L^2}
	\leq
	2\|g-\varphi\|_{L^2}+k\|\nabla\varphi\|_{L^2}.
	\end{equation}
	By density, the left-hand side therefore vanishes as $k\to0$. Since $|v\k|\leq|\k|$, the right-hand side of \eqref{eq: L2 estimate I} is $o(1)$, which yields the $o(k^2)$ remainder in \eqref{mainterm}.
	
	Then,
	\begin{equation}
	\sum_{i,j=1}^3	\frac{k_ik_j}{k^2}\partial_i\partial_j\hat{v}(\p) = \hat{v}''(p)\cos^2\theta +\frac{\hat{v}'(p)}{p}(1-\cos^2\theta),
	\end{equation} where $\theta$ is the angle between $\p$ and $\k$. 
	Thus the main term of \eqref{mainterm}
	is
	\begin{align}&
		-\frac{k^2}{2(2\pi)^3}\int_0^\infty\d p
		f(p)p^2\int_{-1}^1 { \d (\cos\theta)}\int_0^{2\pi}{ \d \varphi}\left(\hat
		v''(p)\cos^2\theta+\frac{\hat v'(p)}{p}(1-\cos^2\theta)\right)\notag
		\\&=\ \eqref{rhs}.
	\end{align}
\end{proof}

\begin{proof}[Proof of Proposition \ref{prop: HF term}]
Let us fix without loss of generality $\hat{v}(0)=1$. On account of Remark \ref{rmk: nu dependence} we will also take $\nu = 1$. We rewrite the Hellmann-Feynman contribution \eqref{eq: finite volume HF} in the thermodynamic limit 
as in Equation \eqref{eq: HF thermodynamic limit}\begin{align}
	&\lim_{L\to+\infty}	\left(\Omega_\Bog\right|\left. [b_\k,[H_{4,2}^L, b_\k^*]]\Omega_\Bog\right) =\notag \\
	=	&
	(c_\k+s_\k)^2\frac{1}{(2\pi)^3}\int \d\p c_\p s_\p
	\hat{v}(\p)\label{eq: HF1}\\
	&	+(c_\k^2+s_\k^2)\frac{1}{(2\pi)^3}\int \d\p s_\p^2\left(-\hat{v}(\p) +\frac{\hat{v}(\p+\k)+\hat{v}(\p-\k)}{2}\right)  \label{eq: HF2}\\
	&   +2 c_\k s_\k \frac{1}{(2\pi)^3}\int\d\p c_\p s_\p\left (-\hat v(\p)+\frac{\hat{v}(\p+\k)+\hat{v}(\p-\k)}{2}\right) \label{eq: HF3}\\
	&	-(c_\k-s_\k)^2 \hat{v}(\k)\frac{1}{(2\pi)^3}\int\d\p s_\p^2.  \label{eq: HFfinal}
\end{align}
Note that the convergence of the Riemann sums is implied by \begin{equation}
	\hat{v}(\p)\in C^1(\mathbb{R}^3)\cap L^2(\mathbb{R}^3),\quad \partial_i \hat{v}(\p)\in L^2(\mathbb{R}^3)
\end{equation} and \begin{equation}
|c_\p s_\p|= O\left(\frac{|\hat{v}(\p)|}{|\p|^2}\right),\qquad s_\p^2 = O\left(\frac{\hat{v}(\p)^2}{|\p|^4}\right), \text{ as }|\p|\to+\infty.
\end{equation}
The first line \eqref{eq: HF1} is the infrared divergent term.  The second and third lines, \eqref{eq: HF2} and \eqref{eq: HF3},  can be treated as in Lemma \ref{lem: L2 convergence}. They give \begin{align}
\eqref{eq: HF2} = &(c_\k^2+s_\k^2)\frac{k^2}{12\pi^2}\int \d p s_p^2
                    (\hat{v}''(p)p^2 +2p\hat{v}'(p)) +o(k),\\
\eqref{eq: HF3} = &c_\k s_\k\frac{k^2}{6\pi^2}\int \d p c_p s_p
                    (\hat{v}''(p)p^2 +2 p\hat{v}'(p))+o(k).
\end{align}

To conclude the proof we just have to re-organize the coefficients $c_\k s_\k$, $(c_\k\pm s_\k)$ and $c_\k^2+s_\k^2$ to recover the expression for $c_4$ and for the infrared term.  Using Equation \eqref{eq: simple form for cksk} we have \begin{align}
                                                                                                                                                                                                                                         c_\k
                                                                                                                                                                                                                                         s_\k
                                                                                                                                                                                                                                         =&
                                                                                                                                                                                                                                            {
                     \frac{\sqrt\nu}{2k}(1+O(k^2))}\\
	(c_\k - s_\k)^2 =& \frac{k}{2\sqrt\nu}\left(1+O(k^2)\right)\\
	(c_\k+s_\k)^2
	=& \frac{2\omega_\Bog(k)}{k^2}\\
	c_\k^2+s_\k^2
	=&
	\frac{\sqrt\nu}{k}
	\left(1+ O(k)\right)
\end{align}

After restoring the correct powers of $\nu$ and $\hat{v}(0)$, we conclude the proof.
\end{proof}


\subsection{Bogoliubov coefficients in momentum coordinates} \label{sec: bog coefficients in momentum rep}
In this subsection, we describe several useful properties and expansions of the Bogoliubov coefficients $c_\k$, $s_\k$, and related functions. The proofs of the results in this subsection are simple but tedious, therfore we will omit them.

\paragraph{\textbf{Regularized Bogoliubov coefficients. }} 

In view of Remark \ref{rmk: nu dependence}, we express all momenta in units of
$\sqrt{\nu}$. Accordingly, throughout this section we carry out
the computation under the normalization $\nu=1$ and $\hat{v}(0)=1$.  Equivalently, we perform the replacements $\hat{v}(p)\equiv \frac{\hat{v}(\sqrt{\nu}p)}{\hat{v}(0)}$, $\omega_\Bog(p)\equiv \frac{\omega_\Bog(\sqrt{\nu }p)}{\nu}$, $s_\p \equiv s_{\sqrt{\nu}\p}$ and $c_{\p}\equiv c_{\sqrt{\nu}\p}$.

The coefficients $c_\p$ and $s_\p$  diverge as $\frac{1}{\sqrt{2 \omega_\Bog(p)}}$ as $p \to 0$. Thus, it will be more convenient to work with regularized versions of the latter coefficients in the following

 \begin{align}
	c(p) : =& \sqrt{2\omega_\Bog(p)}c_\p= \sqrt{\sqrt{\omega_\Bog(p)^2+ \hat v(p)^2} +\omega_\Bog(p)}\notag \\
	=& \sqrt{\frac{p^2}{2}+\hat v(p) +\omega_\Bog(p)}	, \label{eq: regularized c}\\
	s(p) :=& \sqrt{2\omega_\Bog(p)}s_\p=\operatorname{sign}(\hat{v}(p)) \sqrt{\sqrt{\omega_\Bog(p)^2+ \hat v(p)^2} -\omega_\Bog(p)}\notag \\
	=&\operatorname{sign}(\hat{v}(p)) \sqrt{\frac{p^2}{2}+\hat v(p) -\omega_\Bog(p)}	. \label{eq: regularized s}
\end{align} Unlike $c_p$ and $s_p$, the regularized coefficients $c(p)$ and
$s(p)$ extend to $C^5$ functions in a neighborhood of the origin.
In particular,
$
c(0)=s(0)=1.
$
We collect their main properties in the following lemma.

 \begin{lemma}\label{lem: expansion of coefficients}
	Suppose that the potential satisfies Assumptions \ref{ass: primary assumption on the potential} and \ref{ass: secondary hypotheses}. \begin{enumerate}
		\item [a)] The regularized Bogoliubov coefficients satisfy the following relations for all $p\in [0,+\infty[$ \begin{align}
			&c(p)^2-s(p)^2 = 2\omega_\Bog(p)\label{eq: rel1}\\
			& c(p)^2 + s(p)^2 =  p^2 +2\hat v(p)\label{eq: rel2}\\
			& c(p)s(p) = \hat v (p)\label{eq: rel3}\\
			&c(p)-s(p) = p\label{eq: rel4}\\
			&c(p)+s(p) = \frac{2\omega_\Bog(p)}{p}\label{eq: rel5}
		\end{align}
		\item [b)] The regularized Bogoliubov coefficient have the following expansions \begin{align}
			&\hat v(p) = 1 + r(p),\qquad r(p)\in O\left( p^2\right)\label{eq: exp nu}\\
			&c(p) = 1+ \frac{\omega_\Bog(p)}{2} + \frac{\omega_\Bog(p)^2}{8}+\frac{r(p)}{2}+R_c(p) ,\label{eq: exp c}\\
			&s(p) = 1- \frac{\omega_\Bog(p)}{2} + \frac{\omega_\Bog(p)^2}{8}+\frac{r(p)}{2} +R_s(p)  \label{eq: exp s},
		\end{align} as $p\to 0$, where $R_c$ and $R_s$ are $C^2$ and of order $O\left(p^3\right)$ in  neighborhood of $0$.
	\end{enumerate}
\end{lemma}

We next record the large-momentum behavior of $c(p)$, $s(p)$, and
their derivatives.
\begin{lemma}
	\label{lem: asymptotics of coefficients}
	Suppose that the potential satisfies Assumptions \ref{ass: primary assumption on the potential} and \ref{ass: secondary hypotheses}. Then, we have the following \begin{enumerate}
		\item [a)] The rescaled Bogoliubov energy and its first and second derivatives satisfy the following \begin{align}
			&\omega_\Bog(p) = \frac{p^2}{2}\left(1+O\left(\frac{1}{p^2}\right)\right),\\
			&\omega_\Bog'(p) = p\left(1+ O\left(\frac{1}{p}\right)\right),\\
			&\omega_\Bog''(p) = O(1)
		\end{align} as $p\to+\infty$.
		\item [b)] The regularized coefficients $s(p),\;c(p)$ and their  derivatives are continuous functions satisfying \begin{align}
			&c(p) = p\left(1+ O\left(\frac{1}{p^2}\right)\right),\qquad s(p) = \frac{\hat v(p)}{p}\left(1+O\left(\frac{1}{p^2}\right)\right)\\
			&c'(p) = 1+ O\left(\frac{1}{p}\right),\qquad s'(p) = O\left(\frac{|\hat v '(p)|}{p}+\frac{|\hat v(p)|}{p^2}\right)\\
			&c''(p) = O\left(\frac{1}{p}\right),\qquad s''(p) = O\left(\frac{|\hat v''(p)|p^2+|\hat v'(p)|p+ |\hat v(p)|}{p^3}\right)
		\end{align} as $p\to+\infty$.
	\end{enumerate}
\end{lemma}

The symmetric combination of two translated Bogoliubov energies
admits the following Taylor expansion with an exact remainder: \begin{align}\label{eq: Romega}
	&\omega_\Bog(p+\epsilon)+\omega_\Bog(p-\epsilon) = 2\omega_\Bog(p) + \epsilon^2 R_\omega(p,\epsilon),\\
	&R_{\omega}(p,\epsilon) = \int^1_{-1} \d v (1-|v|)\omega_\Bog''(p+v\epsilon)\label{eq: eq for Romega},
\end{align} for $p\geq \epsilon$. Clearly, $R_\omega(p,0) = \omega_\Bog''(p)$, for $p>0$. 

 Lemma \ref{lem: asymptotics of coefficients} immediately yields the following Corollary \begin{cor}\label{cor: Romega}
	Suppose that the potential satisfies Assumptions \ref{ass: primary assumption on the potential} and \ref{ass: secondary hypotheses}. Then \begin{equation}
		|R_\omega(p,\epsilon)| \leq C,
	\end{equation}  for $p\geq \epsilon >0 $ and for some sufficiently large constant $C>0$.
\end{cor}

\paragraph{\textbf{Regularized Fermi golden rule integrands.}}We proceed to define regularized versions of $j(k;p,q)$ and $\kappa(k,p,q)$ in terms of the regularized coefficients $c(p)$ and $s(p)$.  \begin{align}
	&J(k;p, q) = \sqrt{8\omega_\Bog(p)\omega_\Bog(q)\omega_\Bog(k)}j(k; p,q)\label{eq: regularized j}\\
	=& -\hat v(k)(c(k)-s(k))(c(p)s(q)+c(q)s(p)) \notag\\ 
	&+ \hat v(p)(c(p)-s(p))(c(q)c(k)+s(q)s(k))\notag \\ 
	&+\hat v(q)(c(q)-s(q))(c(p)c(k)+s(p)s(k))\notag 
\end{align} and \begin{align}
	&K(k,p, q) = \sqrt{8\omega_\Bog(p)\omega_\Bog(q)\omega_\Bog(k)}\kappa(k,p,q) \label{eq: regularized kappa}\\
	=& -\hat v(k)(c(k)-s(k))(c(p)s(q)+c(q)s(p)) \notag \\ 
	&-\hat v(p)(c(p)-s(p))(s(q)c(k)+c(q)s(k)) \notag\\ 
	&-\hat v(q)(c(q)-s(q))(s(p)c(k)+c(p)s(k)).\notag
\end{align}
 
 From now on we will employ the $(t,s)$ coordinates \ref{def: ts coordinates}.
  
\paragraph{\textbf{Expansions for $J$ and $K$.}} Since $J$ and $K$ are even functions of $s$, and since the potential is twice differentiable \eqref{eq: twice differentiability}, we obtain
\begin{align}
	&J(k;t+sk,t-sk) = J(k;t,t) + k^2s^2S_J(k,t,sk), \label{eq: exp Jtt}\\
	&K(k,t+sk,t-sk) = K(k,t,t) + k^2 s^2S_K(k,t,sk)\label{eq: exp Ktt},
\end{align} where the remainders $S_{J/K}$ are defined as \begin{align}
	S_J(k,t,sk) = \frac{1}{2}\int^{1}_{-1} \d v (1-|v|)\Bigl[ J(k,t+\cdot, t-\cdot)\Bigr]''\Big{|}_{vsk},\label{eq: S remainder for J}\\
	S_K(k,t,sk) = \frac{1}{2}\int^{1}_{-1} \d v (1-|v|)\Bigl[ K(k,t+\cdot, t-\cdot)\Bigr]''\Big{|}_{vsk}.
\end{align} If we were to set $k=0$, we would obtain \begin{align}\label{eq: SJ00}
	S_{J}(0;t,0) = &8\left( \frac{\hat v'(t)\omega_\Bog(t)}{t} + \frac{\hat v(t)\omega_\Bog(t)}{t^2} - \frac{\omega_\Bog'(t)\hat v(t)}{t}\right)  \notag \\ & +2\hat v''(t)\omega_\Bog(t)+2\hat v(t)\omega_\Bog''(t)-4\omega_\Bog'(t)\hat v'(t)  \\
	S_K(0,t,0) =& -S_J(0;t,0).\label{eq: SK00}
\end{align}

We now derive estimates for these remainders \begin{lemma}\label{lemma: estimates SJ and SK}
	Suppose that the potential satisfies Assumptions \ref{ass: primary assumption on the potential} and \ref{ass: secondary hypotheses}. Let us introduce the function \begin{equation}\label{eq: def of Phi}
		\Phi(t):= |\hat{v}(t)|+ t|\hat{v}'(t)|+t^2 |\hat{v}''(t)|
	\end{equation}Then, we can bound the remainders $S_J$ and $S_K$ as \begin{align}
	|S_J(k,t,sk)|+ |S_K(k,t,sk)| \leq C \int^1_{-1}\d v (1-|v|)\Phi(t+vsk)
	\end{align} for $k$ in a sufficiently small neighborhood of $0$. In particular, \begin{equation}
	|S_J(k,t,sk)|+ |S_K(k,t,sk)| \leq C(1+t^2)
	\end{equation} for some sufficiently large $C>0$.
\end{lemma}

We next examine  the diagonal expressions $J(k;t,t)$ and $K(k,t,t)$. The former can be written as \begin{align}
	J(k;t,t)=&2\hat v(t)\Bigl[ (2\omega_\Bog(t)-\omega_\Bog(k))\left(1 +\frac{\omega_\Bog(k)}{4}\frac{t^2}{\omega_\Bog(t)}\right) +  R_J(k,t)\Bigr]
\end{align} where \begin{align}\label{eq: RJ}
	R_J(k,t) = &\frac{\omega_\Bog(k)^2}{4}\frac{t^2}{\omega_\Bog(t)} + 2\omega_\Bog(t)\left(\frac{\omega_\Bog(k)^2}{8} + \frac{r(k)}{2}\right)\notag \\
	&+t(c(t)R_c(k)+s(t)R_s(k)) +(R_s(k)-R_c(k))- r(k)k.
\end{align} Similarly, we have for $K(k,t,t)$ \begin{align}
	K(k,t,t)=&-2\hat v(t)\Bigl[ (2\omega_\Bog(t)+\omega_\Bog(k))\left(1 -\frac{\omega_\Bog(k)}{4}\frac{t^2}{\omega_\Bog(t)}\right) +  R_K(k,t)\Bigr],
\end{align} where 
\begin{align}\label{eq: RK}
	R_K(k,t) = &\frac{\omega_\Bog(k)^2}{4}\frac{t^2}{\omega_\Bog(t)} + 2\omega_\Bog(t)\left(\frac{\omega_\Bog(k)^2}{8} + \frac{r(k)}{2}\right)\notag \\
	&+ t(c(t)R_s(k)+s(t)R_c(k)) -(R_s(k)-R_c(k))+r(k)k.
\end{align} 
The sum and difference of the remainders $R_J$ and $R_K$ are given by  \begin{align}
	R_J+R_K =& \frac{\omega_\Bog(k)^2}{2}\frac{t^2}{\omega_\Bog(t)} + \omega_\Bog(t)\left(\frac{\omega_\Bog(k)^2}{2} +2r(k)\right)\notag \\ &+2\omega_\Bog(t)(R_c(k)+R_s(k)),\label{eq: sum of RJ and RK}\\
	R_J -R_K = & \left(t^2-2\right)(R_c(k)-R_s(k))-2r(k)k.\label{eq: difference of RJ and RK}
\end{align}

Using Equation \eqref{eq: Romega}  we can also write $J(k;t,t)$ and $K(k,t,t)$ in a more useful form:\begin{align}\label{eq: final form for Jtt}
	& 2\hat v(t)\Bigl[(\omega_\Bog(t+sk)+\omega_\Bog(t-sk)-\omega_\Bog(k))\left(1 +\frac{\omega_\Bog(k)}{4}\frac{t^2}{\omega_\Bog(t)}\right) \notag \\
	&- s^2k^2 R_\omega(t,sk) \left(1 +\frac{\omega_\Bog(k)}{4}\frac{t^2}{\omega_\Bog(t)}\right) + R_{J}(k,t) ] = J(k;t,t),
\end{align}  and 
\begin{align}\label{eq: final form for Ktt}
	&- 2\hat v(t)\Bigl[(\omega_\Bog(t+sk)+\omega_\Bog(t-sk)+\omega_\Bog(k))\left(1 -\frac{\omega_\Bog(k)}{4}\frac{t^2}{\omega_\Bog(t)}\right)  \notag \\
	&- s^2k^2 R_\omega(t,sk) \left(1 -\frac{\omega_\Bog(k)}{4}\frac{t^2}{\omega_\Bog(t)}\right) + R_K(k,t) ] = K(k,t,t),
\end{align} where in both equations we have added and subtracted a term $s^2k^2R_\omega(t,sk)$.

 $R_J$ and $R_K$ satisfy the following upper bound:\begin{lemma}\label{eq: upper bound for RJ and RK}
		Suppose that the potential satisfies Assumptions \ref{ass: primary assumption on the potential} and \ref{ass: secondary hypotheses}. Then, we have \begin{equation}
		\max\{|R_J(k,t)|, |R_K(k,t)|\} \leq C  k^2(1+t^2),
	\end{equation} for some sufficiently large $C>0$, as $k\to 0$.
\end{lemma}


\paragraph{\textbf{Density function $h$.}} We further define
\begin{equation}
	h(t) := \frac{t}{\omega_\Bog(t)}.
\end{equation}
This function appears naturally as the density factor in the Fermi golden rule integrals after the change of variables in Definition \ref{def: ts coordinates}. We collect its relevant properties below
\begin{lemma}\label{lem: asymptotic of h}
		Suppose that the potential satisfies Assumptions \ref{ass: primary assumption on the potential} and \ref{ass: secondary hypotheses}. Then $h(t),\; h'(t)$ and $h''(t)$ are continuous and bounded functions. Moreover,  the satisfy \begin{align}
		h(t) = O\left(\frac{1}{t}\right), \qquad h'(t) = O\left(\frac{1}{t^2}\right),\notag  \\ h''(t) = O\left(\frac{1}{t^3}\right) \text{ as }t\to+\infty.
	\end{align}
\end{lemma}

It will be useful to have the following expansion at hand\begin{align}
	h(t+sk)h(t-sk) =& h(t)^2 + s^2k^2 R_h(t,sk),\\
	R_h(t,sk) =& \frac{1}{2}\int^1_{-1}\d v (1-|v|)(h''(t-vsk)h(t+vsk)+h''(t+vsk)h(t-vsk) \notag \\
	&- 2h'(t-vsk)h'(t+vsk)).\label{eq: Rh}
\end{align} 

As a corollary to Lemma \ref{lem: asymptotic of h} we get \begin{cor}\label{lem: corollary for Rh}
	Suppose that the potential satisfies Assumptions \ref{ass: primary assumption on the potential} and \ref{ass: secondary hypotheses}. Then \begin{equation}
	|R_h(t,sk)| \leq C \min\{1,t^{-4}\}
	\end{equation} for some sufficiently large constant $C>0$, $t\in [\frac{k}{2},+\infty[$.
\end{cor}

Setting $k=0$ in \eqref{eq: Rh} we get \begin{align}
	R_h(t,0) =\frac{\omega_\Bog'(t)^2 t^2-\omega_\Bog''(t)\omega_\Bog(t)t^2 -\omega_\Bog(t)^2 }{\omega_\Bog(t)^4}
\end{align}

\subsection{Expansions in the energy coordinates}\label{sec: bog coefficient in energy}
Equations\eqref{eq: positivity of Fourier1}, \eqref{eq: regularity at 0} and \eqref{eq: rot invariace} imply that
$p\mapsto\omega_{\mathrm{bg}}(p)$ is invertible in a sufficiently
small neighborhood $U$ of the origin. We denote its inverse by
$\omega\mapsto p(\omega)$ (or $\omega\mapsto q(\omega)$ ). As $p,q,k\to0$, we can therefore
express $j(k;p,q)$ and $\kappa(k,p,q)$ in terms of the corresponding
Bogoliubov energies. We introduce the following change of variables
\begin{defn}\label{def: change of variables xy} \textbf{$x-y$ coordinates:}
\begin{align}&x := \omega_\Bog(p)+\omega_\Bog(q),\qquad y = \frac{\omega_\Bog(p)-\omega_\Bog(q)}{k}, \notag \\
	& \omega_\Bog(p) = \frac{x+yk}{2},\qquad \omega_\Bog(q) = \frac{x-yk}{2}.\label{eq: change in x and y}
\end{align}
\end{defn}

 With a slight abuse of notation, we use the same symbols $J$ and
 $K$ for the corresponding functions of the energy variables. Thus,
 we write \begin{equation}J\left(k;p\left(\frac{x+yk}{2}\right),q\left(\frac{x-ky}{2}\right)\right) = J(k;x+yk,x-yk).\end{equation}

We then obtain the following expansions, whose proof is essentially
contained in \cite[Sec.3.2]{DP_2026}. \begin{lemma}\label{lem: expansion of J and K}
		Suppose that the potential satisfies Assumptions \ref{ass: primary assumption on the potential} and \ref{ass: secondary hypotheses}.   Then, the following expansions hold \begin{align}
		J(k;x+yk,x-yk) = 2(x-\omega_\Bog(k)) + \ R(k;x+yk,x-yk),\notag \\\text{ as } x,k\to 0,\\
		K(k,x+yk,x-yk) = -2(x+\omega_\Bog(k)) + S(k,x+yk,x-yk) ,\notag \\ \text{ as } x,k\to 0,
	\end{align} where $R$ and $S$ are at least $C^2$ in a neighborhood of $(x,k) = (0,0)$ and are of order $O\left((x^2+y^2k^2)^{\frac{3}{2}}\right)$.
\end{lemma}


We shall also introduce the following  \begin{align}\label{eq: def of f}
	f(\omega) := \frac{\d[ p(\omega)^2]}{\d [\omega^2]} = \left(\frac{\hat{v}(p(\omega))}{\hat{v}(0)}+\frac{1}{2}\left(1+\frac{1}{\hat{v}(0)p(\omega)}\hat{v}'(p(\omega))\right) p(\omega)^2 \right)^{-1} .
\end{align} As a  consequence of Assumptions \eqref{eq: positivity of Fourier1}, \eqref{eq: rot invariace} and \eqref{eq: regularity at 0}, $f$ is bounded in a neighborhood of the origin and it satisfies  \begin{equation}\label{eq: f expansion for small momenta}
	f(\omega) = \left(1+ O\left(\omega^2\right)\right)
\end{equation} as $\omega\to 0$.

  \subsection{Integration domain}\label{sec: integration domain}
In this subsection, we partition $\mathbb{R}^3$ as the disjoint union of three domains $(I_{\k,i})_{i=1}^3$. These will depend on the momentum vector $\k$. However, for sufficiently small
$k=|\mathbf k|$,  their qualitative properties remain unchanged. We also describe their representations in the energy coordinates $(x,y)$ introduced in \eqref{eq: change in x and y} and in the momentum coordinates $(t,s)$ introduced in  \eqref{eq: change in t s}.

We take
\begin{align}
	&	I_{\k,1}:= \left\{ \p \in \mathbb{R}^3 | 2\omega_\Bog\Big(\frac{k}{2}\Big)\leq \omega_\Bog(p)+\omega_\Bog(q) < 2\omega_\Bog(k)-2\omega_\Bog\Big(\frac{k}{2}\Big)\right\},\label{eq: first region}\\
	&	I_{\k,2}:= \left\{ \p \in \mathbb{R}^3 |
   2\omega_\Bog(k)-2\omega_\Bog\Big(\frac{k}{2}\Big)
   \leq\omega_\Bog(p)+\omega_\Bog(q), \quad p+q < 4k \right\},\label{eq: second region}\\
	&I_{\k,3}:= \{ \p \in \mathbb{R}^3 | 4k \leq p+q \leq +\infty\}\label{eq: third region}.
\end{align} Note that in $I_{\k,1}$, the range of $\omega_\Bog(p)+\omega_\Bog(q)$ is symmetric with respect to $\omega_\Bog(k)$. It is clear from their definitions that these sets are pairwise disjoint. \begin{remark}
As will become clear from the estimate of the Fermi golden rule contribution, the decomposition into the domains $I_{\k,1}$, $I_{\k,2}$ and $I_{\k,3}$ is tailored to the coordinate system best suited to each region:
\begin{enumerate}
	\item[$I_{\k,1}$:] 
	The symmetric region around the pole of the Fermi golden rule integral. Energy coordinates are convenient here because they resolve the behavior near the pole.
	
	\item[$I_{\k,2}$:] 
	The transition region between the near-pole domain, whose boundary is at distance $O(k^3)$ from the pole, and the large-momentum region, which is at distance $O(k)$ from the pole. This region is still best treated in energy coordinates.
	
	\item[$I_{\k,3}$:] 
	The large-momentum region. In this case the boundary of the integration domain is simpler in momentum variables.
\end{enumerate}
\end{remark}

The following lemma describes the minimum of
$\omega_{\mathrm{bg}}(p)+\omega_{\mathrm{bg}}(q)$ and its distance
from $\omega_{\mathrm{bg}}(k)$.
\begin{lemma}\label{lemma: estimates for ak}
		Fix $\hat{v}(0) = 1$ and $\nu = 1$. Suppose that the potential satisfies Assumptions \ref{ass: primary assumption on the potential} and \ref{ass: secondary hypotheses}. Then, we have
	\begin{equation} \label{eq: convexity inequality}
		\min_{\p\in\mathbb{R}^3}\big(\omega_\Bog(\p)+\omega_\Bog(\p-\k)\big) = 2\omega_\Bog\left(\frac{\k}{2}\right) < \omega_\Bog(\k).
	\end{equation} Moreover there exist sufficiently large positive constants $D\geq C >0$, such that for all sufficiently small $k>0$ one has\begin{equation}\label{eq: inq for a}
		C k^3\leq \omega_\Bog(k)-2\omega_\Bog\left(\frac{k}{2}\right) \leq  D k^3
	\end{equation}
\end{lemma}
\begin{proof}
	Equation \eqref{eq: convexity inequality} was proved in \cite{DLN_2023} using the strict convexity of $p\to \omega_\Bog(p)$ for sufficiently small $p$. Under our assumptions, strict convexity follows from Equation \eqref{eq: convexity}. Expanding $\omega_\Bog(k)$ at the origin gives \begin{align}
	\omega_\Bog(k)-2\omega_\Bog\left(\frac{k}{2}\right)& =  \frac{3}{16}\left(\hat v''(0)+\frac{1}{2}\right)k^3 + O(k^5)
	\end{align}
\end{proof} 

 Lemma \ref{lemma: estimates for ak} immediately implies that the three domains cover the full integration region \begin{cor}\label{cor: integration domains}		Suppose that the potential satisfies Assumptions \ref{ass: primary assumption on the potential} and \ref{ass: secondary hypotheses}.  Then, for all sufficiently small $k$, \begin{equation}
		I_{\k,1}\cup I_{\k,2}\cup I_{\k,3} = 	\mathbb{R}^3 .
	\end{equation}
\end{cor}
\begin{proof} 
$I_{\k,1}.$ 
Let us parametrize the domain $I_{\k,1}$  in terms of the variables $(x,y)$ in \eqref{eq: change in x and y}. The constraint in \eqref{eq: first region} bounds $x$ as \begin{equation}
	0< 2\omega_\Bog\left(\frac{k}{2}\right)\leq x < 2\omega_\Bog(k) - 2\omega_\Bog\left(\frac{k}{2}\right).
\end{equation} 
Using the small-momentum expansion of the dispersion relation, we
find  \begin{equation}
	ky = \omega_\Bog(p)-\omega_\Bog(q) = (p-q) +  O\left(p^3+q^3\right) ,\text{ as } p,q \to 0,
\end{equation} and taking into account the symmetry $y\leftrightarrow -y$, we have
 \begin{equation}
	-\eta_k(x)\leq y \leq \eta_k(x) ,\quad [-\eta_k(x),\eta_k(x)]\subset [-1-O(k^2),1+O(k^2)]
\end{equation}  for some continuous  function $\eta_k$, uniformly for $x\in[2\omega_\Bog(\frac{k}{2}),2\omega_\Bog(k)-2\omega_\Bog(\frac{k}{2} )[$.

Moreover, the endpoints satisfy \begin{align}
	&p\left(\frac{x+k\eta_k(x)}{2}\right)+p\left(\frac{x-k\eta_k(x)}{2}\right) = k,\quad 2\omega_\Bog\Big(\frac{k}{2}\Big)\leq x\leq \omega_\Bog(k),\\
	&p\left(\frac{x+k\eta_k(x)}{2}\right)-p\left(\frac{x-k\eta_k(x)}{2}\right) = k,\quad \omega_\Bog(k)\leq x,
\end{align} and \begin{equation}
\eta_k(\omega_\Bog(k)) = \frac{\omega_\Bog(k)}{k},\quad \eta_k\left(2\omega_\Bog\Big(\frac{k}{2}\Big)\right) = 0.
\end{equation}

As $p\to \omega_\Bog(p)$ is strictly increasing and convex in a neighborhood of $0$, $\omega\to p(\omega)$ is strictly increasing and concave. Then, we have for $y> 0$ that \begin{align}
	&\partial_y \left( p\left(\frac{x+ky}{2}\right) + p\left(\frac{x-ky}{2}\right) \right) <0,\\ 
	&\partial_y \left( p\left(\frac{x+ky}{2}\right) - p\left(\frac{x-ky}{2}\right) \right)>0,
\end{align} in  $2\omega_\Bog(\frac{k}{2})<x<\omega_\Bog(k)$ and $\omega_\Bog(k)<x$ respectively. Thus, by the implicit function theorem $x\to \eta_k(x)$ is $C^5$ in these intervals. At $x = \omega_\Bog(k)$, $\eta_k(x)$ is continuous but it has different left and right derivatives, namely \begin{equation}
\lim_{\epsilon\to 0}\frac{\eta_{k}(\omega_\Bog(k))-\eta_k(\omega_\Bog(k)-\epsilon)}{\epsilon} = \frac{p'(\omega_\Bog(k))+p'(0)}{k(p'(0)-p'(\omega_\Bog(k)) )}
\end{equation} and
\begin{equation}
	\lim_{\epsilon\to 0}\frac{\eta_{k}(\omega_\Bog(k)+\epsilon)-\eta_k(\omega_\Bog(k) )}{\epsilon} = \frac{p'(0)-p'(\omega_\Bog(k))}{k(p'(0)+p'(\omega_\Bog(k)) )}.
\end{equation} 

$I_{\k,2}.$ The constraint in $I_{\k,2}$ puts an upper bound on $x$\begin{equation}
	2\omega_\Bog(k)-	2\omega_\Bog\left(\frac{k}{2}\right)<x\leq \sup_{p+q<4k}\left(\omega_\Bog(p)+\omega_\Bog(q)\right). 
\end{equation} by expanding the dispersion relation as\begin{align}
	\omega_\Bog(p)+\omega_\Bog(q) &= (p+q)\left(1 +  O\left(p^2+q^2\right)\right) \notag \\
	&= \omega_\Bog(p+q)\left(1 +  O\left(p^2+q^2\right)\right) \text{ as }p,q\to 0
\end{align} one gets \begin{equation}
\sup_{p+q<4k}\left(\omega_\Bog(p)+\omega_\Bog(q)\right) = \omega_\Bog(4k)\left(1 + O\left(k^2\right)\right).
\end{equation} Therefore, the $y$ integration region is of the form \begin{equation}
	-1-O(k^2)\leq y \leq 1+O(k^2)
\end{equation} for all $$x\in \Bigl[2\omega_\Bog(k)-2\omega_\Bog\Big(\frac{k}{2}\Big),  \sup_{p+q<4k}(\omega_\Bog(p)+\omega_\Bog(q))\Bigr[,$$ where the remainder term $O(k^2)$ in the integration region possibly depends on $x$.

$I_{\k,3}.$ Finally, we observe that in terms of the variables $(t,s,\varphi)$, the domain $I_{\k,3}$ simply becomes \begin{equation}
	I_{\k,3}= [2k,+\infty[\times\Bigl[-\frac{1}{2},\frac{1}{2}\Bigr]\times [0,2\pi[.
\end{equation}
\end{proof}

We conclude this subsection by locating the pole of the first Fermi golden rule integral \eqref{eq: real divergence fermi} and \eqref{eq: real divergence fermi2} \begin{prop}\label{prop: position of the pole}
	Suppose that the potential satisfies Assumptions \ref{ass: primary assumption on the potential} and \ref{ass: secondary hypotheses}.  Then, for all sufficiently small $k$, the equation \begin{equation}
		x-\omega_\Bog(k) = \omega_\Bog(p)+\omega_\Bog(q)-\omega_\Bog(k) = 0,
	\end{equation} does not have any solutions if $p>k$.
\end{prop}
\begin{proof}
	By our Assumptions on the potential \eqref{eq: positivity of Fourier1} and \eqref{eq: regularity at 0}, we know that there exists a neighborhood of $p=0$, say $U = [0,c[$ for some $c>0$, such that $p\to\omega_\Bog(p)$ is strictly increasing. Moreover, by \eqref{eq: positivity of Fourier1} we also know that \begin{equation}
		\inf_{p>c}\omega_\Bog(p) > 0.
	\end{equation}
	We can take $k<c$ sufficiently small so that \begin{equation}
		\omega_\Bog(k)< \inf_{p>c}\omega_\Bog(p).
	\end{equation} Therefore, if $p>k$, we have \begin{equation}
		\omega_\Bog(p)+\omega_\Bog(q)>\omega_\Bog(p) > \min\{\inf_{k<p<c}\omega_\Bog(p),\inf_{p> c}\omega_\Bog(p)\} = \omega_\Bog(k).
	\end{equation}

\end{proof}

\subsection{Estimates for the energy integrals}\label{sec: estimates for FG}
   
We  take $k$ sufficiently
small that Corollary \ref{cor: integration domains} applies. We write
\begin{align}
	\delta(I_{\k,i})={}&-\frac12\Re\int_{I_{\k,i}}\frac{\d\p}{(2\pi)^3}
	\Biggl[\frac{j(\k;\p,\p-\k)^2}
	{\omega_\Bog(\p)+\omega_\Bog(\p-\k)-\omega_\Bog(\k)-\i0}\notag\\
	&\hspace{30mm}+\frac{\kappa(\k,\p,\p-\k)^2}
	{\omega_\Bog(\k)+\omega_\Bog(\p)+\omega_\Bog(\p-\k)}\Biggr].
	\label{eq: sigma}
\end{align}
In the energy coordinates of Definition \ref{def: change of variables xy},
\begin{equation}\label{eq: change xy}
	\d\p=\frac{x^2-k^2y^2}{8}
	f\left(\frac{x+ky}{2}\right)f\left(\frac{x-ky}{2}\right)
	\d x\,\d y\,\d\varphi.
\end{equation}

\paragraph{\textbf{Estimate of $\delta(I_{\k,1})$.}}
\begin{prop}\label{prop: FG zero}
	Under Assumptions \ref{ass: primary assumption on the potential} and
	\ref{ass: secondary hypotheses},
	\begin{equation*}
		\delta(I_{\k,1})=\hat v(0)\nu^{3/2}
		O\left(\frac{k^3}{\nu^{3/2}}\right),
		\qquad \frac{k}{\sqrt\nu}\longrightarrow0.
	\end{equation*}
\end{prop}
\begin{proof}
	We first set $\nu=\hat v(0)=1$, as in Remark \ref{rmk: nu dependence}.
	After the angular integration, the contribution containing $J$ is
	\begin{equation}\label{eq: integral for J revised}
		-\frac1{128\pi^2\omega_\Bog(k)}\Re
		\int_{2\omega_\Bog(k/2)}^{2\omega_\Bog(k)-2\omega_\Bog(k/2)}
		\frac{F_k(x)}{x-\omega_\Bog(k)-\i0}\,\d x,
	\end{equation}
	where, using the same endpoint function $\eta_k$,
	\begin{align}
		G(k,x,y)&:=f\left(\frac{x+ky}{2}\right)
		f\left(\frac{x-ky}{2}\right)J(k;x+ky,x-ky)^2,\notag\\
		F_k(x)&:=\int_{-\eta_k(x)}^{\eta_k(x)}G(k,x,y)\,\d y.
		\label{eq: def of F2}
	\end{align}
	On $I_{\k,1}$, Lemmas \ref{lemma: estimates for ak} and
	\ref{lem: expansion of J and K}, and local regularity, give
	\begin{align*}
		&ck^3\leq\omega_\Bog(k)-2\omega_\Bog(k/2)\leq Ck^3,
		\qquad 0\leq\eta_k(x)\leq C,\\
		&|J(k;x+ky,x-ky)|\leq Ck^3,\qquad
		|\partial_xJ(k;x+ky,x-ky)|\leq C,\\
		&|G(k,x,y)|\leq Ck^6,\qquad
		|\partial_xG(k,x,y)|\leq Ck^3,\qquad |F_k(x)|\leq Ck^6.
	\end{align*}
	Here the derivative bound follows from the smooth expression for $J$;
	also $f$ and $f'$ are bounded near zero.
	
	We only need to differentiate $F_k$ near the pole. For
	\begin{equation*}
		\frac{\omega_\Bog(k)+2\omega_\Bog(k/2)}2
		\leq x<\omega_\Bog(k),
	\end{equation*}
	the endpoint relation $p+q=k$, Taylor's formula and
	$\omega_\Bog''(p)=O(p)$ imply
	\begin{equation*}
		ck^3\leq x-2\omega_\Bog(k/2)
		\leq Ck(p-q)^2\leq Ck^3\eta_k(x)^2,
		\qquad \eta_k(x)\geq c.
	\end{equation*}
	The inverse dispersion satisfies
	\begin{align*}
		&p'(\omega)=1+O(\omega^2),\qquad
		p''(\omega)=-\frac34\bigl(1+2\hat v''(0)\bigr)\omega+O(\omega^3),\\
		&p'\left(\frac{x-k\eta_k(x)}2\right)
		-p'\left(\frac{x+k\eta_k(x)}2\right)
		\geq c xk\eta_k(x)\geq ck^2.
	\end{align*}
	Implicit differentiation of the endpoint relation therefore gives
	\begin{equation*}
		|\eta_k'(x)|=
		\frac{p'\left(\frac{x+k\eta_k(x)}2\right)
			+p'\left(\frac{x-k\eta_k(x)}2\right)}
		{k\left|p'\left(\frac{x+k\eta_k(x)}2\right)
			-p'\left(\frac{x-k\eta_k(x)}2\right)\right|}
		\leq Ck^{-3}.
	\end{equation*}
	For $x>\omega_\Bog(k)$ in $I_{\k,1}$ the endpoint relation is $p-q=k$,
	and instead
	\begin{equation*}
		|\eta_k'(x)|=
		\frac{p'\left(\frac{x-k\eta_k(x)}2\right)
			-p'\left(\frac{x+k\eta_k(x)}2\right)}
		{k\Bigl[p'\left(\frac{x-k\eta_k(x)}2\right)
			+p'\left(\frac{x+k\eta_k(x)}2\right)\Bigr]}
		\leq Ck.
	\end{equation*}
	Consequently,
	\begin{align*}
		|F_k'(x)|
		&\leq\int_{-\eta_k(x)}^{\eta_k(x)}|\partial_xG(k,x,y)|\,\d y
		+2|\eta_k'(x)|\sup_{|y|\leq\eta_k(x)}|G(k,x,y)|\leq Ck^3,\\
		&\hspace{8mm}0<|x-\omega_\Bog(k)|
		\leq\frac{\omega_\Bog(k)-2\omega_\Bog(k/2)}2.
	\end{align*}
	Since $F_k$ is continuous at the pole, symmetrization gives
	\begin{align}
		&\Re\int_{2\omega_\Bog(k/2)}^{2\omega_\Bog(k)-2\omega_\Bog(k/2)}
		\frac{F_k(x)}{x-\omega_\Bog(k)-\i\epsilon}\,\d x\notag\\
		&\quad=\int_0^{\omega_\Bog(k)-2\omega_\Bog(k/2)}
		\frac{r\,[F_k(\omega_\Bog(k)+r)-F_k(\omega_\Bog(k)-r)]}
		{r^2+\epsilon^2}\,\d r.
		\label{eq: final difference for J}
	\end{align}
	The right-hand side is bounded, uniformly for $\epsilon>0$, by
	\begin{align*}
		&Ck^3\int_0^{(\omega_\Bog(k)-2\omega_\Bog(k/2))/2}\d r
		+Ck^6\int_{(\omega_\Bog(k)-2\omega_\Bog(k/2))/2}
		^{\omega_\Bog(k)-2\omega_\Bog(k/2)}\frac{\d r}{r}
		\leq Ck^6.
	\end{align*}
	This also supplies an integrable majorant for the limit
	$\epsilon\downarrow0$. Thus \eqref{eq: integral for J revised} is
	$O(k^5)$.
	
	For the second channel, $K(k;x+ky,x-ky)=O(k)$ and
	$x+\omega_\Bog(k)\geq ck$ on $I_{\k,1}$. Hence
	\begin{align}
		&\frac1{128\pi^2\omega_\Bog(k)}
		\int_{2\omega_\Bog(k/2)}^{2\omega_\Bog(k)-2\omega_\Bog(k/2)}\d x
		\int_{-\eta_k(x)}^{\eta_k(x)}\d y\,\notag\\
		&\qquad\times f\left(\frac{x+ky}{2}\right)f\left(\frac{x-ky}{2}\right)
		\frac{K(k;x+ky,x-ky)^2}{x+\omega_\Bog(k)}\notag\\
		&\qquad\leq\frac Ck\,k^3\frac{k^2}{k}=Ck^3.
		\label{eq: integral for K revised}
	\end{align}
	Restoring the factor $\hat v(0)\nu^{3/2}$ and replacing $k$ by
	$\frac{k}{\sqrt\nu}$ proves the claim.
\end{proof}

\paragraph{\textbf{Estimate of $\delta(I_{\k,2})$.}}
\begin{prop}\label{prop: FG small}
	Under Assumption \ref{ass: primary assumption on the potential} and \ref{ass: secondary hypotheses},
	\begin{equation*}
		\delta(I_{\k,2})=-\frac{15\hat v(0)\nu^{3/2}}{16\pi^2}
		\frac{k}{\sqrt\nu}\left(1+O\left(\frac{k^2}{\nu}\right)\right),
		\qquad\frac{k}{\sqrt\nu}\longrightarrow0.
	\end{equation*}
\end{prop}
\begin{proof}
	Again we take $\nu=\hat v(0)=1$. On $I_{\k,2}$,
	\begin{align}
		x-\omega_\Bog(k)&\geq\omega_\Bog(k)-2\omega_\Bog(k/2)\geq ck^3,
		\label{eq: ineq for den J}\\
		x+\omega_\Bog(k)&\geq ck.
		\label{eq: ineq for den K}
	\end{align}
	In particular, there is no pole. Lemma \ref{lem: expansion of J and K}
	gives $R,S=O(k^3)$ and the exact square expansion
	\begin{align}
		&\frac{J(k;x+ky,x-ky)^2}{x-\omega_\Bog(k)}
		+\frac{K(k;x+ky,x-ky)^2}{x+\omega_\Bog(k)}\notag\\
		&=8x+4\bigl(R(k;x+ky,x-ky)-S(k;x+ky,x-ky)\bigr)\notag\\
		&\quad+\frac{R(k;x+ky,x-ky)^2}{x-\omega_\Bog(k)}
		+\frac{S(k;x+ky,x-ky)^2}{x+\omega_\Bog(k)}
		=8x+O(k^3).
		\label{eq: second line xy}
	\end{align}
	We integrate this expression in the momentum coordinates $(t,s)$.
	Uniformly for $k/2\leq t\leq2k$ and $|s|\leq\frac{1}{2}$,
	\begin{align*}
		x&=\omega_\Bog(t+ks)+\omega_\Bog(t-ks)=2t+O(k^3),\\
		h(t+ks)h(t-ks)&=1+O(k^2),\qquad
		\d\p=2(t^2-k^2s^2)\,\d\varphi\,\d t\,\d s.
	\end{align*}
	Thus the momentum integrand on $I_{\k,2}$ satisfies
	\begin{equation*}
		h(t+ks)h(t-ks)\bigl[8x+O(k^3)\bigr]=16t+O(k^3).
	\end{equation*}
	Since the lower energy cutoff is $k+O(k^3)$, the actual domain
	satisfies, after the angular integration,
	\begin{equation*}
		[k/2+Ck^3,2k[\times[-\frac{1}{2},\frac{1}{2}]
		\subset I_{\k,2}\subset[k/2,2k[\times[-\frac{1}{2},\frac{1}{2}].
	\end{equation*}
	The omitted strip has area $O(k^3)$. Its leading integrand is $O(k)$,
	while the integrated remainder is $O(k^3)O(k)=O(k^4)$. Therefore
	\begin{align*}
		\delta(I_{\k,2})
		&=-\frac1{32\pi^2\omega_\Bog(k)}
		\Bigl[\int_{k/2}^{2k}\d t\int_{-\frac{1}{2}}^{\frac{1}{2}}\d s\,16t+O(k^4)\Bigr]\\
		&=-\frac{30k^2+O(k^4)}{32\pi^2\omega_\Bog(k)}
		=-\frac{15}{16\pi^2}k+O(k^3).
	\end{align*}
	The rescaling of Remark \ref{rmk: nu dependence} gives the stated result.
\end{proof}

It remains to estimate $\delta(I_{\k,3})$. For this purpose it is convenient to use the momentum coordinates introduced in Definition  \eqref{eq: change in t s}. With this coordinates, the relevant integral ($\nu=\hat{v}(0)= 1$)becomes \begin{align}
	&\delta(I_{\k,3}) = 	
	-\frac{1}{\omega_\Bog(k)}\frac{1}{32\pi^2 }\int_{2k}^{+\infty} \d t \int^{\frac{1}{2}}_{-\frac{1}{2}}\d s \frac{t^2-k^2s^2}{\omega_\Bog\left(t+sk\right)\omega_\Bog\left(t-sk\right)} \notag \\ &\times\left(\frac{J\left(k; t+sk, t-sk \right)^2}{\omega_\Bog(t+sk)+\omega_\Bog(t-sk)-\omega_\Bog(k)}+ \frac{K\left(k, t+sk, t-sk \right)^2}{\omega_\Bog(t+sk)+\omega_\Bog(t-sk)+\omega_\Bog(k)}\right) \label{eq: real divergence fermi3}
\end{align}

To determine the small $\k$-behavior of this integral, we expand the last line of Equation \eqref{eq: real divergence fermi3}  using  \eqref{eq: final form for Jtt} and \eqref{eq: final form for Ktt}. As a preliminary step, we write  it as \begin{align}
	&\frac{J(k;t,t)^2}{\omega_\Bog(t+sk)+\omega_\Bog(t-sk)-\omega_\Bog(k)} +\frac{K(k,t,t)^2}{\omega_\Bog(t+sk)+\omega_\Bog(t-sk)+\omega_\Bog(k)} \label{eq: most relevant}\\
	&+S_\shf(k,t,sk) \label{eq: relevant 1} + R_3(k,t,sk),
\end{align}
where we have defined the \textit{shift} term
\begin{equation}\label{eq: shift}
	S_\shf(k,t,sk)
	:=4k^2s^2\hat v(t)\bigl(S_J(k,t,sk)-S_K(k,t,sk)\bigr).
\end{equation}
Expanding the squares in \eqref{eq: exp Jtt} and \eqref{eq: exp Ktt},
and then using \eqref{eq: final form for Jtt} and
\eqref{eq: final form for Ktt}, gives the exact remainder
\begin{align}
	R_3(k,t,sk)
	={}&\frac{\hat v(t)\omega_\Bog(k)t^2}{\omega_\Bog(t)}
	k^2s^2\bigl(S_J(k,t,sk)+S_K(k,t,sk)\bigr)\notag\\
	&+\frac{4\hat v(t)k^2s^2S_J(k,t,sk)}
	{\omega_\Bog(t+sk)+\omega_\Bog(t-sk)-\omega_\Bog(k)}\notag\\
	&\qquad\times\Bigl[
	R_J(k,t)-s^2k^2R_\omega(t,sk)
	\left(1+\frac{\omega_\Bog(k)t^2}{4\omega_\Bog(t)}\right)
	\Bigr]\notag\\
	&+\frac{k^4s^4S_J(k,t,sk)^2}
	{\omega_\Bog(t+sk)+\omega_\Bog(t-sk)-\omega_\Bog(k)}
	\label{eq: zero remainder term ts}\\
	&+\frac{4\hat v(t)k^2s^2S_K(k,t,sk)}
	{\omega_\Bog(t+sk)+\omega_\Bog(t-sk)+\omega_\Bog(k)}\notag\\
	&\qquad\times\Bigl[
	s^2k^2R_\omega(t,sk)
	\left(1-\frac{\omega_\Bog(k)t^2}{4\omega_\Bog(t)}\right)
	-R_K(k,t)\Bigr]\notag\\
	&+\frac{k^4s^4S_K(k,t,sk)^2}
	{\omega_\Bog(t+sk)+\omega_\Bog(t-sk)+\omega_\Bog(k)}.
	\label{eq: third line of remainders ts}
\end{align}
To extract the same infrared term as in the Hellmann--Feynman
contribution, we first use $c(k)+s(k)=2\frac{\omega_\Bog(k)}{k}$ in
\eqref{eq: sum of RJ and RK}. This gives the exact identity
\begin{equation*}
	R_J(k,t)+R_K(k,t)
	=\frac{\omega_\Bog(k)^2t^2}{2\omega_\Bog(t)}
	+4\omega_\Bog(t)\left(\frac{\omega_\Bog(k)}k-1\right).
\end{equation*}
Together with
\begin{equation*}
	2\frac{\omega_\Bog(k)}k-1
	=\frac{\omega_\Bog(k)^2}{k^2}
	-\left(\frac{\omega_\Bog(k)}k-1\right)^2,
\end{equation*}
this allows us to absorb the entire sum $R_J+R_K$ into the main
term and a remainder containing $\left(\frac{\omega_\Bog(k)}{k}-1\right)^2$.
The diagonal
contribution \eqref{eq: most relevant} is then decomposed exactly as
\begin{align}
	\eqref{eq: most relevant}
	={}&S_\mai(k,t)\label{eq: most relevant 2}\\
	&-8\hat v(t)^2s^2k^2R_\omega(t,sk)
	+S_3(k,t,sk),\label{eq: relevant J+Ktt}
\end{align}
where
\begin{equation}\label{eq: main term}
	S_\mai(k,t)
	:=16\hat v(t)^2\omega_\Bog(t)\frac{\omega_\Bog(k)^2}{k^2}
	+\frac{\hat v(t)^2\omega_\Bog(k)^2t^4}{\omega_\Bog(t)}
\end{equation}
and
\begin{align}
	S_3(k,t,sk)
	={}&\frac{2\hat v(t)^2t^2}{\omega_\Bog(t)}
	\omega_\Bog(k)\bigl(R_J(k,t)-R_K(k,t)\bigr)\notag\\
	&-\frac{\hat v(t)^2t^4R_\omega(t,sk)}{2\omega_\Bog(t)^2}
	s^2\omega_\Bog(k)^2k^2\notag\\
	&-16\hat v(t)^2\omega_\Bog(t)
	\left(\frac{\omega_\Bog(k)}k-1\right)^2
	\label{eq: remainder tt1}\\
	&+4\hat v(t)^2
	\frac{\Bigl[R_J(k,t)-s^2k^2R_\omega(t,sk)
		\left(1+\frac{\omega_\Bog(k)t^2}{4\omega_\Bog(t)}\right)\Bigr]^2}
	{\omega_\Bog(t+sk)+\omega_\Bog(t-sk)-\omega_\Bog(k)}
	\label{eq: remainder tt2}\\
	&+4\hat v(t)^2
	\frac{\Bigl[R_K(k,t)-s^2k^2R_\omega(t,sk)
		\left(1-\frac{\omega_\Bog(k)t^2}{4\omega_\Bog(t)}\right)\Bigr]^2}
	{\omega_\Bog(t+sk)+\omega_\Bog(t-sk)+\omega_\Bog(k)}.
	\label{eq: remainder tt3}
\end{align}
\paragraph{\textbf{Estimate of $\delta(I_{\k,3})$.}}
We work in the normalization $\nu=\hat v(0)=1$ and retain
Assumptions \ref{ass: primary assumption on the potential} and
\ref{ass: secondary hypotheses}.

The boundedness and continuity of $\hat v$, its behavior at the origin,
and stability imply
\begin{equation*}
	C^{-1}t(1+t)\leq\omega_\Bog(t)\leq Ct(1+t),\qquad t>0.
\end{equation*}
Throughout the estimates, $0<k\leq k_0<$ is sufficiently small,
$t\geq2k$, and $|s|\leq\frac{1}{2}$. Constants may depend on the fixed
potential but are independent of $k,t,s$.

We shall use
\begin{align}
	\omega_\Bog(t+sk)+\omega_\Bog(t-sk)\pm\omega_\Bog(k)
	&\geq C^{-1}t(1+t),\label{eq: uniform denominator I3}\\
	h(t+sk)h(t-sk)&\leq\frac{C}{1+t^2},\notag\\
	\frac{t^2}{\omega_\Bog(t)}&\leq C\min\{t,1\}.\notag
\end{align}
For small $t$, the first inequality follows from
$\omega_\Bog(t)=t+O(t^3)$ and $2k\leq t$.
Away from the origin, the explicit expression of 
$\omega_\Bog$ give the same bound.

We also recall the exact Taylor formula for the density factor:
\begin{equation}
	h(t+sk)h(t-sk)=h(t)^2+s^2k^2R_h(t,sk),
\end{equation}
where
\begin{align}
	R_h(t,sk)
	=&\frac12\int_{-1}^{1}\d v\,(1-|v|)
	\Bigl[h''(t+vsk)h(t-vsk)\notag \\
	&+h(t+vsk)h''(t-vsk)-2h'(t+vsk)h'(t-vsk)\Bigr].
\end{align}
By Lemma \ref{lem: asymptotic of h}
and Corollaries \ref{cor: Romega} and \ref{lem: corollary for Rh}  
\begin{equation}
	|R_h(t,sk)|\leq C\min\{1,t^{-4}\},\qquad
	|R_\omega(t,sk)|\leq C.
\end{equation}

The following four lemmas estimate the main term, $S_3$, the shift term $S_\mathrm{shift}$,
and $R_3$, respectively.

\begin{lemma}\label{prop: main term1}
	Under Assumptions \ref{ass: primary assumption on the potential} and \ref{ass: secondary hypotheses} wuth $\nu = \hat{v}(0)=1$, 
	\begin{align}
		&-\frac1{32\pi^2\omega_\Bog(k)}
		\int_{2k}^{\infty}\d t\int_{-\frac{1}{2}}^{\frac{1}{2}}\d s\,
		h(t+sk)h(t-sk)\Bigl[S_\mai(k,t)\notag\\
		&\hspace{35mm}-8\hat v(t)^2s^2k^2R_\omega(t,sk)\Bigr]
		\label{eq: main term to estimate}\\
		&=-c_\ir(k,1,\hat v)+kc_\mai(\hat v)+o(k),
		\qquad k\to0.
		\label{eq: infrared statement}
	\end{align}
	Here
	\begin{equation}
		c_\ir(k,1,\hat v)
		=\frac{\omega_\Bog(k)}{2\pi^2k^2}
		\int_0^\infty\d t\,\frac{\hat v(t)^2t^2}{\omega_\Bog(t)},
	\end{equation}
	as in \eqref{cir}, and
	\begin{align*}
		c_\mai(\hat v)
		={}&\frac1{\pi^2}
		-\int_0^\infty\frac{\d t}{24\pi^2}\,
		\hat v(t)^2\omega_\Bog(t)R_h(t,0)\\
		&-\int_0^\infty\frac{\d t}{32\pi^2}\,
		\frac{\hat v(t)^2t^6}{\omega_\Bog(t)^3}
		+\int_0^\infty\frac{\d t}{48\pi^2}\,
		\frac{\hat v(t)^2t^2}{\omega_\Bog(t)^2}R_\omega(t,0).
	\end{align*}
\end{lemma}

\begin{proof}
	The first term in $S_\mai$ gives the exact decomposition
	\begin{align}
		&-\frac{\omega_\Bog(k)}{2\pi^2k^2}
		\int_{2k}^\infty\d t\int_{-\frac{1}{2}}^{\frac{1}{2}}\d s\,
		h(t+sk)h(t-sk)\hat v(t)^2\omega_\Bog(t)\notag\\
		&=\frac{\omega_\Bog(k)}{2\pi^2k^2}
		\int_0^{2k}\d t\,\frac{t^2\hat v(t)^2}{\omega_\Bog(t)}
		\label{eq: small momenta}\\
		&\quad-\frac{\omega_\Bog(k)}{2\pi^2}
		\int_{2k}^\infty\d t\int_{-\frac{1}{2}}^{\frac{1}{2}}\d s\,
		s^2R_h(t,sk)\hat v(t)^2\omega_\Bog(t)
		\label{eq: remainder h}\\
		&\quad-c_\ir(k,1,\hat v).
		\label{eq: infrared}
	\end{align}
	Since $\frac{t^2\hat v(t)^2}{\omega_\Bog(t)}=t+O(t^3)$ at the origin,
	\begin{equation*}
		\eqref{eq: small momenta}=\frac{k}{\pi^2}+O(k^3).
	\end{equation*}
	Moreover,
	\begin{equation*}
		|R_h(t,sk)\hat v(t)^2\omega_\Bog(t)|
		\leq C\hat v(t)^2\min\{1,t^{-2}\}.
	\end{equation*}
	This is an integrable upper bound independent of $k,s$. Thus dominated
	convergence, together with $\omega_\Bog(k)=k+O(k^3)$, yields
	\begin{equation*}
		\eqref{eq: remainder h}
		=-\frac{k}{24\pi^2}\int_0^\infty\d t\,
		\hat v(t)^2\omega_\Bog(t)R_h(t,0)+o(k).
	\end{equation*}
	
	For the second term in $S_\mai$, use the exact identity
	\begin{align}
		&\omega_\Bog(k)^2\hat v(t)^2h(t+sk)h(t-sk)
		\frac{t^4}{\omega_\Bog(t)}\notag\\
		&=\omega_\Bog(k)^2\hat v(t)^2h(t)^2
		\frac{t^4}{\omega_\Bog(t)}
		\label{eq: second and third 1}\\
		&\quad+\omega_\Bog(k)^2s^2k^2\hat v(t)^2R_h(t,sk)
		\frac{t^4}{\omega_\Bog(t)}.
		\label{eq: second and third 2}
	\end{align}
	The absolute value of \eqref{eq: second and third 2} is bounded by
	$Ck^4\hat v(t)^2\min\{1,t^{-2}\}$. Its contribution after integration
	and multiplication by $\omega_\Bog(k)^{-1}$ is therefore $O(k^3)$.
	The integrand in \eqref{eq: second and third 1} is $O(t^3)$ at the origin and
	$O(\hat v(t)^2)$ at infinity. Extending its integral from $[2k,\infty[$
	to $[0,\infty[$ therefore changes the energy contribution by $O(k^5)$.
	Replacing the remaining factor $\omega_\Bog(k)$ by $k$ introduces an
	$O(k^3)$ error. Consequently,
	\begin{align*}
		&-\frac{\omega_\Bog(k)}{32\pi^2}
		\int_{2k}^\infty\d t\int_{-\frac{1}{2}}^{\frac{1}{2}}\d s\,
		\hat v(t)^2h(t+sk)h(t-sk)\frac{t^4}{\omega_\Bog(t)}\\
		&\quad=-\frac{k}{32\pi^2}\int_0^\infty\d t\,
		\frac{\hat v(t)^2t^6}{\omega_\Bog(t)^3}+O(k^3).
	\end{align*}
	
	The explicitly retained $R_\omega$ term is handled by dominated
	convergence, using
	$h(t+sk)h(t-sk)|R_\omega(t,sk)|\leq C(1+t^2)^{-1}$:
	\begin{align*}
		&\frac{k^2}{4\pi^2\omega_\Bog(k)}
		\int_{2k}^\infty\d t\int_{-\frac{1}{2}}^{\frac{1}{2}}\d s\,
		s^2\hat v(t)^2h(t+sk)h(t-sk)R_\omega(t,sk)\\
		&\quad=\frac{k}{48\pi^2}\int_0^\infty\d t\,
		\frac{\hat v(t)^2t^2}{\omega_\Bog(t)^2}R_\omega(t,0)+o(k).
	\end{align*}
	
	Combining these three contributions proves the stated formula for $c_\mai$.
\end{proof}

\begin{lemma}\label{prop: remainder 1}
	Under Assumptions \ref{ass: primary assumption on the potential} and \ref{ass: secondary hypotheses} wuth $\nu = \hat{v}(0)=1$, 
	\begin{align*}
		\frac1{\omega_\Bog(k)}
		\int_{2k}^\infty\d t\int_{-\frac{1}{2}}^{\frac{1}{2}}\d s\,
		h(t+sk)h(t-sk)|S_3(k,t,sk)|\leq Ck^3\bigl(1+|\log k|\bigr)=o(k).
	\end{align*}
\end{lemma}

\begin{proof} We first estimate the three terms in Equation \eqref{eq: remainder tt1}.
	By \eqref{eq: difference of RJ and RK},
	$|R_J(k,t)-R_K(k,t)|\leq Ck^3(1+t^2)$.  By corollary $\ref{cor: Romega}$ we have  $$\left| \frac{\hat{v}(t)^2t^4 R_\omega(t,sk)}{2\omega_\Bog(t)^2}s^2\omega_\Bog(k)^2k^2\right|\leq Ck^4 \hat{v}(t)^2.$$
	Moreover, $\frac{\omega_\Bog(k)}{k}-1=O(k^2)$, so that
	\begin{equation*}
		16\hat v(t)^2\omega_\Bog(t)
		\left(\frac{\omega_\Bog(k)}k-1\right)^2
		\leq Ck^4\hat v(t)^2(1+t^2).
	\end{equation*}
	Thus the absolute value of the three terms in
	\eqref{eq: remainder tt1} is at most $Ck^4\hat v(t)^2(1+t^2).$
	Multiplying by the $h(t+sk)h(t-sk)$ and integrating gives an $O(k^3)$
	contribution, because $\hat v\in L^2(0,\infty)$.
	
	For the two square terms \eqref{eq: remainder tt2}, \eqref{eq: remainder tt3}, Lemma \ref{eq: upper bound for RJ and RK}
	and the bounds above give
	\begin{align*}
		&\left|R_J(k,t)-s^2k^2R_\omega(t,sk)
		\left(1+\frac{\omega_\Bog(k)t^2}{4\omega_\Bog(t)}\right)\right|^2
		\leq Ck^4(1+t^2)^2,\\
		&\left|R_K(k,t)-s^2k^2R_\omega(t,sk)
		\left(1-\frac{\omega_\Bog(k)t^2}{4\omega_\Bog(t)}\right)\right|^2
		\leq Ck^4(1+t^2)^2.
	\end{align*}
	It follows from the lower bound in \eqref{eq: uniform denominator I3} that
	\begin{align*}
		&\frac1{\omega_\Bog(k)}
		\int_{2k}^\infty\d t\int_{-\frac{1}{2}}^{\frac{1}{2}}\d s\,
		h(t+sk)h(t-sk)|S_3(k,t,sk)|\\
		&\quad\leq Ck^3\int_0^\infty\hat v(t)^2\,\d t
		+Ck^3\int_{2k}^\infty\d t\,
		\hat v(t)^2\frac{1+t^2}{t(1+t)}\\
		&\quad\leq Ck^3\left(1+|\log k|
		+\int_1^\infty\hat v(t)^2\,\d t\right).
	\end{align*}
	This proves the assertion.
\end{proof}

\begin{lemma}\label{prop: main term 2}
	Under Assumptions \ref{ass: primary assumption on the potential} and \ref{ass: secondary hypotheses} wuth $\nu = \hat{v}(0)=1$, 
	\begin{equation*}
		-\frac1{32\pi^2\omega_\Bog(k)}
		\int_{2k}^\infty\d t\int_{-\frac{1}{2}}^{\frac{1}{2}}\d s\,
		h(t+sk)h(t-sk)S_\shf(k,t,sk)
		=kc_\shf(\hat v)+o(k),
	\end{equation*}
	where
	\begin{align*}
		c_\shf(\hat v)
		=&-\frac1{48\pi^2}\int_0^\infty\d t\,
		\frac{\hat v(t)t^2}{\omega_\Bog(t)^2}S_J(0;t,0)\\
		=&-\frac1{48\pi^2}\int_0^\infty\d t\,
		\Biggl[\frac{8\hat v(t)t^2}{\omega_\Bog(t)^2}
		\left(\frac{\hat v'(t)\omega_\Bog(t)}t
		+\frac{\hat v(t)\omega_\Bog(t)}{t^2}
		-\frac{\omega_\Bog'(t)\hat v(t)}t\right)\\
		&+\frac{2\hat v(t)t^2}{\omega_\Bog(t)^2}
		\Bigl(\hat v''(t)\omega_\Bog(t)
		+\hat v(t)\omega_\Bog''(t)
		-2\omega_\Bog'(t)\hat v'(t)\Bigr)\Biggr].
	\end{align*}
\end{lemma}

\begin{proof}
	We use the function $	\Phi(t)= |\hat{v}(t)|+ t|\hat{v}'(t)|+t^2 |\hat{v}''(t)|$.
	By Lemma \ref{lemma: estimates SJ and SK},
	\begin{align}
		&h(t+sk)h(t-sk)s^2|\hat v(t)S_{J/K}(k,t,sk)|\notag\\
		\leq& C|\hat v(t)|\min\{1,t^{-2}\}
		\int_{-1}^{1}\d v\,(1-|v|)\Phi(t+vsk).\label{eq: RH integral}
	\end{align}
	On any interval $[x,\infty[$, $x>1$, Cauchy--Schwarz bounds, bound the integral of the
	right-hand side by
	\begin{align}
		&C\left(\int_x^\infty\frac{\hat v(t)^2}{t^2}\,\d t\right)^{\frac{1}{2}}
		\left(\int_x^\infty\frac{\d t}{t^2}
		\Bigl[\int_{-1}^{1}\d v\,(1-|v|)\Phi(t+vsk)\Bigr]^2\right)^{\frac{1}{2}}\notag \\ 
		=&C\left(\int_{x}^\infty\frac{\hat v(t)^2}{t^2}\,\d t\right)^{\frac{1}{2}}
		\left(\int_{1-svk}^\infty\frac{\d t}{(t-svk)^2}
		\int_{-1}^{1}\d v\,(1-|v|)\Phi(t+vsk)^2\right)^{\frac{1}{2}}\notag	\\
		\leq &D\left(\int_x^\infty\frac{\hat v(t)^2}{t^2}\,\d t\right)^{\frac{1}{2}} \left(\int_{\frac{1}{2}}^\infty\frac{\Phi(t)^2}{t^2}\,\d t\right)^{\frac{1}{2}}<\infty.
		\label{eq: uniform weighted SJ SK}
	\end{align} where we first used Jensen inequality and then we changed variable $t+vsk \to t$ in the third one. The first factor in \eqref{eq: uniform weighted SJ SK} goes to zero as $x\to+\infty$, and both factors are uniformily bounded in $s,k$.
	
	For each fixed $t>0$, continuity of $S_{J/K}$ 
	gives
	\begin{equation*}
		S_J(k,t,sk)\longrightarrow S_J(0;t,0),\qquad
		S_K(k,t,sk)\longrightarrow S_K(0,t,0).
	\end{equation*}
	On every bounded interval, \eqref{eq: RH integral} is uniformly
	bounded, including near the origin. Dominated convergence there,
	together with uniform smallness of the tails in Equation \eqref{eq: uniform weighted SJ SK}, proves
	\begin{align*}
		\lim_{k\to 0}&\int_{2k}^\infty\d t\int_{-\frac{1}{2}}^{\frac{1}{2}}\d s\,
		h(t+sk)h(t-sk)s^2\hat v(t)S_{J/K}(k,t,sk)\\
		=&
		\frac1{12}\int_0^\infty\d t\,h(t)^2\hat v(t)S_{J/K}(0,t,0).
	\end{align*}
	Using $S_K(0,t,0)=-S_J(0;t,0)$, we consequently obtain
	\begin{align*}
		&\int_{2k}^\infty\d t\int_{-\frac{1}{2}}^{\frac{1}{2}}\d s\,
		h(t+sk)h(t-sk)S_\shf(k,t,sk)\\
		=&\frac{2k^2}{3}\int_0^\infty\d t\,
		h(t)^2\hat v(t)S_J(0;t,0)+o(k^2).
	\end{align*}
	Multiplication by $-\frac{1}{32\pi^2\omega_\Bog(k)}$ proves the assertion.
	The explicit formula for $c_\shf$ follows from \eqref{eq: SJ00}.
\end{proof}

\begin{lemma}\label{prop: remainder  2}
	Under Assumptions \ref{ass: primary assumption on the potential} and \ref{ass: secondary hypotheses} wuth $\nu = \hat{v}(0)=1$, 
	\begin{equation*}
		\frac1{\omega_\Bog(k)}\int_{2k}^\infty\d t
		\int_{-\frac{1}{2}}^{\frac{1}{2}}\d s\,h(t+sk)h(t-sk)|R_3(k,t,sk)|\leq Ck^2.
	\end{equation*}
	
\end{lemma}

\begin{proof}
	The exact formula \eqref{eq: zero remainder term ts}--
	\eqref{eq: third line of remainders ts}, the estimates on $R_J,R_K$
	and $R_\omega$, and \eqref{eq: uniform denominator I3} imply
	\begin{align}
		|R_3(k,t,sk)|
		\leq&Ck^3|\hat v(t)|\frac{t^2}{\omega_\Bog(t)}
		\bigl(|S_J(k,t,sk)|+|S_K(k,t,sk)|\bigr)\notag \\
		&+\frac{Ck^4}{t(1+t)}
		\Bigl[|\hat v(t)|(1+t^2)
		\bigl(|S_J(k,t,sk)|+|S_K(k,t,sk)|\bigr)\notag\\
		&+|S_J(k,t,sk)|^2+|S_K(k,t,sk)|^2\Bigr].
	\end{align}
	For $2k\leq t\leq1$, the potential and $S_J,S_K$ are uniformly
	bounded. Hence
	\begin{equation}
		|R_3(k,t,sk)|\leq Ck^3t+\frac{Ck^4}{t},
	\end{equation}
	whose integral, divided by $\omega_\Bog(k)$, is bounded by
	$Ck^2+Ck^3|\log k|$.
	
	For $t\geq1$, Lemma \ref{lemma: estimates SJ and SK} and
	\eqref{eq: uniform weighted SJ SK} give, uniformly in $k,s$,
	\begin{equation}
		\int_1^\infty\frac{\d t}{t^2}
		\bigl(|S_J(k,t,sk)|+|S_K(k,t,sk)|\bigr)^2\leq C.
	\end{equation}
	Moreover, Cauchy--Schwarz yields
	\begin{align}
		&\int_1^\infty\d t\,|\hat v(t)|
		\bigl(|S_J(k,t,sk)|+|S_K(k,t,sk)|\bigr)\notag \\
		\leq&
		\left(\int_1^\infty t^2\hat v(t)^2\,\d t\right)^{\frac{1}{2}}
		\left(\int_1^\infty\frac{\d t}{t^2}
		\bigl(|S_J(k,t,sk)|+|S_K(k,t,sk)|\bigr)^2\right)^{\frac{1}{2}}
		\leq C.
	\end{align}
	Since $\frac{t^2}{\omega_\Bog(t)}$ and $(1+t^2)(t(1+t))^{-1}$ are bounded
	on this interval, its contribution is at most
	$\frac{C(k^3+k^4)}{\omega_\Bog(k)}\leq Ck^2$. Moreover, The density
	factor is uniformly bounded.
	Combining the two intervals proves the assertion.
\end{proof}

Combining the four lemmas with \eqref{eq: real divergence fermi3},
we obtain, in normalized units,
\begin{equation*}
	\delta(I_{\k,3})
	=-c_\ir(k,1,\hat v)
	+k\bigl(c_\mai(\hat v)+c_\shf(\hat v)\bigr)+o(k).
\end{equation*}
Since the contributions from $I_{\k,1}$ and $I_{\k,2}$ are, respectively,
$o(k)$ and $-\frac{15k}{16\pi^2}+o(k)$, this convention gives
\begin{equation*}
	c_3(1,\hat v)
	=c_\mai(\hat v)+c_\shf(\hat v)-\frac{15}{16\pi^2}.
\end{equation*} which can be seen to coincide with the coefficient in \eqref{eq: combined FGR channels..1}-\eqref{eq: combined FGR channels..} after restoring $\nu$ and $\hat v(0)$ as in
Remark \ref{rmk: nu dependence}.

	\appendix

\section{Scattering length}\label{app: born series} In this
  appedix we
define the scattering length and
derive the Born series, which allows to compute it,  essentially
following \cite{ErdosSchleinYau2008}, with our notation and
conventions.
Note that  our definition differs from the
corresponding one in \cite{ErdosSchleinYau2008} by a factor of $2$ in
front of the potential.

We start from a Schr\"odinger operator $-\Delta+\kappa v(x)$ on $L^2(\rr^3)$
with $v$ sufficiently fast decaying and spherically symmetric. First
we look for a zero-energy generalized eigenfunction of this operator. We make an
ansatz saying that this eigenfunction has the form $1-w(x)$, where
$w$ is of the order $\kappa$, and has the asymptotic form
\beq w(x)\simeq \frac{\mathfrak{a}}{|x|},\quad |x|\to\infty,\eeq
where $\mathfrak{a}$ is called the {\em scattering length}.
Now the eigenvalue equation
\beq\big(-\Delta+\kappa v(\x)\big)\big(1-w(\x)\big)=0\eeq
can be rewritten as
\beq\kappa v=\big(1+\kappa v(-\Delta)^{-1}\big)(-\Delta
w).\eeq
Hence
\beq
-\Delta w=\sum_{n=0}^\infty(-1)^n\kappa^{n+1}\big(v(-\Delta)^{-1}\big)^n v.\eeq
In the momentum representation this can be rewritten as
	\begin{align} &-\widehat{\Delta w}(\p)\,=\,\kappa\hat{v}(\p) \label{eq: scattering legnth
                        2} \\ \notag&+\sum_{k=1}^{+\infty}(-1)^k
                                      \kappa^{k+1}\int
                                      \frac{\d\p_1\dots
                                      \d\p_k}{((2\pi)^{3})^k}\frac{\hat{v}(\p-\p_1)\hat{v}(\p_1-\p_2)\dots
                                      \hat{v}(\p_{k-1}-\p_k)\hat{v}(\p_k)}{|\p_1|^2\dots
                                      |\p_k|^2},\end{align}
                                  Now
                                  \begin{align}
                                 -\widehat{\Delta w}(\0)=  - \int\Delta
                                    w(\x)\d\x&=-\lim_{R\to+\infty}\int_{|\x|<R}\Delta
                                    w(\x)\d
                                    \x\\=\lim_{R\to+\infty}\int_{|\x|=R}\nabla 
                                    w(\x)\cdot\d\vec\Omega_R&=\lim_{R\to+\infty}\int_{|\x|=R} 
                                    \frac{\mathfrak{a}}{R^2}\d\Omega_R=4\pi\mathfrak{a},\end{align}where
                                  $\Omega_R$ is the usual measure on
                                  the sphere of radius $R$ and
                                  $\vec{\Omega}_R$ is the same measure
                                  multiplied by the outward normal vector.
                                  Setting $\p=\0$ we obtain
                                    	\begin{align}\label{eq: scattering legnth 2}
          4\pi\mathfrak{a} = &\kappa\hat{v}(\0) \notag \\ &+\sum_{k=1}^{+\infty}(-1)^k \kappa^{k+1}\int \frac{\d\p_1\dots \d\p_k}{((2\pi)^{3})^k}\frac{\hat{v}(\p_1)\hat{v}(\p_1-\p_2)\dots \hat{v}(\p_{k-1}-\p_k)\hat{v}(\p_k)}{|\p_1|^2\dots |\p_k|^2},\notag\end{align}

			\vspace{5mm}

\paragraph{\textbf{Acknowledgements.}} L. P. is grateful for the support of the National Group of Mathematical Physics (GNFM-INdAM).
\vspace{5mm}

\paragraph{\textbf{Data availability statement.}}
Data sharing is not applicable to this article as no new data were created or analysed in this study.
\vspace{5mm}

\paragraph{\textbf{Conflict of interest statement.}} The authors certify that they have no affiliations with or involvement in any
organization or entity with any financial interest or non-financial interest in
the subject matter discussed in this manuscript.

\end{document}